\documentclass[draftclsnofoot, onecolumn, 12pt]{IEEEtran}

\usepackage{cite}
\usepackage{amsmath,amssymb,amsfonts,acronym}
\usepackage{algorithmic}
\usepackage{textcomp}
\usepackage[usenames]{color}
\usepackage{graphicx}
\usepackage{subcaption}
\usepackage{arydshln}
\usepackage{bm}
\usepackage{url}
\usepackage[table]{xcolor}

\usepackage{graphicx}
\usepackage{pstool}

\definecolor{aliceblue}{rgb}{0.94, 0.97, 1.0}
\definecolor{aliceblue2}{rgb}{0.84, 0.87, 1.0}

\definecolor{green-yellow}{rgb}{0.68, 1.0, 0.18}
\definecolor{almond}{rgb}{0.94, 0.87, 0.8}

\acrodef{3D}{three dimensional}
\acrodef{AP}{access point}
\acrodef{AoD}{angle-of-departure}
\acrodef{D2D}{device-to-device}
\acrodef{DFT}{discrete Fourier transform}
\acrodef{MPC}{multipath component}
\acrodef{TDL}{time delay line}

\acrodef{2D}{two dimensional}
\acrodef{OEB}{orientation error bound}
\acrodef{PEB}{Position Error Bound}

\acrodef{5G}{fifth generation}
\acrodef{6G}{sixth generation}
\acrodef{TX}{transmitter}
\acrodef{RX}{receiver}

\acrodef{AF}{ambiguity function}

\acrodef{FFT}{fast Fourier transform}
\acrodef{TTD}{true-time delay}

\acrodef{CR}{channel response}

\acrodef{GNSS}{Global Navigation Satellite System}

\acrodef{MAP}{maximum a posteriori probability}

\acrodef{MIMO}{multiple-input multiple-output}
\acrodef{mmW}{millimeter-wave}

\acrodef{SIMO}{single-input multiple-output}

\acrodef{MISO}{multiple-input single-output}

\acrodef{SISO}{single-input single-output}
\acrodef{TDMA}{time division multiple access}
\acrodef{FDMA}{frequency division multiple access}
\acrodef{CRLB}{Cramer-Rao Lower Bound}

\acrodef{SCAs}{small cell access points}

\acrodef{SCA}{small cell access point}

\acrodef{BS}{base station}

\acrodef{DF}{detect \& forward}

\acrodef{JF}{just forward}

\acrodef{DS}{delay spread}

\acrodef{CSI}{channel state information}

\acrodef{ACK}{acknowledge}
\acrodef{ADC}{analog-to-digital converter}
\acrodef{AWGN}{additive white Gaussian noise}
\acrodef{BPZF}{band-pass zonal filter}
\acrodef{CDF}{cumulative distribution function}
\acrodef{ch.f.}{characteristic function}
\acrodef{CIR}{channel impulse response}
\acrodef{CRB}{Cram\'{e}r-Rao bound}
\acrodef{DP}{direct path}
\acrodef{ED}{energy detector}
\acrodef{EM}{electromagnetic}
\acrodef{FCC}{Federal Communications Commission}
\acrodef{FIM}{Fisher Information Matrix}
\acrodef{GDOP}{geometric dilution of precision}
\acrodef{GLRT}{generalized likelihood ratio test}
\acrodef{GPS}{Global Positioning System}
\acrodef{INR}{interference-to-noise ratio}
\acrodef{IoT}{Internet of Things}
\acrodef{IR-UWB}{impulse radio UWB}
\acrodef{i.i.d.}{independent, identically distributed}
\acrodef{LIS}{large intelligent surface}
\acrodef{LRT}{likelihood ratio test}
\acrodef{LOS}{line-of-sight}
\acrodef{LLR}{log-likelihood ratio}
\acrodef{LRT}{likelihood ratio test}
\acrodef{LS}{least squares}
\acrodef{MAC}{medium access control}
\acrodef{MRC}{maximal-ratio combining}
\acrodef{MF}{matched filter}
\acrodef{MLE}{maximum likelihood estimator}
\acrodef{mm-wave}{millimeter-waves}
\acrodef{MMSE}{minimum-mean-square-error}
\acrodef{MSE}{mean square error}
\acrodef{MU}{multi-user}
\acrodef{MUI}{multi-user interference}
\acrodef{MUR}{Multistatic RADAR}
\acrodef{NBI}{narrowband interference}
\acrodef{no-lens}{no lens}
\acrodef{NL}{nonlinear}
\acrodef{NR-lens}{non-reconfigurable lens}
\acrodef{NLOS}{non-line-of-sight}
\acrodef{OAM}{Orbital Angular Momentum}
\acrodef{PAM}{pulse amplitude modulation}
\acrodef{PEB}{position error bound}
\acrodef{PDF}{probability distribution function}
\acrodef{PDP}{power delay profile}
\acrodef{PFA}{probability of false alarm}
\acrodef{ppm}{part-per-million}
\acrodef{PPM}{pulse position modulation}
\acrodef{PPP}{Poisson point process}
\acrodef{R-lens}{reconfigurable lens}
\acrodef{RFID}{radio frequency identification}
\acrodef{RIS}{reconfigurable intelligent surface}
\acrodef{RMSE}{root mean square error}
\acrodef{RSS}{received signal strength}
\acrodef{RSSI}{received signal strength indicator}
\acrodef{RTT}{round-trip time}
\acrodef{RV}{random variable}
\acrodef{SIR}{signal-to-interference ratio}
\acrodef{SNR}{signal-to-noise ratio}
\acrodef{SU}{single-user}
\acrodef{TH}{time-hopping}
\acrodef{THz}{terahertz}
\acrodef{TNR}{threshold-to-noise ratio}
\acrodef{UCA}{uniform circular array}
\acrodef{ULA}{uniform linear array}
\acrodef{UWB}{impulse-radio ultrawide bandwidth}
\acrodef{WBI}{wideband interference}
\acrodef{WPAN}{wireless personal area networks}
\acrodef{WSN}{Wireless Sensor Network}
\acrodef{WWLB}{Weiss-Weinstein lower bound}

\acrodef{CW}{continuous wave}

\acrodef{RF}{radiofrequency}

\acrodef{FCC}{Federal Communications Commission}

\acrodef{EIRP}{effective radiated isotropic power}

\acrodef{RCS}{radar cross section}

\acrodef{BAV}{balanced antipodal Vivaldi}

\acrodef{PRake}{partial Rake}

\acrodef{RTLS}{Real-time locating systems}

\acrodef{EFI}{equivalent Fisher information matrix}

\acrodef{SPEB}{squared position error bound}

\acrodef{SOEB}{squared orientation error bound}

\acrodef{Hi-RADIAL}{High-accuracy RAdio Detection, Identification,
And Localization}

\acrodef{HCRB}{hybrid Cram\'{e}r-Rao bound}
\acrodef{HFIM}{hybrid Fisher Information Matrix}

\acrodef{ZZB}{Ziv-Zakai bound}

\acrodef{TOA}{time-of-arrival}
\acrodef{DOA}{direction-of-arrival}

\acrodef{ToF}{time-of-flight}

\acrodef{WSN}{wireless sensor network}

\acrodef{MAC}{medium access control}

\acrodef{RSS}{received signal strength}

\acrodef{TDoA}{time difference-of-arrival}

\acrodef{RF}{radiofrequency}

\acrodef{PSD}{power spectral density}

\acrodef{RTT}{round-trip time}

\acrodef{AOA}{angle-of-arrival}

\acrodef{MF}{matched filter}

\acrodef{ED}{energy detector}

\acrodef{ML}{maximum likelihood}

\acrodef{MUR}{Multistatic radar}

\acrodef{HDSA}{high-definition situation-aware}

\acrodef{RRC}{root raised cosine}

\acrodef{OFDM}{orthogonal frequency division multiplexing}

\acrodef{IF}{intermediate frequency}

\acrodef{PHY}{physical layer}

\acrodef{S-V}{Saleh-Valenzuela}

\acrodef{UHF}{ultra-high frequency}

\acrodef{PR}{pseudo-random}

\acrodef{SoC}{System on Chip}

\acrodef{SoP}{System on Package}

\acrodef{SPMF}{Single-Path Matched Filter}

\acrodef{IMF}{Ideal Matched Filter}

\acrodef{SCR}{signal-to-clutter ratio}

\acrodef{BEP}{bit error probability}

\acrodef{BER}{bit error rate}

\acrodef{WSR}{wireless sensor radar}

\acrodef{HPBW}{half power beam width}

\acrodef{LEO}{localization error outage}

\acrodef{SLAM}{simultaneous localization and mapping}
\acrodef{std}{standard deviation}
\acrodef{WPT}{wireless power transfer}
\acrodef{EV}{electric vehicle}

\def\rm{r_{\hat{m} \hat{k}}}

\def\Narray{N_\text{A}}

\newcommand{\Sur} {y,z}

\newcommand{\tbw}{\mathbf{w}}
\newcommand{\tw}{{w}}

\newcommand{\Ae}{A_\mathrm{e}}
\newcommand{\pu}{\mathbf{p}_\mathrm{u}}
\newcommand{\pin}{\mathbf{p}_i}

\newcommand{\etai}{\eta_\mathrm{int}}
\newcommand{\etau}{\eta_\mathrm{u}}
\newcommand{\etaw}{\eta_\mathrm{w}}

\newcommand{\df} {d_\text{F}}

\newcommand{\Fp}{F_\text{p}}

\newcommand{\p}{\mathbf{p}}
\newcommand{\chitest}{\chi_\mathrm{t}}
\newcommand{\chiu}{\chi_\mathrm{u}}
\newcommand{\ptest}{\mathbf{p}_\mathrm{t}}
\newcommand{\Af}{{A}_\mathrm{f}}

\newcommand{\Drho}{D_\rho}

\newcommand{\Nint}{N_\mathrm{int}}

\newcommand{\ft}{\mathcal{F}}

\makeatletter
\let\old@ps@headings\ps@headings
\let\old@ps@IEEEtitlepagestyle\ps@IEEEtitlepagestyle
\def\confheader#1{%
\def\ps@headings{%
\old@ps@headings%
\def\@oddhead{\strut\hfill#1\hfill\strut}%
\def\@evenhead{\strut\hfill#1\hfill\strut}%
}%
\def\ps@IEEEtitlepagestyle{%
\old@ps@IEEEtitlepagestyle%
\def\@oddhead{\strut\hfill#1\hfill\strut}%
\def\@evenhead{\strut\hfill#1\hfill\strut}%
}%
\ps@headings%
}
\makeatother

\makeatletter
\begin{document}

%\markboth{submitted for Review to IEEE Transactions on Wireless Communications}{F. Guidi and D. Dardari: ...}

\title{Radio Positioning with EM Processing \\ of the Spherical Wavefront}

\author{
    \IEEEauthorblockN{Francesco Guidi~\IEEEmembership{Member,~IEEE,} and Davide Dardari~\IEEEmembership{Senior,~IEEE}  }
\IEEEcompsocitemizethanks{\IEEEcompsocthanksitem F. Guidi is with the National Research Council (CNR) of Italy, Institute of Electronics, Computer and Telecommunication Engineering (IEIIT), via del Risorgimento 2, 40136 Bologna, Italy. 

D. Dardari is with 
the 
%Dipartimento di Ingegneria dell'Energia Elettrica e dell'Informazione 
Department of Electrical, Electronic, and Information Engineering
``Guglielmo Marconi" - DEI, University of Bologna,Via Venezia 52, 47521 Cesena, ITALY. 
(e-mail: francesco.guidi@ieiit.cnr.it, davide.dardari@unibo.it).}
\thanks{The work has been partially funded by the European Commission under
H2020 project XCYCLE (grant nr. 635975). This is also part of ATTRACT that has received funding from the European Union's Horizon 2020 Research and Innovation Programme.}

}
%% use for special paper notices


\maketitle

\begin{abstract}
%\boldmath
%\textcolor{red}{Da fare alla fine}

Next 5G and beyond applications have attracted a tremendous interest towards systems using
antenna arrays with an extremely large number of antennas where the technology conceived
for communication might also be exploited for high-accuracy positioning applications. In this paper we
investigate the possibility to infer the position of a single antenna transmitter using a single asynchronous
receiving node by retrieving information from the incident spherical wavefront. To this end, we consider
the adoption of a suitable mix of processing at electromagnetic {(EM)} and signal levels, as a lower
complexity alternative to classical massive array systems where the processing is done entirely at signal
level. Thus, we first introduce a dedicated general model for different EM processing architectures,
and successively we investigate {the attainable positioning performance according to the use or not of a lens that can have either a reconfigurable or a fixed
phase profile.} The effect of the interference is also investigated to evaluate the robustness of the
considered system to the presence of multiple simultaneous transmitting sources. Results, obtained for
different apertures of the exploited lens/array, confirm the possibility to achieve interesting
positioning performance using a single antenna array with limited aperture.
\end{abstract}


\bstctlcite{IEEEexample:BSTcontrol}

\IEEEpeerreviewmaketitle

\vspace{7pt}

\section{Introduction}
Indoor positioning systems have attracted a great interest in a large variety of scenarios because of the possibility of performing high accuracy {users'} localization even when \ac{GNSS} signal is not {available} and alternative positioning systems are required \cite{win2011network,win2018theoretical,DarCloDju:J15}. 
%In fact, even if \ac{GNSS} is recognized to be one of the most accurate sources of position information, it is often not available in indoors, a.
Currently, there is a large variety of ad-hoc solutions for indoor localization and tracking \cite{GueDarDju:J20-2,ConEtAl:J19,liu2007survey,GueDarDju:J20}, spanning from the systems based on ultrasounds or to more recent \ac{IR-UWB} techniques \cite{WymEtAl:J12,YuShen:J18,liu2007survey,van2015machine}.  
%In addition, solutions
%for \ac{SLAM} have been tackled, mainly using laser and camera-based technologies \cite{durrant2006simultaneous}.
Unfortunately, most of such solutions require that a mobile node is detected at least from three {reference nodes (multiple anchor nodes) located in known positions}, with the need to realize ad-hoc and often redundant infrastructures that might become {not convenient in many indoor scenarios}.  {Consequently, it would be of great help if the networks deployed for communication  could be also used for indoor localization, but such networks usually consist of a single access point designed to guarantee single-anchor coverage.}
%or they fail in scarce visibility conditions (e.g., in presence of smoke). {\color{blue}?? }

{In this context, next \ac{6G} mobile wireless networks will introduce new technologies based on the massive deployment of antennas into small areas, allowing to boost not only communication but also single-anchor localization capabilities at an unprecedented scale of accuracy \cite{guidi2018indoor,RohEtAl:J14,GuiGueDar:J16,witrisal2016high,GuiEtAl:J17,abu2020single,GuerraBackscattering18}. In general,  direct positioning approaches with single antenna arrays provide better performance than two-step localization algorithms as the latter may be suboptimal  according to the data processing inequality \cite{CheWanShe:C17,zhao2018optimal,garcia2017direct}.
Nevertheless, traditional two-step localization algorithms,  based on the estimation of intermediate quantities such as \ac{AOA} and \ac{TOA}, are often implemented in practice as they are more pragmatic and less complex \cite{taponecco2011joint,SheWin:J10,WanWuShe:J19,WanWuShe:J19-2,stoica2005spectral,chuang2015high}. As an example, both ESPRIT and MUSIC-based approaches have  been widely proposed for \ac{AOA} estimate \cite{KoLee:C18}.}
{Unfortunately, such solutions require multiple interactions between transmitter and receiver as well as an extremely precise system synchronization \cite{GueGuiDar:J18}, which could reduce the available bandwidth for communication and make the system  still costly, especially if addressed to \ac{IoT} applications}. 

%Indeed, direct positioning is a timely solution that might help in reducing such issues. 

A possible alternative solution is to infer the transmitter position from the spherical wavefront associated to the signal {transmitted} by the mobile node (source), which is possible in all those situations where the wavefront curvature is significant with respect to the antenna aperture in relation to the wavelength. 
{In fact, while in far-field propagation regime the wavefront is plane and only the \ac{AOA} information can be inferred using an antenna array, when operating in near-field regime (Fresnel region) the wavefront tends to be spherical and also the distance information, e.g. the position,  can be inferred from it.}

This concept is not new, and it has been widely exploited for acoustic waves \cite{LeC:J95,FerWyb:C01} or at microwaves only considering very short distances or using very large (often not practical) antennas \cite{DELMAS2016117}. In \cite{LiaLiu:J10}, the curvature information has been exploited, with a moving source approaching to the receiver so that, entering in the Fresnel region, the incoming wave cannot be regarded as plane anymore. For instance, in \cite{HadSacFas:C17}, an approach using multi-tone signalling and multi-arrays is described, whereas in \cite{ElKetAl:C09-2}, a MUSIC-based method is proposed and an extensive analysis on the attainable fundamental localization limits is derived in near-field propagation conditions \cite{ElKetAl:J13}. 
In addition, other previous works apply the Fresnel approximation to arrays with special geometries, e.g. uniform linear arrays \cite{GroAbYin:J05,Chen2004,DenYinWan:C07,YinEtAl:J17}, and account for the model mismatch while evaluating the positioning performance \cite{SinWanCha:J17}.

With the advent of \ac{mm-wave}-based solutions, direct positioning is in principle possible even with antenna arrays with limited aperture \cite{zhang2018spherical} or by the exploitation of distributed architectures \cite{vukmirovic2018position,vukmirovic2019direct}.
%Several architectures can be of particular interest, exploiting or not the presence of a lens performing the \ac{EM} processing of the incident wavefront.
{However, this requires the presence of a huge amount of antenna elements, each connected to a RF chain, which could lead to an extremely complex receiver, as they usually contain several electronic components, such as analog to digital converters (ADC), mixers, local oscillators etc. \cite{ShaTon:J19}. 
\\
A way to reduce the complexity is by performing some (pre-)processing of the incident wavefront directly at \ac{EM} level through the introduction of \ac{EM} lens, {as fostered by the advent of next \ac{6G} \cite{Samsung6g}}.    
Such \ac{EM} lenses can be either reconfigurable, namely \ac{R-lens}, or not, namely \ac{NR-lens}. According to \ac{NR-lens}, recent studies have investigated the possibility to exploit \ac{EM} lens-based massive arrays operating at \ac{mm-wave} as a promising solution for drastically reducing the overall system complexity \cite{ZenZha:J16}. \ac{NR-lens} can be implemented with different techniques, either including dieletric materials with ad-hoc designed surfaces (e.g., convex lenses), or 
antenna arrays or sub-wavelength spatial phase shifters \cite{ZenZha:J17,ZenZha:J16}.}
By adopting a lens to collimate the beams in precise directions, it is possible to spatially discriminate signals in the analog domain \cite{ZenZha:J16,zeng2014electromagnetic,Gui:C18}. Consequently, thanks to the lens, there is a unique relation between the incident and the output angles of the impinging and refracted waves, respectively. This operation allows reducing the number of antennas with respect to traditional massive arrays, and to move from discrete beamforming architectures towards continuous-aperture phased arrays. Indeed, the use of such lenses has already been investigated only in far-field \cite{ZenZha:J16,zeng2014electromagnetic,Gui:C18}, and their performance has not yet been characterized in the Fresnel region.

{In addition, while {NR-lenses} are passive and thus not reconfigurable, \ac{R-lens} based solutions are programmable in real-time. In this sense, \ac{6G} solutions are moving towards the realization of holographic phase profiles, or of \acp{RIS}, {to put in place a} programmable smart radio environment \cite{hu2018beyond,HuRusEdf:J18,bjornson2019massive,zappone2019wireless,basar2019wireless,di2019smart,ntontin2019reconfigurable}.
{According} to this vision, metamaterials represent an important and recent solution for the realization of \ac{R-lens} \cite{HaiEtAl:J19}}
thanks to the possibility to achieve a flexible control of the EM wavefront \cite{masini2020use,dardari2020communicating,elzanaty2020reconfigurable} while guaranteeing, whenever required, compact size \cite{PfeGrb:J13, PanBalBir:J15}. 
%An interesting solution is  represented by the new concept of \textit{HyperSurfaces}, referring to software-controlled
%metamaterials embedded in any surface in the environment \cite{liaskos2015design}. In this way, it is foreseen a future where walls and objects in the environments can be equipped with tailored hypersurface that interact with the environment, so that a programmable indoor wireless environment can be created \cite{liaskos2018using}.
Other interesting opportunities account for the execution of mathematical operations with layers of metamaterials \cite{silva2014performing}, the realization of electronically reconfigurable transmitarray \cite{DiPEtAl:J17}, or reconfigurable reflectarray technology \cite{HumPer:J14}.

%Recently, the possibility to localize users and targets by means of massive arrays at mm-wave through beamsteering operation as well as ad-hoc techniques has been demonstrated. 

%\paragraph{Realization of Reconfigurable Lenses}

%So far we have assumed the availability of technologies that allow to have a full and perfect control of the phase profile.

%Moving a step-forward, the objective of this manuscript is to exploit the direct estimate of the position of the source at mm-waves through the wavefront curvature by means of both traditional arrays and arrays with an embedded lens, in order to investigate the attainable performance with such a scheme. In particular, thanks to the reduced wavelength, we take advantage of the use of arrays with length shorter than at lower frequencies.
%In this way, it becomes straightforward to determine the capability to spatially discriminate two users in the environment only by retrieving the information from the wavefront curvature.

%
\begin{figure}[t!]
\psfrag{em}[c][c][0.7]{{EM Processing}}
\psfrag{pp}[c][c][0.7]{{(e.g., lens)}}
\psfrag{s}[c][c][0.7]{TX source}
\psfrag{a}[c][c][0.7]{Antenna Array}
\psfrag{sc}[c][c][0.7]{Configuration}
\psfrag{ph}[c][c][0.7]{$\phi_{oyz}$}
\psfrag{x}[c][c][0.7]{$x$}
\psfrag{y}[c][c][0.7]{$y$}
\psfrag{z}[c][c][0.7]{$z$}
\psfrag{o}[c][c][0.7]{$0$}
\psfrag{t}[c][c][0.7]{$\theta$}
\psfrag{t2}[c][c][0.7]{$\theta_{oyz}$}
\psfrag{l}[c][c][0.7]{$z$}
\psfrag{Dy}[c][c][0.7]{$D_y$}
\psfrag{Dz}[c][c][0.7]{$D_z$}
\psfrag{d}[c][c][0.7]{\!\!$d$}
\psfrag{p}[c][c][0.7]{\quad\quad\quad$\p(d,\Theta=(\theta,0))$}
\psfrag{N}[c][c][0.7]{$N_\text{a}$}
\psfrag{d2}[c][c][0.7]{\,\,\,\,\,$d+a$}
\psfrag{d1}[c][c][0.7]{\!\!\!\!$d_{0yz}$}
\centerline{
   \includegraphics[width=0.2\linewidth,draft=false]
    {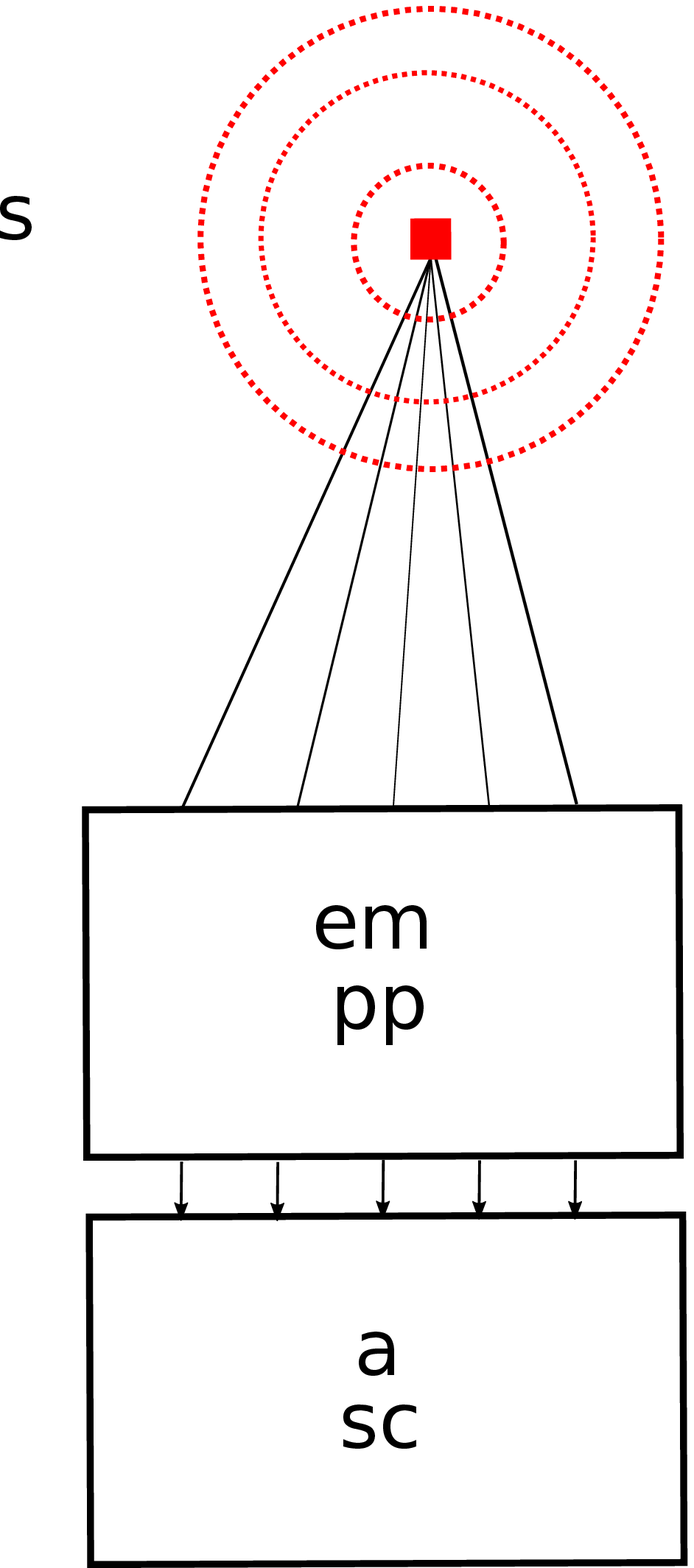}\quad \quad\quad \quad \includegraphics[width=0.35\linewidth,draft=false]
    {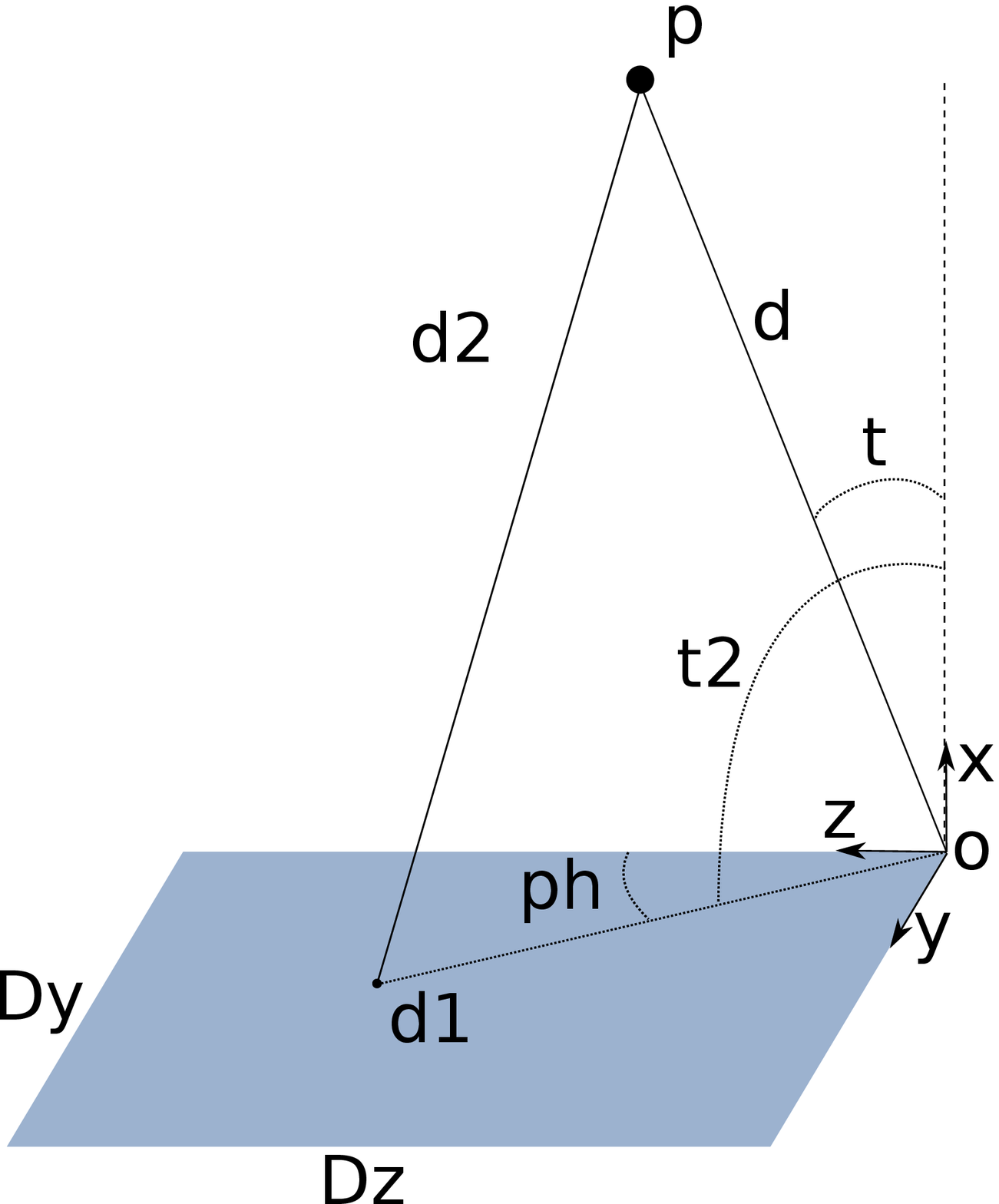}}\caption{Left: general scheme composed of the \ac{EM} processing scheme that focuses the incident wave towards the antenna array. Right: incident wavefront on the EM processing surface.} \label{fig:genericscheme}
\end{figure}

In this context, starting from our analysis in \cite{Gui:C18,dardari2018direct,guerra2020}, here we investigate the radio positioning capabilities of a \ac{mm-wave} source by introducing a generic architecture composed of an \ac{EM} processing section, which can directly operate on the spherical wavefront, and of an array architecture that collects the impinging signal, as shown in Fig.~\ref{fig:genericscheme}.\footnote{{For the sake of simplicity, we assume that it holds $\Theta=(\theta,0)$. The extension to $\phi \neq 0$ is straightforward.}}
More specifically, differently from the state of the art, we first introduce a general model that entails the presence (if any) of an \ac{EM} lens performing the processing at \ac{EM} level before the antenna array. Successively, we compare different architectures according to the presence or absence of the lens, that can be either reconfigurable or not, and to the number of employed antennas. 
%, as a more advanced lens design permits to reduce the number of required antennas. 
{Such a trade-off is then investigated by analyzing the attainable positioning accuracy  and interference rejection capability when the aperture is varied. }
%Thus, we put together in a unique work an analysis comprising the interference rejection and the positioning capabilities for different architectures.

The main contributions of the manuscript can be summarized as follows.
\begin{itemize}
\item 
We provide a generalized framework for retrieving positioning information from the wavefront curvature when \ac{EM} processing is performed prior the processing of signals at each antenna. {Differently from \cite{HuRusEdf:J18}, we propose a general framework that encompasses the kind of antenna array adopted with the add-on (if any) of an external lens capable to partially (or completely) perform the post-processing.}
\item {We consider different practical schemes, employing or not the use of an \ac{EM} lens. In particular, we investigate the effect on the positioning accuracy when the required processing is split between EM level, performed either by using \ac{R-lens} or \ac{NR-lens}, and signal level, i.e., processing the RF signal after an antenna element. 
}
\item 
We evaluate, {through an extensive simulation analysis}, the positioning performance, comprising a differential approach that might unburden the {signal processing} complexity in practical systems when \ac{no-lens} is employed;
\item 
{We propose an ad-hoc multi-user scheme and we evaluate the related impact of the interference} by determining the capability of the proposed architectures to spatially discriminate multiple transmitters {while localizing them}.
%On the other side, when a \ac{NR-lens} based configuration is adopted, we consider the \ac{ML} on the trajectory drawn on the antennas, so that a reduced number of antennas can be considered.
\end{itemize}

The remainder of the paper is organized as follows. Sec.~\ref{sec:trad} contains insights on how to gather position information from the signal wavefront, and Sec.~\ref{sec:positioning} shows positioning techniques. Sec.~\ref{sec:int} reports considerations on the interference, and Sec.~\ref{sec:res} describes the achieved results. Conclusions are finally drawn in Sec.~\ref{sec:con}.
%Sec.~\ref{sec:ml} reports the \ac{ML} estimator derivation, while in Sec.~\ref{sec:res} results are reported. Finally, conclusions are drawn in Sec.~\ref{sec:con}.

\section{Position Information in the Spherical Wavefront}\label{sec:trad}
\subsection{Operating Frequency Impact}

Before introducing the considered architectures, we investigate the trade-off between the size of the array and the operating frequency to determine the region where the impact of the wavefront curvature is appreciable and hence exploitable for positioning. As usually done, we here consider as a delimiter the Fraunhofer distance $\df=2\,D^2/\lambda$, with $\lambda$ indicating the wavelength, and $D$ the antenna diameter.
By assuming antennas spaced apart of $\lambda/2$, we have $\Narray=2\,D/\lambda$.
\begin{figure}[t!]
\psfrag{x}[c][c][0.8]{$f_0$ [GHz]}
\psfrag{y}[c][c][0.8]{$\df$ [m]}
\psfrag{data1111111111111111}[c][c][0.7]{ $D=10\,$cm}
\psfrag{data2}[c][c][0.7]{\quad \quad\quad \quad  \quad \,\,\,$D=20\,$cm}
\psfrag{data3}[c][c][0.7]{\quad \quad\quad \quad \quad \,\,\,$D=30\,$cm}
\psfrag{data4}[c][c][0.7]{\quad \quad\quad \quad \quad \,\,\,$D=40\,$cm}
\psfrag{data5}[c][c][0.7]{\quad \quad\quad \quad \quad \,\,\,$D=50\,$cm}
\psfrag{rata11111111111111}[c][c][0.7]{ $\Narray=100$}
\psfrag{rata2}[c][c][0.7]{\quad \quad\quad \quad\,\, $\Narray=200$}
\psfrag{rata3}[c][c][0.7]{\quad \quad\quad \quad \,\, $\Narray=300$}
\psfrag{rata4}[c][c][0.7]{\quad \quad\quad \quad \,\,\,\,$\Narray=400$}
\psfrag{rata5}[c][c][0.7]{\quad \quad\quad \quad \,\,\,\,$\Narray=500$}
\psfrag{d}[c][c][0.7]{$d$}
\psfrag{p}[c][c][0.7]{\quad$\p$}
\psfrag{N}[c][c][0.7]{$N_\text{a}$}
\psfrag{d2}[c][c][0.7]{\!\!\!\!$d+a$}
\begin{subfigure}[b]{0.5\textwidth}
   \centering 
      \includegraphics[width=0.98\linewidth,draft=false]
    {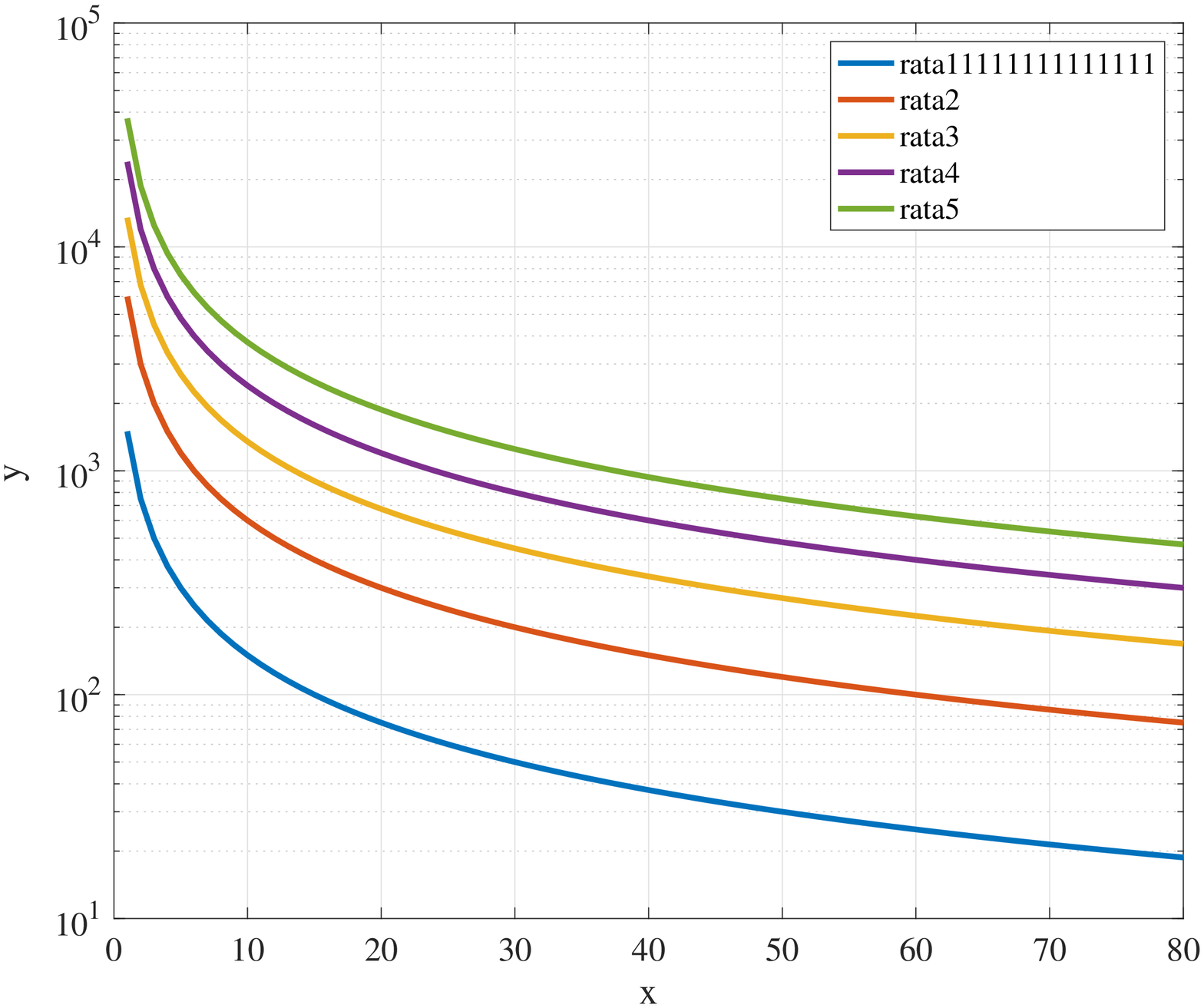}
   \caption{Fixed number of antennas.} 
   \label{fig:NASA_Logo_Sub}
\end{subfigure}% 
\begin{subfigure}[b]{0.5\textwidth}
   \centering 
   \includegraphics[width=0.98\linewidth,draft=false]
    {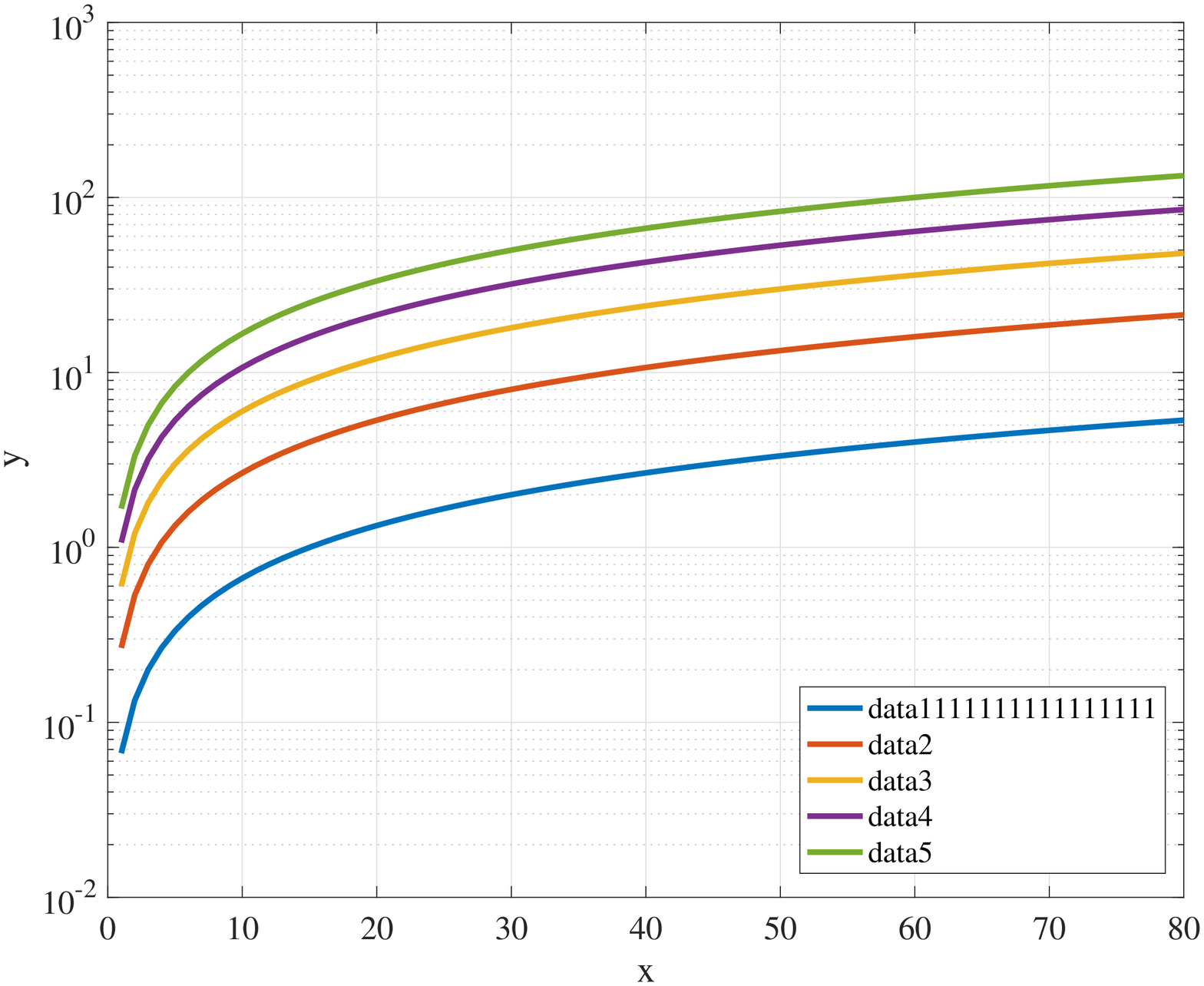}
    \caption{Fixed array size.} 
    \end{subfigure}%
\caption{Starting point of the Fraunhofer region for different frequencies and different array sizes.}\label{fig:fraunhofer} 
\end{figure}

The impact of the operating frequency is reported in Fig.~\ref{fig:fraunhofer} for two scenarios. In the first one, the number of employed antennas is kept constant regardless the frequency or the array size. In the latter, the constraint is on the array dimension, as it happens in practical set up where antenna's size limitations are often present.  
{If the constraint is on $\Narray$ (i.e., on RF chains),} the curvature effect is more appreciable at lower frequencies as reported in Fig.~\ref{fig:fraunhofer}-left. On the other side, if the size of the array is constrained to a certain value, a higher frequency allows to gather a better information on the spherical wavefront (see Fig.~\ref{fig:fraunhofer}-right). As an example, for $D=50\,$cm, the Fraunhofer region, for which the wavefront is considered planar, starts at $d \simeq10\,$m for $f_0=5\,$GHz, and at $d \simeq100\,$m for $f_0=60\,$GHz.
Such considerations confirm that the impact of the wavefront curvature is not negligible when large antenna arrays, operating at high frequency, are employed.

In the following, after describing the considered architectures,  we will illustrate how positioning is possible with only a single antenna array.

%
%\begin{figure}[t!]
%\psfrag{x}[c][c][0.7]{}
%\psfrag{y}[c][c][0.7]{}
%\psfrag{z}[c][c][0.7]{\!\!\!\!\! %\!\!\!\!\!\!\!\!\!\!  flat lens}
%\psfrag{o}[c][c][0.7]{$0$}
%\psfrag{t}[c][c][0.7]{$\theta$}
%\psfrag{l}[c][c][0.7]{$z$}
%\psfrag{a}[c][c][0.7]{array}
%\psfrag{d}[c][c][0.7]{$d$}
%\psfrag{p}[c][c][0.7]{\quad$\p$}
%\psfrag{N}[c][c][0.7]{$D_z$}
%\psfrag{d2}[c][c][0.7]{\!\!\!\!$d+a$}
%\centerline{
%   \includegraphics[width=0.5\linewidth,draft=false]
%    {figures/scenario_holo.eps}}
%\caption{Flat lens retrieving the information on %the position $\p$ from the curvature %wavefront.}\label{fig:scenario} 
%\end{figure}

\subsection{Architectures}
\begin{figure}[t!]
\psfrag{rl}[c][c][0.8]{\textit{reconfigurable lens}}
\psfrag{nrl}[c][c][0.8]{\textit{non-reconfigurable lens}}
\psfrag{nemp}[c][c][0.8]{}
\psfrag{pa}[c][c][0.8]{\textit{planar antenna array}}
\psfrag{x}[c][c][0.6]{$x$}
\psfrag{y}[c][c][0.6]{$y$}
\psfrag{z}[c][c][0.6]{$z$}
\psfrag{ant}[c][c][0.8]{antennas}
\psfrag{fa}[c][c][0.8]{on focal arc}
\psfrag{s}[c][c][0.8]{single}
\psfrag{a}[c][c][0.8]{antenna}
\psfrag{pt}[c][c][0.8]{\,\,phase}
\psfrag{pt2}[c][c][0.8]{profile}
\psfrag{rf}[c][c][0.7]{RF}
\psfrag{s1}[c][c][0.9]{(a) R-lens}
\psfrag{s2}[c][c][0.9]{(b) NR-lens}
\psfrag{s3}[c][c][0.9]{(c) no-lens}
\centerline{
   \includegraphics[width=1\linewidth,draft=false]
    {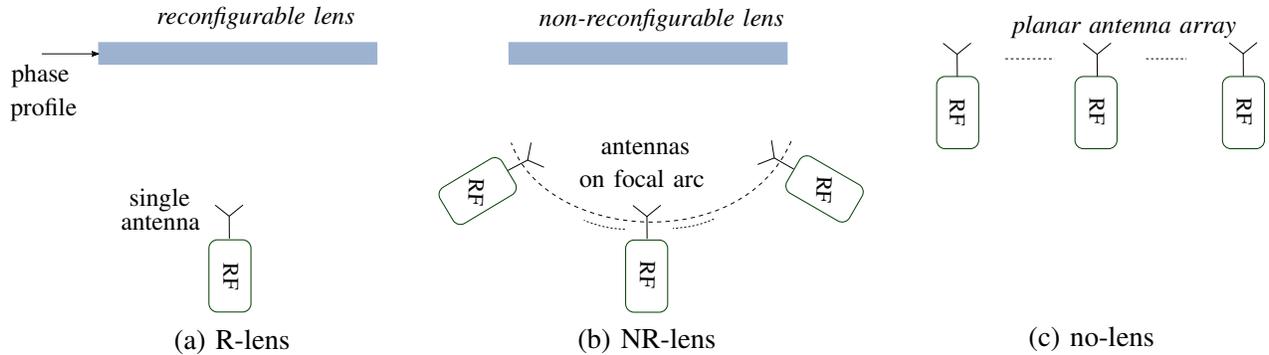}}\caption{{Lateral view of the considered architectures.}} \label{fig:architectures}
\end{figure}

In order to exploit the wavefront curvature for positioning, in this paper we consider three different architectures, reported in Fig.~\ref{fig:architectures}, {involving a device performing the processing at EM level (e.g. a lens) and one or more antenna elements.  
\\
In particular, we aim at investigating the effect on the positioning accuracy when the required processing is split between processing at EM level (e.g., using programmable lens/metasurfaces) and processing at signal level (i.e., processing the RF signal after an antenna element). 
A different balance between EM level and signal level processing (i.e., a different architecture) translates into a different balance in complexity. To one extreme, where no processing is done at EM level, such as in classical antenna arrays, a large number of antenna elements and RF chains (e.g., LNA, ADC, ..) is required and the complexity is completely put on the processing at RF level. 
To the other extreme, a \ac{R-lens} is used and only one antenna/RF chain is required. In this case the complexity is mainly moved to the \ac{EM} part of the system, i.e., the reconfigurable lens. Between these two extremes, we have chosen a \ac{NR-lens} that represents a compromise in terms of number of RF chains ($\Narray$ is lower with respect to the no-lens case) and lens fabrication (once the phase profile has been initially realized, it cannot be reconfigured). 
}

%Despite the three schemes present a different number of antennas, we are interested in investigating the trade-off concerning where the complexity is present in the hardware fabrication. For an architecture entailing only the use of antennas (i.e., \ac{no-lens}), the level of complexity is determined by $\Narray$, i.e., by the RF chains used, each entailing a significant number of electronic components (e.g., low noise amplifier and analog-to-digital-converter) \cite{ShaTon:J19}. 
%On the other side, when the lens is employed, part of such a complexity is reduced and it is moved to the lens realization, according to the fact that is reconfigurable or not. More specifically, the \ac{R-lens} allows to dramatically reduce the number of required RF chains by focusing the incident \ac{EM} wave towards a unique antenna thanks to its reconfigurability. This is paid at the prize of a high complexity required in the lens (comprising the network control module) fabrication. On the contrary, the \ac{no-lens} configuration, which does not make use of any lens and it is adopted as a baseline for performance comparison, entails a large number of antennas, and processing chains, in order to accurately process the received signal (see Fig.~\ref{fig:architectures}).
%\\
%In between, there is the trade-off offered by the \ac{NR-lens}, which makes use of a less complex lens with respect to the \ac{R-lens}, and of a lower number of antennas with respect to the \ac{no-lens}.

In the following, we introduce an ad-hoc general model
%, valid for the presence (if any) of an \ac{EM} lens, 
to retrieve the position information from the incident spherical wavefront.

\subsection{Signal Model}
As previously stated, the \ac{EM} lens can be realized with different techniques (e.g., metamaterials), and it allows to retrieve the position from an omnidirectional source by exploiting the wavefront curvature. 

To this purpose, consider a source transmitting a signal at $f_0$ and located at position $\p$, which is at distance $d$ from the reference point of the RX located in $(0,\,0,\,0)$ ({as shown in Fig.~\ref{fig:genericscheme}-right}), and denote with $\Theta=(\theta,\phi)$ the incident angle.  

Suppose there are $\Narray$ receiving antennas in positions
$\{ \p_n \}$, $n=0, 1, \ldots \Narray-1$. 
Define the surface of the EM lens (if any) lying on the $YZ$-plane as  $\mathcal{S}$, with $z \in [0,\,D_z]$, $y \in [0,\,D_y]$ and {$\Af=D_y\,D_z$}.
Then, we denote with $\mathbf{r} =\left[r_0,\,\ldots ,r_n,\, \ldots r_{\Narray-1} \right]^T$
the vector containing the equivalent complex baseband signal received at each antenna, defined as
%\mathbf{h} &= \left[h_0 (\p),\, h_1(\p), \,\ldots ,h_n (\p),\, \ldots h_{\Narray-1} (\p ) \right]  \nonumber \\
%
\begin{align}\label{eq:genmodel}
\mathbf{r}=\mathbf{s} +  \tbw =\ft_{\mathbf{s}}({h}(\mathcal{S},\p)) +  \tbw \,,
\end{align}
where {$ \tbw = \left[\tw_0 , \,\ldots ,\tw_n,\, \ldots \tw_{\Narray-1} \right]^T $ is the vector containing the \ac{AWGN}, i.e., $\tw_n \sim \mathcal{CN}\left(0,\,\sigma^2\right)$, whereas the useful signal carrying positioning information is}
\begin{align}
\mathbf{s} =\left[s_0, \,\ldots ,s_n,\, \ldots s_{\Narray-1} \right]^T=\left[\ft_{s_0}({h}(\mathcal{S},\p)), \,\ldots ,\ft_{s_n}({h}(\mathcal{S},\p)),\, \ldots \ft_{s_{\Narray-1}}({h}(\mathcal{S},\p)) \right]^T
\end{align}
{with $\ft_{\mathbf{s}}(\cdot)$ being the functional representing the operation performed by the \ac{EM} processing on the signal ${h}(\Sur,\p)$, observed on each location $(y,z)$ of the surface due to a source located in $\p$, that allows to receive the signal vector $\mathbf{s}$ at each corresponding antenna.}
Notably, $h(\Sur,\p)$ is given by
\begin{align}\label{eq:h}
h (\Sur,\p)=  A_\mathrm{pl}\,e^{-j\chi} \, e^{-j 2\pi f_0 \tau (\Sur,\p)}  = x_0\,e^{-j 2\pi f_0 \tau (\Sur,\p)} \,,
\end{align}
where $A_\mathrm{pl}$ denotes the received signal amplitude,\footnote{{Due to the considered environment geometry, {where the receiver antenna is much smaller than the distance,} the received signal amplitude assumes approximately the same value at each antenna.}} {the phase $\chi \!\sim \!\mathcal{U}\left[0,2\pi\right)$ assumes the same value at each antenna and it is uniformly distributed between $0$ and $2\,\pi$ to model the lack of synchronization between the source and the receiver}, $f_0$ is the central frequency and
{
\begin{align}\label{eq:tau}
\tau (\Sur,\p) =\frac{||\p-\p_l(y,z)||-||\p-\p_l (0,0)||}{c} =\frac{a(\Sur,\p)}{c} \,,
\end{align}
where $\p_l(y,z)$ and $\p_l(0,0)$ are the generic and the reference points on the EM processing section, respectively, and $c$ is the speed of light. {Thus, the phase profile experienced by the array is composed of an unknown offset $\chi$ which is common to all the elements of the  antenna plus a element-dependent phase shift which depends on the position of the source. Therefore, a suitable processing of the phase profile is sufficient to infer the position of the source  provided that also the offset $\chi$ is estimated, even if it does not carry any information. }}

{According to the reference system of Fig.~\ref{fig:genericscheme}-right, the term $a(\Sur,\p)$ is the extra-distance travelled by the wave, which is given by
\begin{align}\label{eq:a}
& a(\Sur,\p)= -d + d \sqrt{1+ \frac{d_{0yz}^2}{d^2}-2\cdot \frac{d_{0yz}}{d} g(\Theta,\Theta_{oyz})} \,,
%& b(z,\p) = -d + d \sqrt{{1}+ \frac{ y^2}{d^2}}  \,.
\end{align}
where $d_{0yz}=\sqrt{y^2+z^2}$ is the distance of the point with coordinates $(0,y,z)$ in the surface from its reference point located in $(0,0)$. The term $g\left( \Theta,\Theta_{oyz}\right)$ is given by 
\begin{align}\label{eq:g}
  g\left( \Theta,\Theta_{oyz}\right)&=   \sin\left( \theta \right)   \cos\left( \phi_{0yz}  \right)   \,,
\end{align}
with $\Theta=\left(\theta, \phi=0 \right)$ and $\Theta_{0yz}=\left(\theta_{0yz}=\frac{\pi}{2}, \phi_{0yz} \right)$ referring to the target and to the generic point of the aperture with coordinates $(0,y,z)$, respectively.}\footnote{Note that for a planar aperture lying on the $YZ$-plane, it holds $\theta_{oyz}=90^\circ$.}

 The random phase $\chi$ includes the complete uncertainty on the received signal phase, since the transmitter and receiver are supposed to be not synchronized and no information can be retrieved from the phase and the \ac{TOA} of the received signal.

Differently from classical antenna array theory which assumes planar wavefronts (far field condition), here function $h (\Sur,\p)$ depends not only on the \ac{AOA} $\Theta$ but also on the distance $d$, i.e., on the position $\p$. Notably, if we consider the signal in the generic $(0,y,z)$ position of the flat lens performing the \ac{EM} processing, $h(\Sur,\p)$ contains the information on the extra distance traveled by the \ac{EM} wave to reach the generic coordinate $(0, y,\,z)$ of the RX flat lens.

Note that if $z \ll d$, according to the {second order Taylor-McLaurin series expansion with respect to $\sqrt{1+ \frac{d_{0yz}^2}{d^2}-2\cdot \frac{d_{0yz}}{d} g(\Theta,\Theta_{oyz})}$}, it is 
{
\begin{align}\label{eq:happrox}
a(\Sur,\p)
%&=-d+d\sqrt{1+\frac{d_{0yz}^2}{d^2}- 2 \frac{d_{0yz}}{d} g(\Theta,\Theta_{oyz})} & \nonumber \\
\approx  -{d_{0yz}}g(\Theta,\Theta_{oyz})+  \frac{d_{0yz}^2}{2\,d}\left(1-g^2(\Theta,\Theta_{oyz}) \right)  \,,
\end{align}
{where we have neglected terms including $1/d^2$ and $1/d^3$. In particular,}} the first term refers to the traditional array phase term containing the \ac{AOA} information, whereas the second term includes information on the source distance which becomes negligible for large distances.

In the following, we analyze some possible solutions to realize the architectures herein investigated in \ac{LOS} scenario, by discussing their impact on the \ac{EM} processing or on the array geometry.

\subsubsection{EM Processing with a \ac{R-lens}}

%\textcolor{red}{Commento generale: anche nel testo sarei per evitare di EM processing, perche' gia il termine EM lascia implicitamente pensare che ci sia una sorta di pre-processamento in analogico del fronte d'onda.} {\color{blue} a me non spiace per renderlo piu' appealing }
%\rednote{Non si puÃ² fare 2D anche questo cosÃ¬ da renderlo meglio confrontabile con la lente fissa?
%}
The use of a \ac{R-lens} that allows to achieve a full control of the phase profile of the signal, represents an extreme case where the complexity is entirely put on the \ac{EM} processing, and the array is reduced to a single antenna, {i.e., $\Narray=1$}. 
{Examples of such lenses are reported in \cite{NepBuf:J17}. Indeed, despite the technology is not yet mature for the realization of a continuous (e.g., holographic) phase profile, this solution represents a benchmark in complexity entirely put in the lens design.}

\begin{figure}[t!]
\psfrag{dy}[c][c][0.7]{$D_\text{y}$}
\psfrag{y}[c][c][0.7]{$y$}
\psfrag{x}[c][c][0.7]{$x$}
\psfrag{z}[c][c][0.65]{$z$}
\psfrag{a}[c][c][0.7]{$n = 0$}
\psfrag{b}[c][c][0.7]{\!\!\!\!\!\!\!\!\!\!\!\!\!\!\!\!\!\!\!\!\!\!\!\!\!\!\!\!\!\!\!\!\!\!\!\!\!\!\!\!$n=-\frac{(\Narray-1)}{2}$th ant.}
\psfrag{c}[c][c][0.7]{$n$th ant.}
\psfrag{s}[c][c][0.7]{$\theta_n$}
\psfrag{t}[c][c][0.7]{$\theta$}
\psfrag{p}[c][c][0.7]{$\phi$}
\psfrag{ps}[c][c][0.7]{$\p$}
\psfrag{dd}[c][c][0.7]{\!\!\!\!\!\!$d$}
\psfrag{ddd}[c][c][0.7]{\!\quad$d+a$}
\psfrag{f}[c][c][0.7]{\,\,$\Fp$}
\psfrag{P1}[c][c][0.7]{$\mathcal{P}_1$}
\psfrag{P2}[c][c][0.7]{$\mathcal{P}_2$}
\psfrag{l}[c][c][0.7]{EM focusing lens}
\centerline{
   \includegraphics[width=0.332\linewidth,draft=false]
    {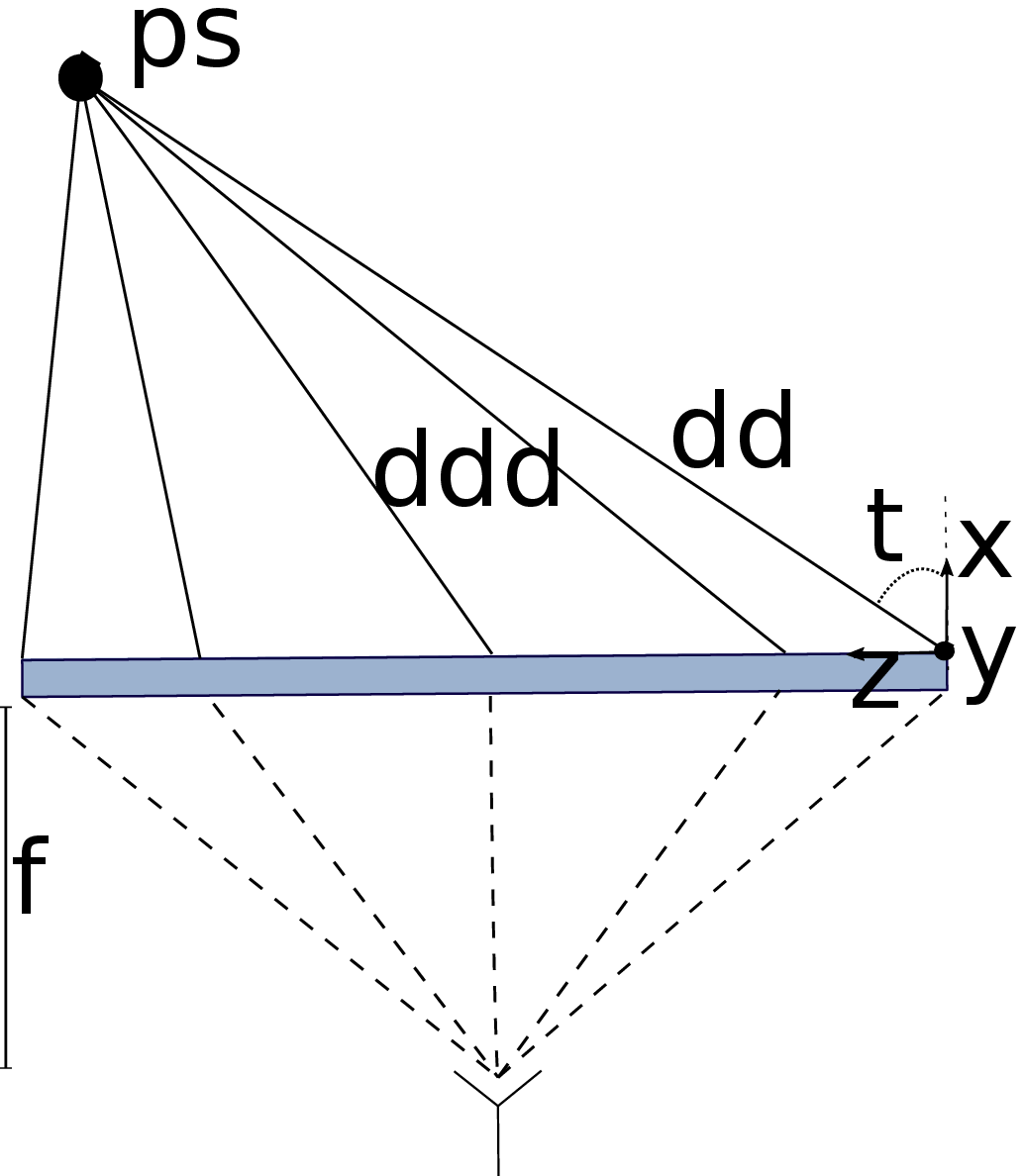}\quad \quad \quad\quad \quad \quad \includegraphics[width=0.28\linewidth,draft=false]
    {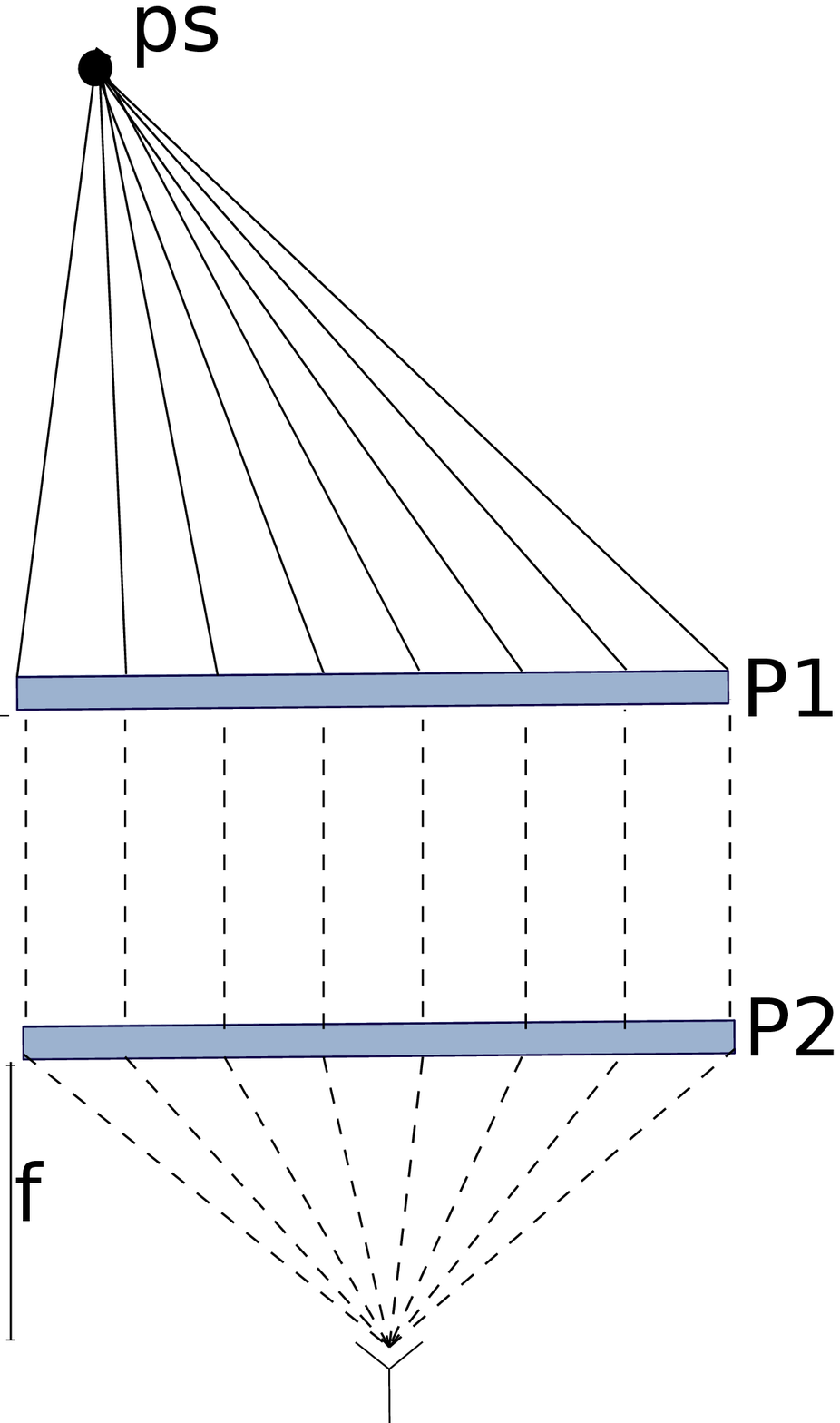}}
\caption{Left: Lateral-view of the \ac{R-lens} scenario where the \ac{EM} processing, realized with a reconfigurable lens, focuses the entire signal towards a single antenna. Right: equivalent processing performed by one \ac{R-lens} followed by a \ac{NR-lens} through $\mathcal{P}_1$ and $\mathcal{P}_2$.}
\label{fig:reconfiglens} 
\end{figure}

In particular, the objective of the \ac{R-lens} is to focus the impinging wave towards one point, placed in position $\p_0=\left[-\Fp,\frac{D_y}{2},\frac{D_z}{2}\right]$ ({refer to the coordinates system in Fig.~\ref{fig:genericscheme}-right)} at distance $\Fp$ from the lens surface, as shown in Fig.~\ref{fig:reconfiglens}. 
{Notably, we can equivalently represent the re-phasing procedure operated by the flat lens by a double-lens system performing two processing operations, namely $\mathcal{P}_1$ and $\mathcal{P}_2$,} as depicted in Fig.~\ref{fig:reconfiglens}-right. More specifically, through $\mathcal{P}_1$, the first lens converts a spherical wavefront into a planar one, with normal direction at its output. On the other side, through the operation $\mathcal{P}_2$, the second lens focuses the normal planar incident wavefront into one point, where an antenna is located. Thus, $\mathcal{P}_2$ does not depend on the source position $\p$, i.e., the second lens is a \ac{NR-lens}, whereas $\mathcal{P}_1$ concerns the reconfigurable processing that varies with the source position.

{According to such considerations on the processing operated by the \ac{R-lens}, the reconfigurable lens phase profile can be described by a function {$\kappa(\Sur,\chitest, {\ptest})$},\footnote{{Differently from $h(y,z,\p)$, in $\kappa(y,z,\chitest,\ptest)$ we make explicit also the dependence on $\chitest$.}} depending on the assumed target position $\ptest$ and on the test offset $\chitest$, as {
\begin{align}
    \kappa(\Sur,\chitest,{\ptest}) &=\mathcal{P}_1(\Sur,\chitest,{\ptest})\cdot \mathcal{P}_2(\Sur) \,.
\end{align}}
}
In the following, we illustrate the conditions to properly design the phase profiles of the two lenses, that is $\mathcal{P}_1$ and $\mathcal{P}_2$. Note that the distance of the two lenses can be considered negligible. 
\paragraph{Design of $\mathcal{P}_1$}
At the output of the first lens, it is required to obtain a planar wavefront, 
i.e., the phase of the output \ac{EM} wave is independent of $y$ and $z$.
%thus the phase of the signal should be the same, despite the considered $z$th position on the lens. 
Due to this consideration, $\mathcal{P}_1(\Sur,\chitest,{\ptest})$ is chosen so that, when the actual source position is in $\p$ {with offset $\chi$,} it is,
\begin{align}
    \mathcal{P}_1(\Sur,\chitest,{\ptest})\cdot {h}(\Sur,\p) = e^{-j\Psi_{0_1}} \,,
\end{align}
where $\Psi_{0_1}$ is a constant term, {so that ideally it should be
\begin{align}
    \mathcal{P}_1(\Sur,\chitest,\ptest) =e^{j\left(2\pi f_0 \tau (\Sur,\p)-\Psi_{0_1}\right)} \,.
\end{align}}
%
%In the following, we will assume $\Psi_{0_1}=0$.
%Operating like that, now we have a planar normal wavefront incident to the second lens.
\paragraph{Design of $\mathcal{P}_2$}
Assuming now that the wavefront is planar, $\mathcal{P}_2(\Sur)$ is designed so that the following condition holds
\begin{align}
\mathcal{P}_2(\Sur) \cdot e^{-j\Psi_0(\Sur)} =e^{-j\Psi_{0_2}}\,,
\end{align}
where $\Psi_{0_2}$ is a constant (irrelevant) term and ${\Psi_0} (\Sur)$ accounts for the travelled distance from a generic point in $(0,y,z)$ on the second \ac{NR-lens} surface to the focal point $\Fp$, that is
\begin{align}
\Psi_0(\Sur)=\frac{2\pi}{\lambda}\sqrt{\Fp^2+ \left(y-\frac{D_y}{2}\right)^2+\left(z-\frac{D_z}{2}\right)^2} \,.
\end{align}
%
%where ${\Psi_{0_2}}$ indicates a constant phase term.
Thus, $\mathcal{P}_2$ indicates the dephasing term introduced by the \ac{NR-lens} such that all rays with normal incidence arrive at $\Fp$ with identical phase for constructive superposition. It follows that
\begin{align}
\mathcal{P}_2(\Sur)  =e^{j\left(\frac{2\pi}{\lambda}\sqrt{\Fp^2+ d_{cyz}^2}-\Psi_{0_2}\right)} \,,
\end{align}
{where $d^2_{cyz}= \left(y-{D_y}/{2}\right)^2+\left(z-{D_z}/{2}\right)^2$}.
Without loss of generality, in the rest of the manuscript we will assume ${\Psi_{0_1}}={\Psi_{0_2}} = 0$. Thus, the signal received at the antenna, after the \ac{EM} processing operated by the \ac{R-lens} response (e.g., by the double-lens scheme), can be expressed as
\begin{align}\label{eq:rrlens}
r=r_0 &=s_0
%+ {w}_0\!= \!\frac{1}{\lambda\sqrt{ \, D_y\, D_z}}\int_{D_z}\int_{D_y}  \!\!\kappa(y,z,\hat{\p}) \!\cdot h(y,z ,\p)  e^{-j\Psi_0 (y,z)} dy dz 
+ {w}_0\,,
\end{align}
with 
{
\begin{align}\label{eq:fhreconfigu}
s_0=\ft_{s_0}(h(\mathcal{S},\p))= \frac{1}{\lambda\sqrt{ \, D_y\, D_z}}\int_{D_z}\int_{D_y}   \kappa(\Sur,\chitest,{\ptest}) \cdot h(\Sur ,\p)  e^{-j\Psi_0 (\Sur)} dy dz  \,.
\end{align}
}
In case {${\ptest}=\p$} {and $\chitest=\chi$}, it holds $\kappa(\Sur,\chi,\p) \cdot h(\Sur ,\p)  e^{-j\Psi_0 (\Sur)}=1$, and the signal received at the antenna simply reduces to
\begin{align}\label{eq:rlens_final}
 r_0= \sqrt{\Ae}\cdot x_0+ {w}_0 \,,
\end{align}
where $\Ae={\frac{D_y\,D_z}{\lambda^2}}$ represents the normalized aperture of the \ac{R-lens}.
%exploited for the realization of the \ac{EM} processing.

\subsubsection{\ac{EM} processing with a \ac{NR-lens}}
\label{sec:lens}
\begin{figure}[t!]
\centering
\psfragfig[width=0.35\linewidth]{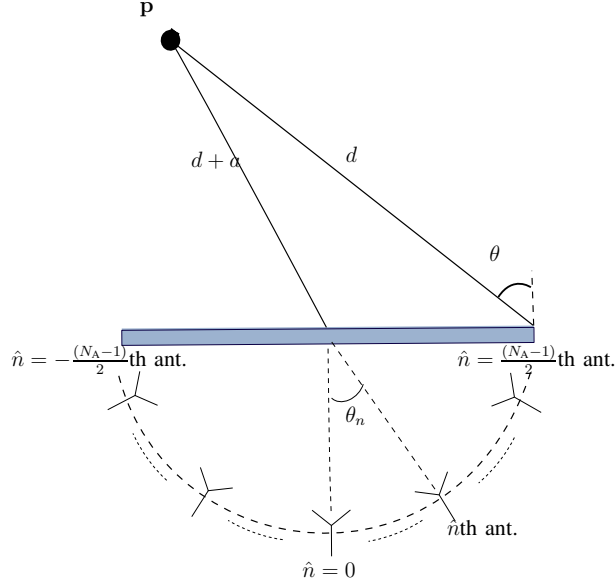}{
\psfrag{dy}[c][c][0.7]{$D_\text{y}$}
\psfrag{y}[c][c][0.7]{$y$}
\psfrag{x}[c][c][0.7]{$x$}
\psfrag{z}[c][c][0.7]{$z$}
\psfrag{a}[c][c][0.7]{$\hat{n} = 0$}
\psfrag{b}[c][c][0.7]{\!\!\!\!\!\!\!\!\!\!\!\!\!\!\!\!\!\!\!\!\!\!\!\!\!\!\!\!\!\!\!\!\!\!\!\!\!\!\!\!$\hat{n}=-\frac{(\Narray-1)}{2}$th ant.}
\psfrag{c}[c][c][0.7]{$\hat{n}$th ant.}
\psfrag{s}[c][c][0.7]{$\theta_n$}
\psfrag{t}[c][c][0.7]{$\theta_\text{out}$}
\psfrag{t1}[b][b][0.7]{\!\!\!\!\!$\theta$}
\psfrag{p}[c][c][0.7]{$\theta_0$}
\psfrag{ps}[c][c][0.7]{$\p$}
\psfrag{dd}[c][c][0.7]{\!\!\!\!\!\!\!\!\!\!\!\!$d$}
\psfrag{ddd}[c][c][0.7]{\quad\quad\quad\quad \!\!\!\! $d+a$}
\psfrag{d}[c][c][0.7]{\quad\quad\quad\quad$\hat{n}=\frac{(\Narray-1)}{2}$th ant.}
\psfrag{l}[c][c][0.7]{EM focusing lens}
}
\caption{Top-view of the \ac{EM} processing, realized with a \ac{NR-lens}. The use of the lens allows to preserve the number of employed antennas affordable \cite{Gui:C18}.}
\label{fig:lens} 
\end{figure}

We now consider the scenario where the \ac{EM} processing is realized by means of a \ac{NR-lens}, {placed along the $yz$-plane, with size $D_y \times D_z$}, and a linear antenna array is employed to collect the processed signal, as shown in Fig.~\ref{fig:lens}.\footnote{{We intentionally consider a 2D model in order not to complicate too much the notation. Nevertheless, the extension to a $3$D model is straightforward, as done in \cite{YanEtAl:C19,ZenZha:J17}.}}
%\footnote{If planar arrays are employed, the analysis can be extended according to \cite{ZenZha:J17}.} 
In particular, the lens is introduced to collimate the impinging wave in specific directions in analog domain, without the need of complex ad-hoc digital signal processing or metamaterials.
%In addition, thanks to thelens, the same aperture is achieved with a larger number of antennas when a traditional antenna array is considered.
%On the other side, the use of only one antenna as before is no more affordable in terms of capability to gather the signal after the lens, as will be detailed in the following.

{
According to the guidelines given in \cite{ZenZha:J16}, the array is equipped with $\Narray$ antennas located on the focal arc of the lens in {$\left\{\mathbf{p}_n \right\}=\left\{(-F_p \cos \theta_n , \frac{D_y}{2}, \frac{D_z}{2}+F_p \sin \theta_n )\right\}$}, with $\theta_n \in [-\pi/2, \pi/2]$.
Then, {we consider 
$\Narray=1+  \lfloor 2\,\tilde{D_z} \rfloor $},
where $\tilde{D}_z = D_z/\lambda$, and with $\lfloor \cdot \rfloor$ denoting the largest integer no greater than its argument \cite{ZenZha:J16}.}
{By accounting for a source located on the $xz$-plane, the signal received on the focal arc, in the position $\p_n$, can be expressed as
\begin{align}\label{eq:rn_nrlens}
r_n=r (\mathbf{p}_n)& =\ft_{s_n}(h(\mathcal{S},\mathbf{p}))+w_n 
=s_n(\p)+w_n \nonumber \\
&=\frac{1}{\lambda\,\sqrt{D_y\,D_z}}\int_{0}^{D_y}\int_{0}^{D_z} {h}(\Sur,\p) \, e^{j \Psi_n(y,z)} dz dy +{w}_n\,,
\end{align}
where {
\begin{align}
 \Psi_n (y,z) = \Phi_0- k_0\left(z-\frac{D_z}{2}\right)\sin\theta_n\,,
\end{align}}
is the dephasing term given by the lens towards the antenna in $\p_n$, according to the analysis reported in \cite{ZenZha:J16} for an incident planar wavefront, and the normalization term $1/(\lambda\,\sqrt{D_y\,D_z})$ is chosen to guarantee that the
overall power intercepted by the lens is proportional to its normalized aperture $\Ae=(D_y\,D_z)/\lambda^2$ \cite{ZenZha:J16}.
Moreover, $\Phi_0$ represents a phase term constant at each antenna. In the following, we assume $\Phi_0=2\,\pi\,b$, with $b $ any integer.
Then, we can write {
\begin{align}\label{eq:lensfinal}
&{r}_n \!=s_n(\p)+w_n=\! \frac{x_0}{\lambda\,\sqrt{D_y\,D_z}}\int_{0}^{D_y}\int_{0}^{D_z} \! \!\! \!e^{-j 2\pi  a(\Sur, \p)} e^{j 2\pi  \tilde z  \sin \theta_n} d y\, d  z+\! {w}_n \,,
\end{align}
where $\tilde z=\frac{D_z}{2}-z$}, the expression of $a(\Sur, \p)$ is given by \eqref{eq:a}, and $s_n(\p)$ can be expressed by
\begin{align}\label{eq:sn_nrlens}
s_n(\p)=\frac{x_0}{\lambda\,\sqrt{D_y\,D_z}}\int_{0}^{D_y}\int_{0}^{D_z} \! \!\! \!e^{-j 2\pi  a(\Sur, \p)} e^{j 2\pi  \tilde z  \sin \theta_n} d y\, d  z \,.
\end{align}
}
Solving \eqref{eq:lensfinal} allows to study the impact of the wavefront curvature at the receiver's antennas. Note that in case the wavefront is planar, the solution of \eqref{eq:lensfinal}  reduces the one given in \cite{ZenZha:J16}.

\subsubsection{Post-processing with \ac{no-lens}}\label{subsec:nolens}
For comparison purposes, assume now that the \ac{EM} processing is avoided (baseline scenario), which corresponds to the extreme case with respect to \ac{R-lens} where all the complexity is entirely put into the receiving array
%This corresponds to an asymptotic scenario but, differently from the use of \ac{R-lens}, where the architecture complexity is located in the \ac{EM} processing, here the complexity is entirely put into the receiving array 
with $\Narray=N_z\times N_y$ antennas placed at an inter-distance of $\lambda/2$. 
\begin{figure}[t!]
\psfrag{x}[c][c][0.7]{$x$}
\psfrag{y}[c][c][0.7]{$y$}
\psfrag{z}[c][c][0.7]{$z$}
\psfrag{o}[c][c][0.7]{}
\psfrag{t}[c][c][0.7]{$\theta$}
\psfrag{t2}[c][c][0.7]{$\theta_{n0}$}
\psfrag{l}[c][c][0.7]{$z$}
\psfrag{a}[c][c][0.7]{array}
\psfrag{d}[c][c][0.7]{$d$}
\psfrag{p}[c][c][0.7]{\quad$\p$}
\psfrag{N}[c][c][0.7]{$N_{z}$}
\psfrag{M}[c][c][0.7]{$N_{y}$}
\psfrag{d2}[c][c][0.7]{\!\!\!\!$d+a$}
\psfrag{ph}[c][c][0.7]{$\phi_{n0}$}
\centerline{
   \includegraphics[width=0.35\linewidth,draft=false]
    {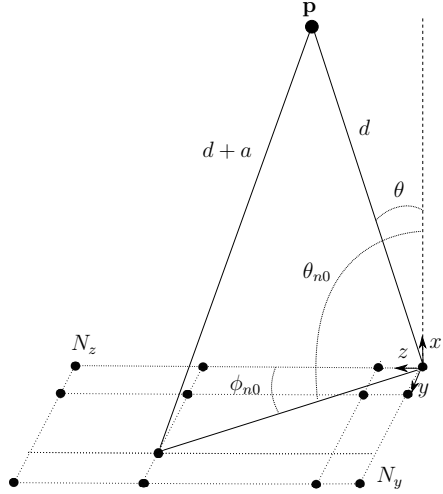}}
\caption{Scenario for the \ac{no-lens}, that is, in absence of \ac{EM} processing. %The information on the position $\p$ is retrieved from the curvature wavefront at each antenna element.
}\label{fig:scenario} 
\end{figure}
{We consider such antennas positioned in $\mathbf{p}_n=(0,n_y\,\lambda/2, n_z \,\lambda/2)$ on the receiving surface, with $n_y=\lfloor \frac{n}{N_z}\rfloor$ and $n_z=(n\mod{N_z})$, for $n=0,\,\ldots,\,\Narray-1$,} and it holds $    \ft_{s_n}(h(\mathcal{S},\p))=h(\p_n,\p) \,.$
The RX signal at the $n$th antenna element ${r}_n $ is simply given by \cite{ZenZha:J16}
\begin{align}\label{eq:trad}
{r}_n =r(\mathbf{p}_n)=s_n(\p)+w_n=h_n +w_n=x_0 \, e^{-j 2\pi f_0 \tau (\mathbf{p}_n,\p)} + {w}_n \,.
\end{align}
{
where
\begin{align}\label{eq:stradfinal}
    s_n(\p)=x_0 \, e^{-j 2\pi f_0 \tau (\mathbf{p}_n,\p)} \,.
\end{align}
}
Again, here the information on $\p$ is embedded in the received signal and, by properly designing the receiver, it is possible to directly infer the position of the transmitter avoiding a prior synchronization phase to align the source and receiving array clocks.

In the following, we derive possible approaches to directly estimate the source location while using the architectures here introduced.

\section{Position Estimation}\label{sec:positioning}
As previously stated, the source positioning task can be achieved by exploiting the curvature of the wavefront, by overcoming the need to tightly synchronize the transmitter and the receiver, which is often unfeasible due to the required interactions for estimating the \ac{TOA}. 

{In the following, we describe the adopted positioning approach according to each architecture, in the presence of one source in \ac{LOS}.}

\subsubsection{{Positioning with R-lens}}
{
In case a \ac{R-lens} is adopted, for each test position $\ptest$, the lens has to be reconfigured. {This problem is similar to that of beamforming for phased arrays as in \cite{stoica2005spectral}, where the signal is available only after the recombination procedure, and one possibility is to pick the configuration that maximizes the output energy.} 
%From one side, this operation requires a complex EM processing, but it drastically simplifies the processing at signal level.
Thus, we can define
\begin{align}
{\ell}(\ptest)=\left|r_0(\ptest) \right|^2, 
\end{align}
where $r_0(\ptest)$ is given by \eqref{eq:fhreconfigu} and \eqref{eq:rrlens}, and the position can be estimated as
\begin{align}\label{eq:Posrlens}
\hat{\p}= {\arg} \underset{\ptest}{\max}  \quad {\ell(\ptest)} \,.
\end{align}
Notably, $\ptest$ is varied within a set of test positions $\mathcal{P}$ and, consequently, for each $\ptest$, a different $r_0$ is obtained.
According to the aforementioned operations, thanks to the processing operated at EM level, the position of the source can be retrieved by using only one antenna and with simple processing operations. Nevertheless, this requires that, for each $\ptest$, a different received signal is collected as the lens needs to be reconfigured.
}
\subsubsection{{Positioning with NR-lens}}
%
%\subsection{\ac{ML}-based positioning for \ac{no-lens}}\label{sec:ml}

{In case a \ac{NR-lens} is employed, we consider the \ac{ML} estimator.} In our scenario, the likelihood function (depending on the spherical wavefront) is taken with respect to the position $\p$ of the transmitter and the unknown phase $\chi$ {and it is given by}
\begin{align}\label{eq:ML}
\!\Lambda (\p,\chi)=p(\mathbf{r}|\mathbf{p})\propto\prod_{n=1}^{{\Narray}} \exp \left\{- \frac{1}{2\,\sigma^2} \bigg\lVert {r}_n - s_n(\p,\chi) \bigg\rVert^2 \right\} \,,
\end{align}
with $\sigma^2 = \mathbb{E}\left[w_n^2 \right]$, 
%with $N_0$ representing the noise \ac{PSD} at each antenna,
and where we have made explicit the dependence of $s_n(\p,\chi)$ on the position $\p$ and on the phase offset $\chi$. Taking the logarithm and discarding all the terms that do not bring contribution for maximizing $\p$, the log-likelihood function reduces to
\begin{align}\label{eq:logML}
\ell(\p,\chi) =& \sum_{n=1}^{{\Narray}} \left\{\Re \bigg\{{r}_n \cdot s_n^*(\p,\chi) \bigg\} -\frac{1}{2} E_\text{rx} (\p)\right\} \,,
\end{align}
where $E_\text{rx} (\p)=  \sum_{n=1}^{\Narray} E_n (\p)$, with $E_n (\p)= |s_n (\p)|^2 $ being the received energy per antenna. 
%where $E_\text{rx} (\p)=  \sum_{n=1}^{\Narray} \lVert\sn(\p)\rVert^2$ being the received energy per antenna. 
\\
Finally, the \ac{ML} estimate of the distance can be expressed as
\begin{align}\label{eq:hatp}
\hat{\p}= \arg \underset{\p,\, \chi}{\max} \,\,\, \ell (\p,\chi)  \,,
\end{align}
that, in accordance with the previous derivation, yields to
\begin{align}\label{eq:MLpos}
\hat{\p}= {\arg} \underset{\p,\,\chi}{\max} \left\{\sum_{n=1}^{{\Narray}}  \Re \bigg\{{r}_n \cdot s_n^*(\p,\chi) \bigg\}  - \frac{1}{2} E_\text{rx} (\p)\right\}\,.
\end{align}
{According to the analysis in Sec.~\ref{sec:trad}, for \eqref{eq:MLpos} we have that  $s_n(\p,\chi)$ and ${r}_n$  are given by \eqref{eq:sn_nrlens} and \eqref{eq:lensfinal}, respectively. }

\subsubsection{{Positioning with no-lens}}
{
For the \ac{no-lens} case, it is possible to adopt the \ac{ML} as before in \eqref{eq:MLpos} for retrieving the source position, by using \eqref{eq:stradfinal} and \eqref{eq:trad} for $s_n(\p,\chi)$ and ${r}_n$, respectively. Notably, since the EM processing is absent in this case, the number of employed antennas $\Narray$ is higher with respect to the previous case.
\\
The maximization in \eqref{eq:MLpos} could involve a certain level of complexity if $\Narray$ is large, therefore in the following we propose a positioning approach entailing a lower complexity for practical systems, and that can be applied only to the \ac{no-lens} scenario. }

\paragraph{Differential Approach}
\label{subsec:ortho}

Consider a uniform planar array where, by adopting the Taylor-McLaurin series expansion in \eqref{eq:happrox}, $h_n$ reported in \eqref{eq:trad} can be written as 
{
\begin{align}\label{eq:nolens-hn}
h_n &\approx x_0\,\text{exp}\left\{-j \frac{2\pi}{\lambda}\left[ \frac{d^2_{n0}}{2\,d}\bigg(1-g^2(\theta,\phi_{n0})\bigg)-d_{n0}\,g(\theta,\phi_{n0})\right]\right\}\nonumber \\ 
&\approx x_0\,\text{exp}\left\{-j 2 \pi \frac{\left[n_y^2+n_z^2\cos^2(\theta)\right] \lambda}{8\,d} \right\} \exp\left\{ j  \pi n_z\,\sin \theta \right\}    \,,
\end{align}
with $d_{n0}=\sqrt{\left(n_y\,\lambda/2\right)^2+\left(n_z\,\lambda/2\right)^2}$ according to Sec.~\ref{subsec:nolens}, and $\phi_{n0}$ is graphically reported in Fig.~\ref{fig:scenario}.
Then, for $n\geq 1$, we can write
\begin{align}
 \!\! h_{n}  h^\star_{ n-1}\!=\!|x_0|^2\,\text{exp}\left\{-j  \pi \frac{\left[(n_y^2 -m_y^2) + ( n_z^2-m_z^2)\cos^2(\theta)\right] \lambda}{4\,d}  \right\} \exp\left\{ j  \pi \,(n_z-m_z)\sin \theta \right\} ,
\end{align}
with $|x_0|^2=A_{\mathrm{pl}}^2$ and $m_y=\lfloor \frac{n-1}{N_z}\rfloor$ and $m_z=((n-1)\mod{N_z})$.}
Thus, the offset term has been removed, and it holds
\begin{align}\label{eq:trad_diff}
r^{\mathrm{d}}_n=r_{n} r^\star_{n-1}= h_n\, h^\star_{n-1}  + w^{\mathrm{d}}_n =s^{\mathrm{d}}_n + w^{\mathrm{d}}_n  \,, \quad n=2,\ldots, \Narray\,,
\end{align}
with $s^{\mathrm{d}}_n=h_n\, h^\star_{n-1}$, and $w^{\mathrm{d}}_n=w_n\,w^\star_{n-1}+w_n\,h^\star_{n-1}+w^\star_{n-1}\,h_{n}$ contains the noise terms at the output of the differential operation.
The position can now be estimated as follows
\begin{align}\label{eq:mldiff}
\hat{\p}=\underset{\p}{\arg\max} \left\{\sum_{n=2}^{\Narray}\bigg( r^{\mathrm{d}}_n\cdot \left[s^{\mathrm{d}}_n \right]^\star \bigg) - \frac{1}{2} E^{\mathrm{d}}_\text{rx} (\p)\right\} \,,
\end{align}
{where $E^{\mathrm{d}}_\text{rx}=\sum_{n=2}^{\Narray}|s^{\mathrm{d}}_n|^2$}.
{
In a worst-case in terms of complexity, when an exhaustive search is used, the \ac{ML} only compares a number of test positions and selects the maximum as an estimate of the position, without searching over all possible phase offset $\chi$, thanks to the use of the proposed differential scheme.
%In this case, when brute-search approaches are employed, the \ac{ML} search is reduced only to the possible test positions $N_{\ptest}$, avoiding the need to search over possible test phase offset $\chi$. 
\\    
In fact, the approach in \eqref{eq:MLpos} entails the construction of a 3D matrix with size $\left(N_\text{array}\times N_{\ptest} \times N_{{\chi}_\text{t}}\right)$
where $N_{\ptest}$ and $N_{{\chi}_\text{t}}$ are the number of tests for position and offset, respectively. Contrarily, with the proposed approach, {the search in \eqref{eq:mldiff} reduces to a matrix with size $\left(N_\text{array}\times N_{\mathbf{\ptest}} \right)$.}}
%This matrix reduction is paid at the prize of $N_\text{array}-1$ multiplications, and of a small performance degradation, since the noise is enhanced through the performed operations.

In the numerical results the performance of the described approaches will be evaluated.

%Note that by naming $m = 2n-1$ and $d_m=m d_x$, it can be found that the relation between the distance and the antenna position is  
%
%\begin{align}
%&{P}^{\mathcal{O}}_m= \mathrm{b}_m\,  \left(A_\mathrm{pl}\right)^2 e^{ -j \pi \,\lambda \,\sin \theta } \,e^{-j 2 \pi \, d_m \, f(\p) }\,  ,\quad \quad m = 3,\ldots, {M}
%\end{align}
%with $M=2(\Narray-1)$ and $ \mathrm{b}_m =  \left(m \, \mod \,2\right) $ 
%\begin{align}
%& \text{$\mathrm{b}_m = 0$  \quad \quad \,\, if $\left(\mathrm{b}_m \, \text{mod} \,2\right) = 0$} \nonumber \\
%& \text{$\mathrm{b}_m =  \left(A_\mathrm{pl}\right)^2 \, $ \,  if $\left(\mathrm{b}_m \, \text{mod} \,2\right)  = 1$}
%\end{align}
%
%and $d_z = \lambda/2$ and $f(\p)=\frac{\cos^2\theta}{d}$.

%Notably, the same operation can be extended to a planar array, where the differential operation is performed row by row.

\section{Multi-Source Interference}\label{sec:int}
%\textcolor{red}{Invertirei l'ordine delle sezioni 3 e 4}\\
%{\color{blue} D'accordo }
%\\
{The intent of the previous section was to establish how the position of a source can be determined with the proposed signal model. On the other side, real environments are usually characterized by the presence of multiple simultaneously transmitting sources. To this purpose, we investigate here the impact of the interference when discriminating an intended useful source from one or more interference transmitters. Indeed, the established performance is related to the positioning resolution of the system. }

Let us now assume that interference sources, in addition to the intended useful one, are simultaneously transmitting in the same environment. In this case, the received signal is 
\begin{align}\label{eq:sir1}
\mathbf{r}= \mathbf{s}_\mathrm{u}+\mathbf{s}_\mathrm{int}+\mathbf{w}=\ft_{\mathbf{s}}({h}(\mathcal{S},\p_\mathrm{u}))  + \sum_{i=1}^{\Nint} \ft_{\mathbf{s}}({h}(\mathcal{S},\p_i))+ \mathbf{w}\,,
\end{align}
where $\mathbf{s_u}$ and ${\mathbf{s}_\mathrm{int}}$ refer to the contribution from the intended useful source, located in $\p_\mathrm{u}$ with phase offset $\chi_\mathrm{u}$, and from the interfering $i$th source, located in $\p_i$ and with phase offset $\chi_i$, respectively. The signal received at the $n$th antenna is
\begin{align}
    r_n=s_{\mathrm{u},n} + \sum_{i=1}^{N_\mathrm{int}}{s}_{i,n} +w_n \,.
\end{align}
%
%As previously stated, since we do not consider here a tight synchronization between the transmitter and the receiver, signals are received with a different phase offset $\chi$.
%, whereas the amplitude $A_\text{pl}$ is assumed to be ideally the same due to perfect control approaches. 
%This means that $h(y,z,\mathbf{p})$   and $h(y,z,\mathbf{p}^{(i)})$ in \eqref{eq:sir1} have independent $\chi$ and the same amplitude $A_\text{pl}=A$.

{
In the following, in order to evaluate the impact of the interference, we first consider an ideal scenario where the position of the intended useful source is assumed to be perfectly estimated (e.g., according to the approach of Sec.~\ref{sec:positioning}). Notably, this corresponds to a phase profile matched to the received phase profile, that is $\hat{\p}=\p_\mathrm{u}$ and $\chi=\chi_\mathrm{u}$, which corresponds to a \ac{MRC}. Successively, we evaluate the \ac{SIR} in order to determine the performance degradation due to the interference. The output of the \ac{MRC} can be expressed as}
\begin{align}
\ell_\text{MRC}(\pu,\chi_\mathrm{u})=   \etau  + \etai + \etaw \,,
\end{align}
%
%with $\p_\mathrm{int}$ being the vector containing the interference source positions, and 
{where $\etau$, $\etai$ and $\etaw$ correspond to the processed versions of the useful, interference and noise contribution, respectively, and they are defined in the following for each architecture.} Nevertheless, 
%
%\begin{align}\label{eq:etam}
%    \etau &=\sum_{n=1}^{{\Narray}} {s}_{\mathrm{u},n} \cdot s_n^*(\p_\mathrm{u},\chi_\mathrm{u}) \,, \nonumber \\
%    \etai &=\sum_{n=1}^{{\Narray}} \sum_{i=1}^{N_\mathrm{int}}{s}_{i,n} \cdot s_n^*(\p_\mathrm{u},\chi_\mathrm{u}) \,, \nonumber \\
%     \etaw &=\sum_{n=1}^{{\Narray}} {w}_n \cdot s_n^*(\p_\mathrm{u},\chi_\mathrm{u})   \,.
%\end{align}
%
according to the considered model, we define the \ac{SIR} as
\begin{align}\label{eq:sir0}
\mathrm{SIR}=\frac{| \etau|}{ |\etai |} \quad \,.
\end{align}
In the following, considerations are drawn for each architecture.

\subsection{{SIR} using \ac{R-lens}}\label{sec:sirRlens}
%We first consider the \ac{R-lens} scenario. 
{Due to the aforementioned considerations, in case of a \ac{R-lens}, for the \ac{MRC} we have $\kappa(y,z,\chitest,\ptest)=\kappa(y,z,\chiu,\pu)$. Thus, for this case it holds $\etau=w$ and,}
 assuming that the phase-profile of the \ac{R-lens} is perfectly matched to the received signal phase, according to \eqref{eq:rrlens}, \eqref{eq:fhreconfigu} and \eqref{eq:rlens_final} we have 
%$\Fu=\Hu \rvert_{\hat{\p}=\p}$ and $\Fi=\Hi \rvert_{\hat{\p}=\p}$, so that
%
\begin{align}
    \etau=\frac{\Af}{\lambda \sqrt{D_y\,D_z}} \,x_0 \,,
\end{align}
with $\Af=D_y\,D_z$, and $\etai$ is given by
%\begin{align}
%    \Fi = \frac{1}{\lambda \sqrt{ \, D_y\, D_z}}\int_{D_z}\int_{D_y}    e^{-j 2\pi f_0 \tau (y,z,\p^{(i)})} dy dz 
%\end{align}
%
%so that it is possible to write
%
\begin{align}
   \!\! \etai &= \frac{x_0}{\lambda \sqrt{ \, D_y\, D_z}}\cdot  \sum_{i=1}^{\Nint}\eta_i \,,
\end{align}
where
\begin{align}
\eta_i=  e^{j(\chiu-\chi_i)}\,\int_{D_z}\int_{D_y}    e^{j 2\pi f_0 \left(\tau (\Sur,\pu)-\tau (\Sur,\pin)\right)}\, dy dz \,.
\end{align}
Thus, the \ac{SIR} for the \ac{R-lens} can be expressed as
\begin{align}\label{eq:sirgeneral-rlens}
    \mathrm{SIR}_\text{R-lens}
    %=\frac{ \Af}{\left|\sum_{i=1}^{\Nint} e^{j(\chi_u-\chi_i)}\,\int_{D_z}\int_{D_y}    e^{j 2\pi f_0 \left(\tau (y,z,\pu)-\tau (y,z,\pin)\right)} dy dz \right|}
    =\frac{ \Af}{\left|\sum_{i=1}^{\Nint} \eta_i \right|} \, .
\end{align}

Then, assume the presence of only one generic interfering user, with $\chi_i=\chiu$, i.e., a worst case scenario.
For a given $\pu$, i.e., $d$ and $\theta$, we can evaluate the impact of the interference according to $\pin=(d+\Delta d,\,\theta+\Delta \theta)$, where the couple $(\Delta d,\,\Delta \theta)$ represents the variation of $\pin$ from $\pu$ along $d$ and $\theta$.
Thus, by expliciting the dependence of $\etai$ on $\Delta d$ and $\Delta \theta$, we can write
\begin{align}\label{eq:sirarea}
 &\eta_i(\Delta d, \Delta \theta)= \int_{D_z}\int_{D_y}    e^{j 2\pi f_0 \left[\tau (\Sur,d,\theta)-\tau (\Sur,d+\Delta d,\theta+\Delta \theta)\right]}dy dz  \,.
\end{align}
To this purpose, in order to make use of the expression in \eqref{eq:tau}, we consider the approximation reported in \eqref{eq:happrox}, in the following way
%
%\begin{align}
%{d_{0yz}}g(\Theta,\Theta_{oyz})+  \frac{d_{0yz}^2}{2\,d}\left(1-g^2(\Theta,\Theta_{oyz})\right)=\sqrt{y^2+z^2}\sin(\theta)\,\cos\phi_{oyz} +\frac{y^2+z^2}{2\,d}\,(1-\sin^2(\theta)\cos^2(\phi_{oyz}))
%\end{align}
%
%Passando a coordinate polari, e tornando a quelle cartesiane, possiamo scrivere:
%
\begin{align}\label{eq:aint}
&a(\Sur,\pu)\approx -z\sin\theta +\frac{y^2}{2\,d} +\frac{z^2\,\cos^2(\theta)}{2\,d}\,, \nonumber \\
&a(\Sur,\pin)\approx
-z\sin\left(\theta+\Delta \theta\right) +\frac{y^2}{2\,(d+\Delta d)} +\frac{z^2\,\cos^2(\theta+\Delta \theta)}{2\,(d+\Delta d)} \,,
\end{align}
that gives to
\begin{align}\label{eq:eta_nrlens}
    &\eta_i(\Delta d,\Delta \theta)= \nonumber \\
    &\!\int_{D_z}\!\int_{D_y}    \!\!\!\!\!\text{exp}\!\left\{j \frac{2\pi}{\lambda}\! \! \left(\!z\left[\sin(\theta\!+\!\Delta \theta)\!-\!\sin \theta \right]\!+\!\frac{y^2}{2\,d}\!\!\left(1-\frac{1}{1\!+\!\frac{\Delta d}{d}}\! \!\!\right)\!\!+\!\frac{z^2}{2\,d}\!\left[\!\cos^2\theta\!- \frac{\!\cos^2(\theta+\Delta \theta)}{1+\frac{\Delta d}{d}}\!\right]\right)\!\right\}dy dz.
\end{align}
From \eqref{eq:eta_nrlens} it is not easy to get insights about the impact of various parameters on the \ac{SIR}. 
%The solution of \eqref{eq:eta_nrlens} does not allow to easily make considerations. 
A possibility to simplify the model is to consider only distance variations $\Delta d$, with $\Delta \theta=0$, by assuming $\cos \theta \approx 1$, which is true when users are close to the boresight direction.
Thus, we can write
\begin{align}
   \eta_i(\Delta d,0) %=\int_{D_z}\int_{D_y}    \text{exp}\left\{j \frac{2\pi}{\lambda}  \left(\frac{y^2}{2\,d}\left(\frac{1}{1+\frac{\Delta d}{d}}-1\right)+\frac{z^2}{2\,d}\left[ \frac{\cos^2(\theta)}{1+\frac{\Delta d}{d}}-\cos^2\theta\right]\right)\right\}dy dz \nonumber \\
 \approx \int_{D_z}\int_{D_y}    \text{exp}\left\{j \frac{2\pi}{\lambda}  \left(\frac{y^2+z^2}{2\,d}\left(1-\frac{1}{1+\frac{\Delta d}{d}}\right)\right)\right\}dy dz \,,
\end{align}
and, in order to get an approximate closed-form of \eqref{eq:eta_nrlens}, we change from Cartesian to polar coordinates $(\rho, {\varphi})$, indicating a point on the surface, and {we approximate the surface of the lens with a quadrant of a circle with radius $D_{\rho}$, such that {$\frac{\pi D_\rho^2}{4} =D_z\, D_y$, thus obtaining}
\begin{align}
\eta_i(\Delta d,0)&\approx \int_{0}^{\Drho}\int_{0}^{\pi/2}     \text{exp}\left\{-j \frac{2\pi}{\lambda}  \left(\frac{\rho^2}{2\,d}\left(\frac{1}{1+\frac{\Delta d}{d}}-1\right)\right)\right\}\rho \,d\rho \,d\varphi \nonumber \\
&\approx \Af \!\left[e^{-j\,\frac{\,\pi}{\lambda}\,\Drho^2\,\gamma}\text{sinc}\left(\frac{\Drho^2\,\gamma}{\lambda}\right)\right],
\end{align}
}
where 
%$D_\rho$ is the diameter of the approximating circle and
%
\begin{align}\label{eq:kappa}
    %& \beta= \sin(\theta\!+\!\Delta \theta)\!-\sin \theta \,, \quad \quad  
    \gamma=\frac{1}{2\,d}\left(\frac{1}{1\!+\!\frac{\Delta d}{d}}-1\right)  \,.
    %\gamma= \frac{1}{2\,d}\left(\frac{\cos^2(\theta+\Delta \theta)}{1+\frac{\Delta d}{d}}\!-\!\cos^2\theta \right) .
\end{align}
Notably, when $\Delta d=0$, i.e., $\gamma=0$, it is $\eta_i(0,0)=\Af$. 
Coming back to \eqref{eq:sirgeneral-rlens}, we can now determine when the \ac{SIR} is above a certain threshold $\xi^\star$, that is, %the following condition has to be satisfied
\begin{align}\label{eq:vulnRLENS}
   \mathrm{SIR}_\text{R-lens}= \frac{\etau }{|\eta_i(\Delta d,0)|}=\frac{1}{\left|\text{sinc}\left({\frac{\Drho^2}{2\,\lambda\,d}\,\left(\frac{1}{1+\frac{\Delta d}{d}}-1\right)}\right)\right|} >\xi^\star\,.
\end{align}
%
%%
%\begin{align}\label{eq:sirrlens}
%     &\frac{1}{\left|\text{sinc}\left({\frac{\Drho^2}{2\,\lambda\,d}\,\left(\frac{1}{1+\frac{\Delta d}{d}}-1\right)}\right)\right|} >\xi^\star \,.
%\end{align}
%
Result in \eqref{eq:vulnRLENS} is useful to determine the distance interval $\Delta d$ which is vulnerable to the interference, in other words, the minimum spacing $\Delta d$ between users such that {it holds} $\mathrm{SIR}>\xi^\star$. 
In particular, when $d \gg \Drho$, i.e., {$d \gg D_y,\,d\gg\,D_z$}, the argument of the $\text{sinc}$-function tends to be $0$, and consequently the \ac{SIR} to $1$, {representing a worst case scenario. Thus, in such conditions the wavefront curvature becomes weak and it becomes difficult to discriminate the position of two transmitters.} 
 {To better picture these considerations, note that since $\text{sinc}(x)$ has a $1/(\pi x)$ envelope, \eqref{eq:vulnRLENS} is surely satisfied for
\begin{align}\label{eq:sir-rlens-final}
    &\left|\frac{2\,\Af}{\lambda\,d}\,\left(\frac{\Delta d}{d+\Delta d}\right)\right| >{\xi^\star} \,.
\end{align}
%
%that yields to the condition
%
%\begin{align}\label{eq:deltadcond}
%\begin{cases}
%    \Delta d < -d \frac{\lambda\,\xi^\star}{\lambda\,\xi^\star+2\,\frac{\Af}{d}}\,, \\
%    \Delta d > d \frac{\lambda\,\xi^\star}{\lambda\,\xi^\star-2\,\frac{\Af}{d}} \,.
%    \end{cases}
%\end{align}
%
Notably, according to \eqref{eq:sir-rlens-final}, the relation of $d$ with the physical aperture ${\Af}$ and with $\Delta d$ plays an important role in determining the capability to discriminate the useful source from an interference one. Indeed, large values of $2\Af/(\lambda\,d)$ allow to relax the requirement on $\Delta d$ for attaining a SIR above $\xi^\star$.}
%according to \eqref{eq:deltadcond}

%In addition, note that $\mathrm{SIR}_\text{R-lens}\rightarrow \infty$ when $\frac{\Drho^2}{2\,\lambda\,d}\,\left(\frac{1}{1+\frac{\Delta d}{d}}-1\right)=1$, that is when
%
%\begin{align}
   %&\frac{\Drho^2}{2\,\lambda\,d}\left(\frac{1}{1+\frac{\Delta d}{d}}-1\right)= 1 \nonumber \\
%   & \Delta d \rightarrow -\frac{d}{1  + \frac{\Drho^2}{2\lambda\,d}}  \,
%\end{align}
%
%and thus the interference effect is negligible.

\subsection{\ac{SIR} using \ac{NR-lens}}
{
Differently from before, in this case the ouput of the \ac{MRC} is given by
\begin{align}\label{eq:etam}
    \etau &=\sum_{n=1}^{{\Narray}} {s}_{\mathrm{u},n} \cdot s_n^*(\p_\mathrm{u},\chi_\mathrm{u}), \quad \etai =\sum_{n=1}^{{\Narray}} \sum_{i=1}^{N_\mathrm{int}}{s}_{i,n} \cdot s_n^*(\p_\mathrm{u},\chi_\mathrm{u}) \,,\quad 
     \etaw =\sum_{n=1}^{{\Narray}} {w}_n \cdot s_n^*(\p_\mathrm{u},\chi_\mathrm{u})   \,,
\end{align}
%
%the phase profile of the \ac{NR-lens} can not be matched to the phase of the intended useful transmitter.
%Now, the signals $s_{\mathrm{u},n}$ and $s_{i,n}$ of \eqref{eq:etam} are given by \eqref{eq:lensfinal} as {
with
\begin{align}\label{eq:snrlens-multint}
 s_{\mathrm{u},n} &= \frac{x_0}{\lambda\,\sqrt{D_y\,D_z}}\int_{0}^{D_y}\int_{0}^{D_z} \! \!\! \!e^{-j 2\pi  a(\Sur, \pu)} e^{j 2\pi  \tilde z  \sin \theta_n} d y\, d  z  \,\nonumber \\
s_{i,n} & =\frac{x_0}{\lambda\,\sqrt{D_y\,D_z}}\int_{0}^{D_y}\int_{0}^{D_z} \! \!\! \!e^{-j 2\pi  a(\Sur, \pin)} e^{j 2\pi  \tilde z  \sin \theta_n} d y\, d  z \,,
\end{align}
} and they are characterized by the same lens response, regardless the value of $\pu$, since the lens is designed with a fixed phase profile. Indeed, the source position is inferred according to the focusing of the received signal towards the array placed behind the lens \cite{ZenZha:J16}. {By means of the above equations, the \ac{SIR} for multiple interference can be found by using \eqref{eq:sir0}.}

Then, for the single interference case, by assuming again the approximation of $(\theta,\Delta \theta)=(0,\,0)$, and by injecting \eqref{eq:aint} and \eqref{eq:snrlens-multint} into \eqref{eq:etam}, the evaluation of $\etau$ and {$\etai=\!\frac{x_0^2}{\lambda^2\Af}\eta_i(\Delta d,0)$} now requires the resolution of the following equations
{
\begin{align}
\etau &=\frac{x_0^2}{\lambda^2\,\Af}\sum_{n=1}^{\Narray}\left|\int_{D_z}\int_{D_y}    e^{-j \frac{2\pi}{\lambda}  \left[\frac{y^2+z^2}{2\,d}- \tilde z  \sin \theta_n\right]} dy dz \right|^2  \nonumber \\
\eta_i(\Delta d,0)&\!=\sum_{n=1}^{\Narray}\left\{\!\left[\int_{D_z}\int_{D_y} \!   e^{-j \frac{2\pi}{\lambda}  \left[\frac{y^2+z^2}{2\,(d+\Delta d)}- \tilde z  \sin \theta_n\right]} dy dz \right]\!\!\left[\int_{D_z}\int_{D_y}  \!\!  e^{-j \frac{2\pi}{\lambda}  \left[\frac{y^2+z^2}{2\,d}- \tilde z  \sin \theta_n\right]} dy dz\right]^{\!*}\right\} ,
\end{align}
which give
\begin{align}\label{eq:sirnrlens_general}
    \!\!\!\mathrm{SIR}_\text{NR-lens}\!=\frac{\sum_{n=1}^{\Narray}\left|\int_{D_z}\int_{D_y}    e^{-j \frac{2\pi}{\lambda}  \left[\frac{y^2+z^2}{2\,d}- \tilde z  \sin \theta_n\right]} dy dz \right|^2 }{\left|\eta_i(\Delta d,0) \right|}>\xi^\star\,.
\end{align}
}
In this case it is not possible to obtain simple expressions but their numerical solution allows to determine the values of the $\mathrm{SIR}_\text{NR-lens}$ for different values of $d,\,\theta, \Delta d$, {as shown in Sec.~\ref{sec:res}.}

\subsection{\ac{SIR} using \ac{no-lens}}
The same {expressions as in \eqref{eq:snrlens-multint}} hold also for the no-lens scenario. In particular, in this case we have
\begin{align}
    \etau &= x_0^2\,\Narray, \quad \quad \etai=x_0^2\sum_{i=1}^{\Nint} \eta_i\,,
\end{align}
with
\begin{align}
    \eta_i=e^{j(\chi_u-\chi_i)}\sum_{n=1}^{\Narray}   e^{j 2\pi f_0 \left(\tau (\p_n,\pu)-\tau (\p_n,\pin)\right)} \,,
\end{align}
that allows us to write
\begin{align}\label{eq:sirNolens}
\mathrm{SIR}_\text{no-lens}=
%\frac{\Narray}{\left|\sum_{i=1}^{\Nint}e^{j(\chi_u-\chi_i)}\sum_{i=1}^{\Narray}   e^{j 2\pi f_0 \left(\tau (\p_n,\pu)-\tau (\p_n,\pin)\right)}  \right|}=
\frac{\Narray}{\left|\sum_{i=1}^{\Nint}\eta_i\right|} \,.
\end{align}
%
%with $\eta(d,\theta,\Delta d_i,\Delta \theta_i)=\sum_{i=1}^{\Narray}   e^{j 2\pi f_0 \left(\tau (\p_n,\p)-\tau (\p_n,\p^{(i)})\right)} $. 
By making the approximations $\etai=\eta_i(\Delta d,0)$ as in Sec.~\ref{sec:sirRlens}, we can write \eqref{eq:sir0} as
\begin{align}\label{eq:sirnemp}
     \mathrm{SIR}_\text{no-lens}=\frac{\Narray}{\left|\sum_{n=1}^{\Narray} e^{-j \, \frac{2\,\pi}{\lambda}\,\gamma \, d^2_{n0}}\right|} >\xi^\star \,,
\end{align}
where $d_{n0}$ is defined below \eqref{eq:nolens-hn}, $n_y$ and $n_z$ are given in Sec.~\ref{subsec:nolens}, and $\gamma$ is given by \eqref{eq:kappa}.
Notably, since 
\begin{align}
\left|\sum_{n=1}^{\Narray} e^{-j \, \frac{2\,\pi}{\lambda}\,\gamma \, d^2_{n0}}\right|\leq \sum_{n=1}^{\Narray} \left|e^{-j \, \frac{2\,\pi}{\lambda}\,\gamma \, d^2_{n0}}\right|=\Narray \,,
\end{align}
we obtain $\mathrm{SIR}_\text{no-lens}\geq 1$. 

Again, the values of $\Delta d$ satisfying \eqref{eq:sirnemp} represent the areas where the system is robust with respect to the interference. As an example, if $\Delta d /d \ll 1$, $\gamma \approx 0$ that yields to $\mathrm{SIR}_\text{no-lens} \approx 0\,$dB.
%Indeed, the increase of $\Narray$ is expected to improve the $\mathrm{SIR}$,

%In the following, the performance of the approaches herein described are compared and discussed.

\section{Numerical Results}\label{sec:res}
\label{sec:results}

\begin{figure}[t!]
\psfrag{x}[c][c][0.7]{$\Ae$}
\psfrag{y}[c][c][0.7]{$\Narray$}
\psfrag{z}[c][c][0.7]{NR-lens}
\psfrag{data11111111111111}[c][c][0.7]{\!\!\!\!\!\! \!\!\!\!\!\!\!\!\!\! \!\!\!\!\!\!\!\!\!\! \!\!\!\!\!\!\!\!\!\!\!\!\!\!\! R-lens}
\psfrag{data2}[c][c][0.7]{\quad \!\!\! no-lens}
\psfrag{data3}[c][c][0.7]{\quad \quad\quad \quad\quad \, NR-lens, $D_z=D_y$}
\psfrag{data4}[c][c][0.7]{\quad \quad\quad \quad\quad\quad NR-lens, $D_z=2\,{D_y}$}
\psfrag{data5}[c][c][0.7]{\quad \quad\quad \quad\quad\quad NR-lens, $D_z=4\,{D_y}$}
\psfrag{data6}[c][c][0.7]{\quad \quad\quad \quad\quad\quad  NR-lens, $D_z=6\,{D_y}$}
\psfrag{data7}[c][c][0.7]{\quad \quad\quad\quad \quad\quad  NR-lens, $D_z=8\,{D_y}$}
\centerline{
     \includegraphics[width=0.5\linewidth,draft=false]
    {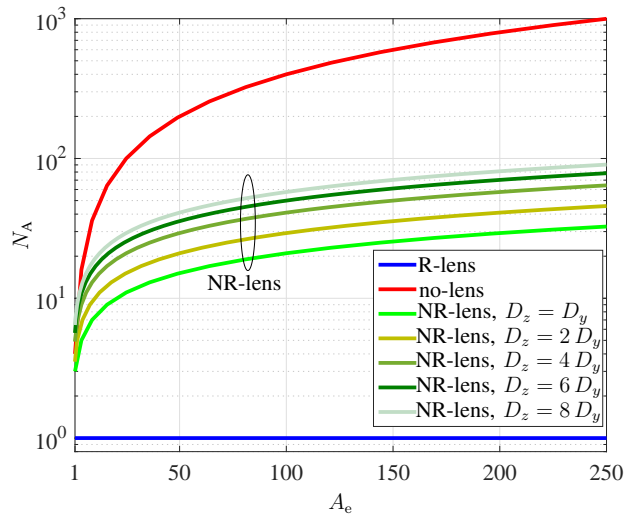}}
\caption{Number of required antennas for each configuration.}\label{fig:reqAnt} 
\end{figure}

In this section, the different architectures are compared in terms of positioning performance and interference rejection in some specific configurations. {In particular, we first consider a single user scenario and, successively, we move to the multi-user case in order to appreciate the capability to discriminate an intended useful source in presence of interference. In all the configurations we discuss about the complexity trade-off, evidencing the cost in terms of performance loss when the number of RF chains is reduced. 
In fact, for a fixed aperture $\Ae$, a different number of antennas (and hence RF processing chains) is required to each architecture, as shown in Table.~\ref{tab:tradeoff}.
Thus, there are two competing effects. From one side, a higher complexity in the processing at \ac{EM} level allows to reduce the number of RF chains. In fact, in the extreme scenario of the \ac{R-lens}, it holds $\Narray=1$, regardless the choice of the aperture (see Fig.~\ref{fig:reqAnt}) but a reconfigurable phase profile should be designed for the lens. When no \ac{EM} processing is adopted, there is the need to employ large $\Narray$ (i.e., a large number of RF chains), as for the \ac{no-lens}. In between, there is the trade-off offered by the \ac{NR-lens}, where the number of antennas depends on the geometry, i.e., on the choice of $D_z$, but it is required to perform some processing in the analog domain with a \ac{NR-lens}.}

\begin{table}[t!]
\centering
\caption{Complexity trade-off as a function of $\Ae$.}
\label{tab:tradeoff}
\begin{tabular}{l|l|l|}
\rowcolor{aliceblue2} \textbf{Architecture} &  \textbf{Required RF Chains} & \textbf{EM Processing} \\ \hline
R-lens & $\Narray=1$ & reconfigurable lens  \\ \hline
  NR-lens &  $\Narray=1+\big\lfloor 2\,\frac{\Ae \lambda}{D_y}\big\rfloor$ & non-reconfigurable \\ \hline
no-lens &$\Narray=\lfloor 4\,\Ae\rfloor$ & absent  \\\hline
\end{tabular}
\end{table}
 
%Since usually \ac{NR-lens} have a bi-dimensional aperture, we considered also a physical aperture of $D_y=D_z$ along the $y$-axis. Notably, in agreement with the analysis and comparison of the architectures conducted in \cite{ZenZha:J16}, this implies a higher normalized aperture  $\Ae$ when the lens is considered, thanks to its capacity to collect and collimate the wavefront of the impinging wave.
%When instead the \ac{R-lens} is accounted for, only one antenna is used, but the \ac{EM} processing is supposed to have the same physical length $D_\text{z}$ as its counterparts.

In the following, if not otherwise indicated, we consider rectangular areas with $D_y=2.5$\,cm in all scenarios, and $D_z=10\,$cm, $D_z=15\,$cm and $D_z=20\,$cm that correspond to normalized areas of $\Ae=D_y\,D_z/\lambda^2=100$, $\Ae=150$ and $\Ae=200$, respectively.

\subsection{{Single-User Scenario} }\label{sec:pos}

\paragraph{{Simulation Scenario}} 
{In accordance with the analysis of Sec.~\ref{sec:positioning}, we now evaluate the position estimation performance in a single-user scenario.} 
We consider a transmitter sending a signal centered at $f_0 \!=\! 60\,$GHz, with an \ac{EIRP} of $23\,$dBm. 
%Notably, such a value is still in line with the \ac{FCC} regulations in the relative frequency bandwidths \cite{FCC60r}.
At the receiver, we account for a noise power of $-106\,$dBW, and the parameter $A_\mathrm{pl}$ of \eqref{eq:h} is obtained from the link budget in free-space condition.

Results are expressed in terms of the \ac{RMSE} of the position estimate, which is evaluated as
\begin{align}
\mathsf{RMSE} \left({\p} \right)= \sqrt{\frac{1}{N_\text{c}}\sum_{m=1}^{N_\text{c}} \lVert  \hat{\p}_{m} - \p \rVert^2} \,,
\end{align}
where $N_\text{c}$ is the number of Monte Carlo iterations considered in simulations and $\hat{\p}_{m}$ is the position estimate at the $m$th iteration. In particular, for each iteration, a different noise realization is generated according to $\sigma^2$, as well as a different realization of phase $\chi$, which is kept the same for all the antennas. In this way, the random phase models a complete clock mismatch between the transmitter and the receiver.

\begin{figure}[t!]
\psfrag{x}[c][c][0.7]{$d \,$[m]}
\psfrag{y}[c][c][0.7]{$\mathsf{RMSE} ({\p})$ [m]}
\psfrag{data11111111111111111111}[c][c][0.7]{\!\!\!\!\!\!\ac{R-lens}, $\Ae=100$}
\psfrag{data2}[c][c][0.7]{\!\!\quad \quad\quad \quad\quad \quad\ac{R-lens}, $\Ae=150$}
\psfrag{data3}[c][c][0.7]{\!\!\quad \quad\quad \quad\quad \quad\ac{R-lens}, $\Ae=200$}
\psfrag{data4}[c][c][0.7]{\,\quad \quad\quad \quad\quad \quad\ac{NR-lens}, $\Ae=100$}
\psfrag{data5}[c][c][0.7]{\,\quad \quad\quad \quad\quad \quad\ac{NR-lens}, $\Ae=150$}
\psfrag{data6}[c][c][0.7]{\,\quad \quad\quad \quad\quad \quad\ac{NR-lens}, $\Ae=200$}
\psfrag{data7}[c][c][0.7]{\!\quad \quad\quad \quad\quad \quad\ac{no-lens}, $\Ae=100$}
\psfrag{data8}[c][c][0.7]{\!\!\quad \quad\quad \quad\quad \quad\,\,\ac{no-lens}, $\Ae=150$}
\psfrag{data9}[c][c][0.7]{\!\!\quad \quad\quad \quad\quad \quad\,\,\ac{no-lens}, $\Ae=200$}
\psfrag{data1111111111111111}[c][c][0.7]{ \!\!\!\!\!\!\!\!\!\!\!\!\!\!\!\!\!\!\!\!\!\! $\Ae=100$}
\psfrag{data22}[c][c][0.7]{ \quad \, $\Ae=150$}
\psfrag{data33}[c][c][0.7]{ \quad \, $\Ae=200$}
\psfrag{data44}[c][c][0.7]{\quad \, \quad \quad \quad\!\!\!\! $\Ae=100$, diff.}
\psfrag{data55}[c][c][0.7]{\quad \, \quad \quad \quad\!\!\!\! $\Ae=150$, diff.}
\psfrag{data66}[c][c][0.7]{\quad \! \quad \quad \quad\!\! $\Ae=200$, diff.}
\centerline{     \includegraphics[width=0.5\linewidth,draft=false]
    {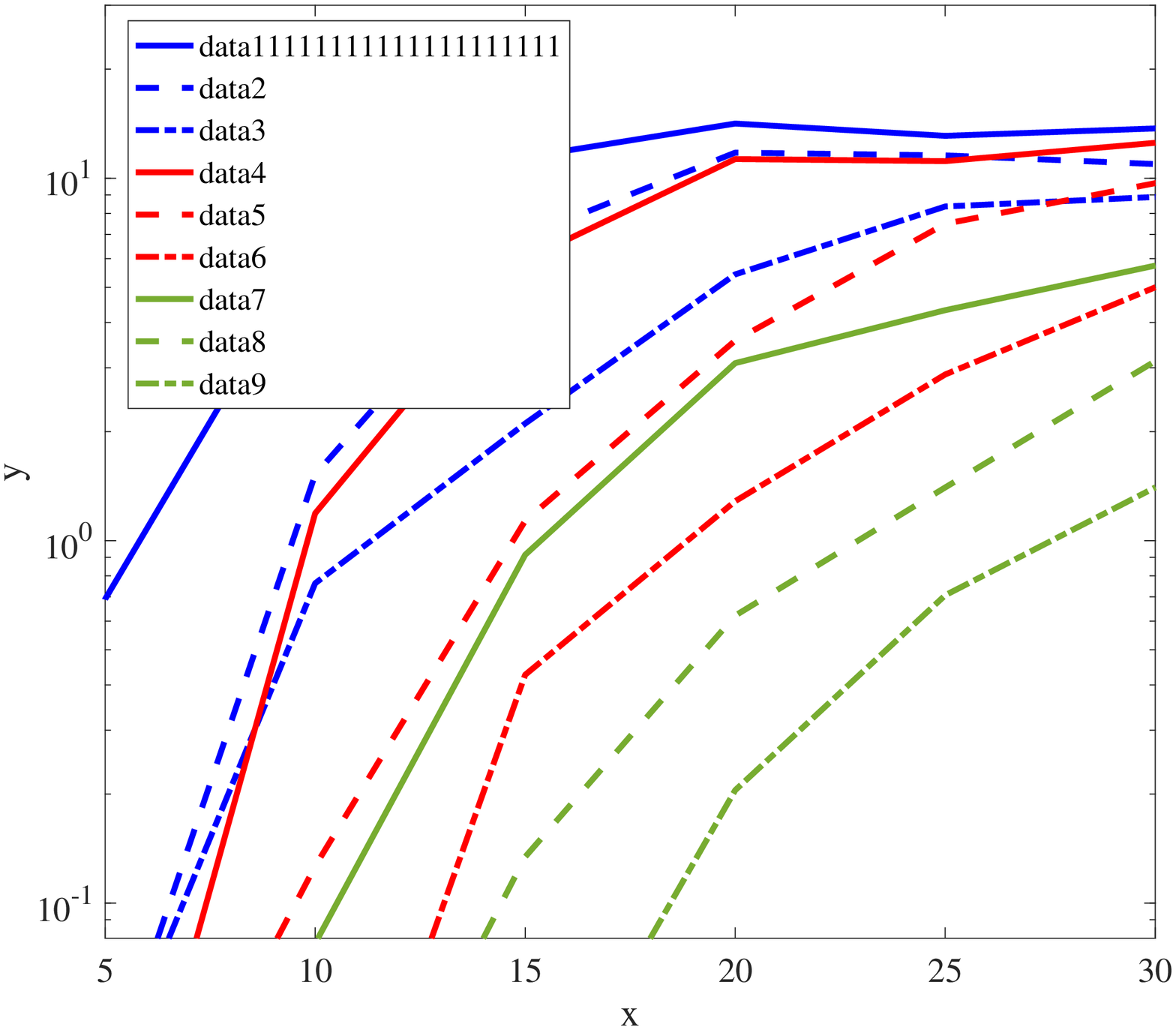}
   \includegraphics[width=0.5\linewidth,draft=false]
    {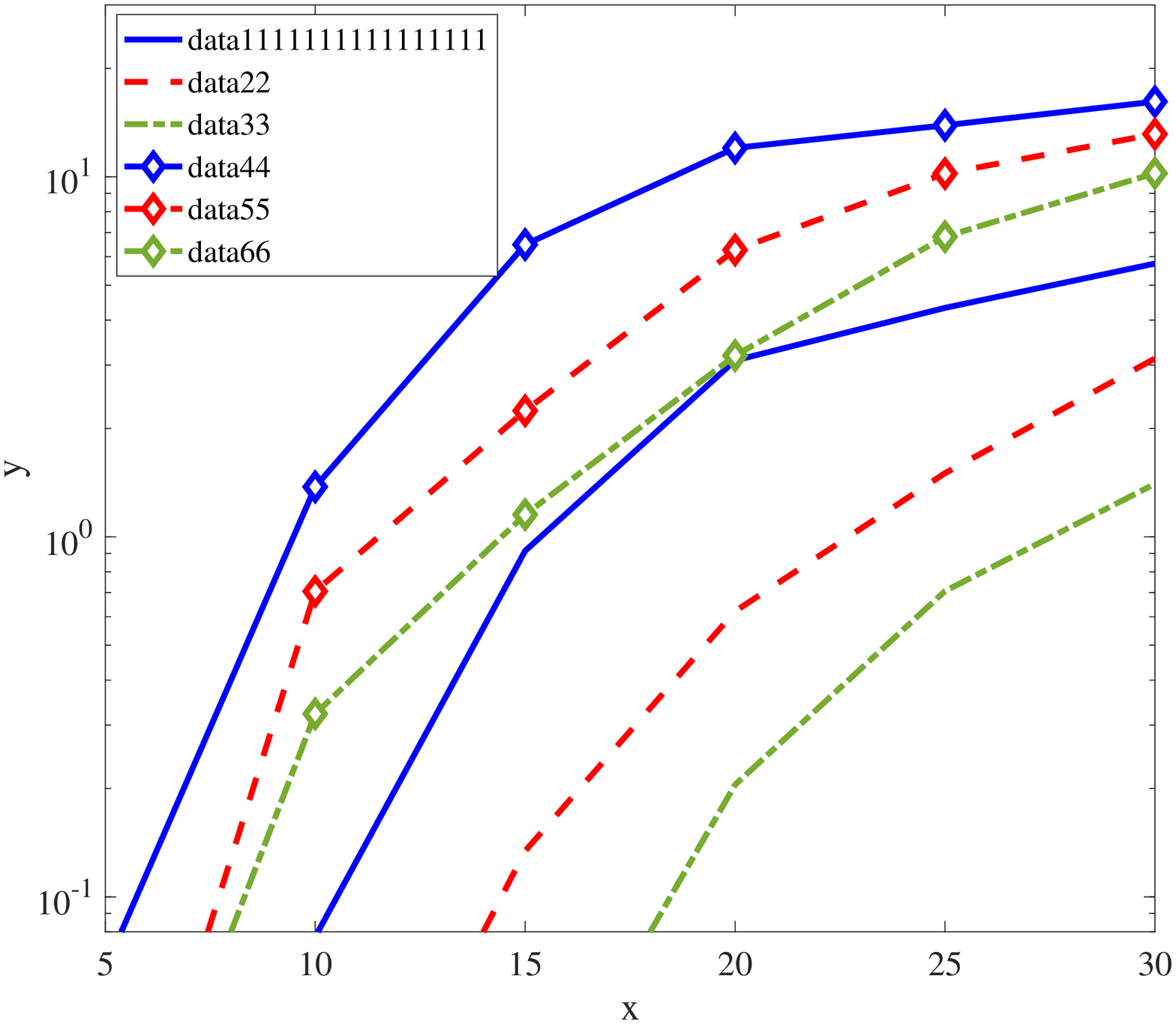}}
\caption{{RMSE} as a function of the TX-RX distance $d$, fixed $\text{AOA}=0^\circ$ and different architectures (left), and comparison between ML and differential estimation for no-lens (right).}\label{fig:RMSE-D}
\end{figure}

\paragraph{{RMSE for Different Source Distance}}
Figure~\ref{fig:RMSE-D}-left reports the obtained results for the different schemes and $\Ae$, {using the \ac{ML} position estimator in \eqref{eq:MLpos} for the \ac{NR-lens} and \ac{no-lens}, and the position estimator in \eqref{eq:Posrlens} for the R-lens.}
We initially fixed the \ac{AOA} to $0^\circ$ by varying only the TX-RX distance from $5\,$m to $30\,$m.
As evidenced in the figure, the larger is $\Ae$ the better is the position estimate thanks to the increased physical area that permit to collect a larger amount of power and exploit better the wavefront curvature. 

{The \ac{no-lens}-based solution provides the best performance thanks to the employment of $\Narray$ antennas (and RF chains), at the expense of a  much higher complexity at RF level. Nevertheless, the use of a \ac{R-lens} with  $\Narray=1$ (only one RF chain)} allows to attain a sub-meter \ac{RMSE} for $d=10\,$m. 
\begin{figure}[t!]
\psfrag{x}[c][c][0.7]{$x$ [m]}
\psfrag{y}[c][l][0.7]{$y$ [m]}
\psfrag{50}[c][c][0.55]{$50$}
\psfrag{45}[c][c][0.55]{45}
\psfrag{40}[c][c][0.55]{$40$}
\psfrag{35}[c][c][0.55]{35}
\psfrag{30}[c][c][0.55]{$30$}
\psfrag{25}[c][c][0.55]{25}
\psfrag{20}[c][c][0.55]{$20$}
\psfrag{15}[c][c][0.55]{15}
\psfrag{10}[c][c][0.55]{$10$}
\psfrag{18}[c][c][0.55]{$18$}
\psfrag{16}[c][c][0.55]{$16$}
\psfrag{14}[c][c][0.55]{$14$}
\psfrag{12}[c][c][0.55]{$12$}
\psfrag{8}[c][c][0.55]{$8$}
\psfrag{6}[c][c][0.55]{$6$}
\psfrag{5}[c][c][0.55]{$5$}
\psfrag{4}[c][c][0.55]{$4$}
\psfrag{2}[c][c][0.55]{$2$}
\psfrag{0}[c][c][0.55]{$0$}
\centerline{
   \includegraphics[width=0.35\linewidth,draft=false]
    {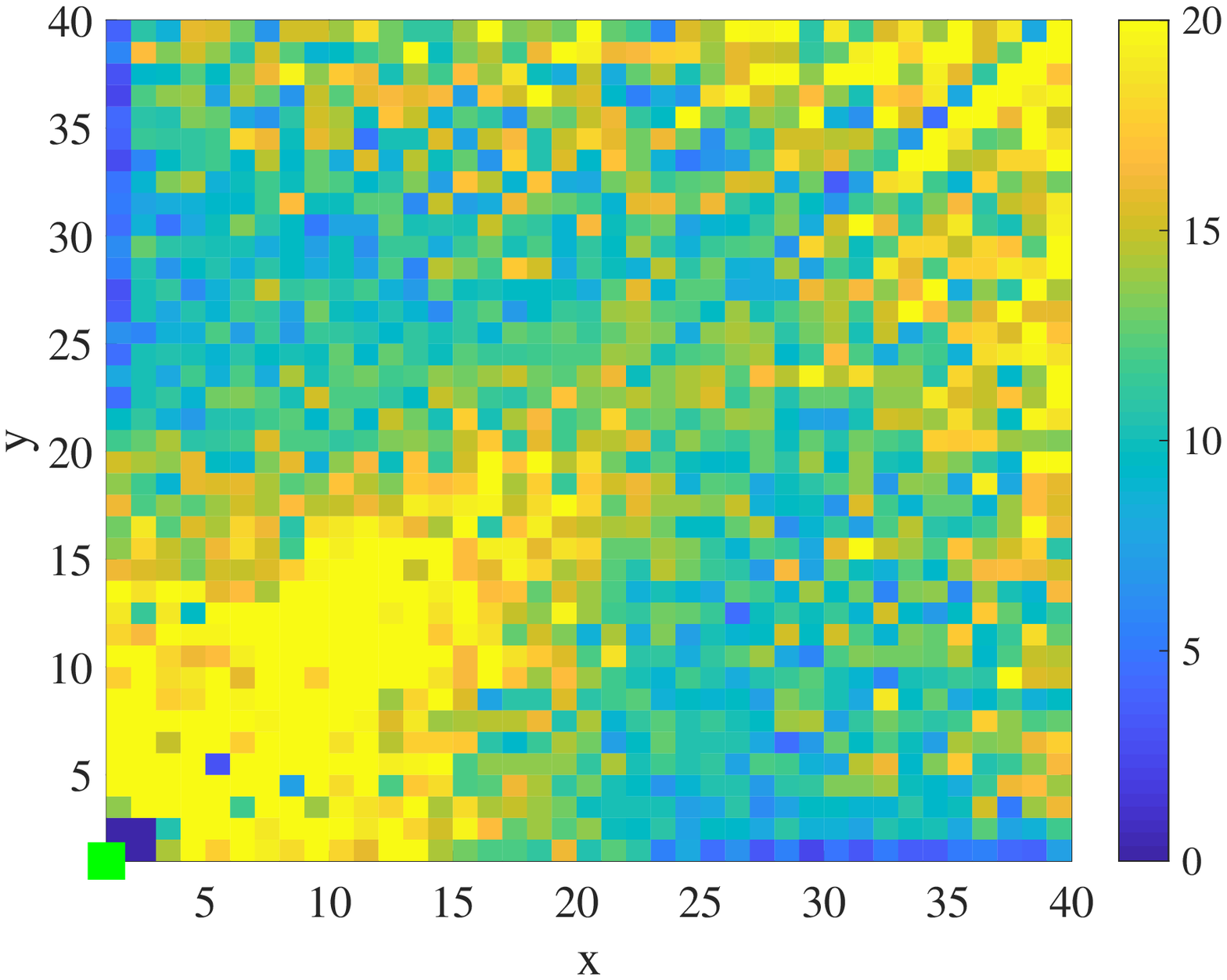}   \includegraphics[width=0.35\linewidth,draft=false]
    {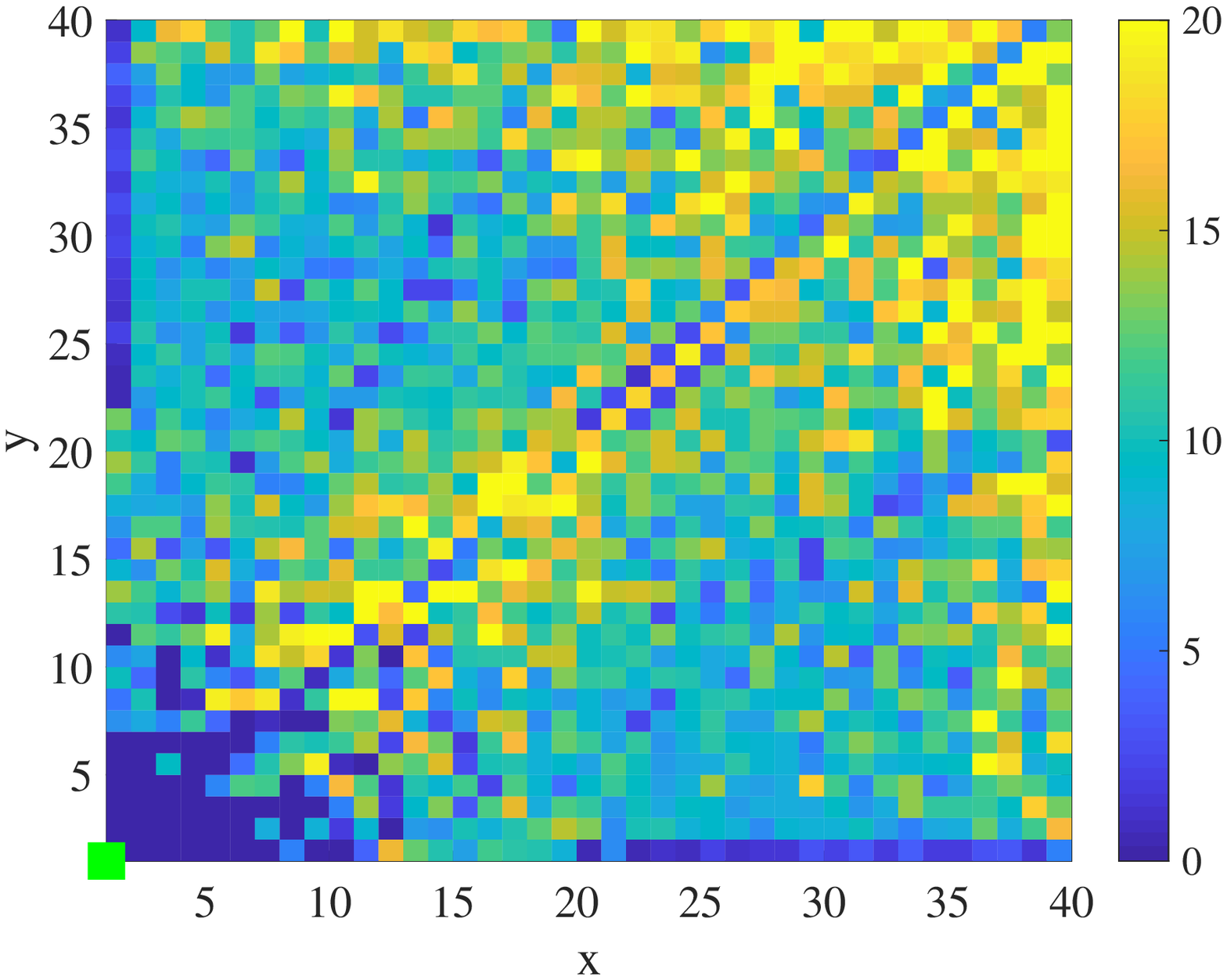}\includegraphics[width=0.35\linewidth,draft=false]
    {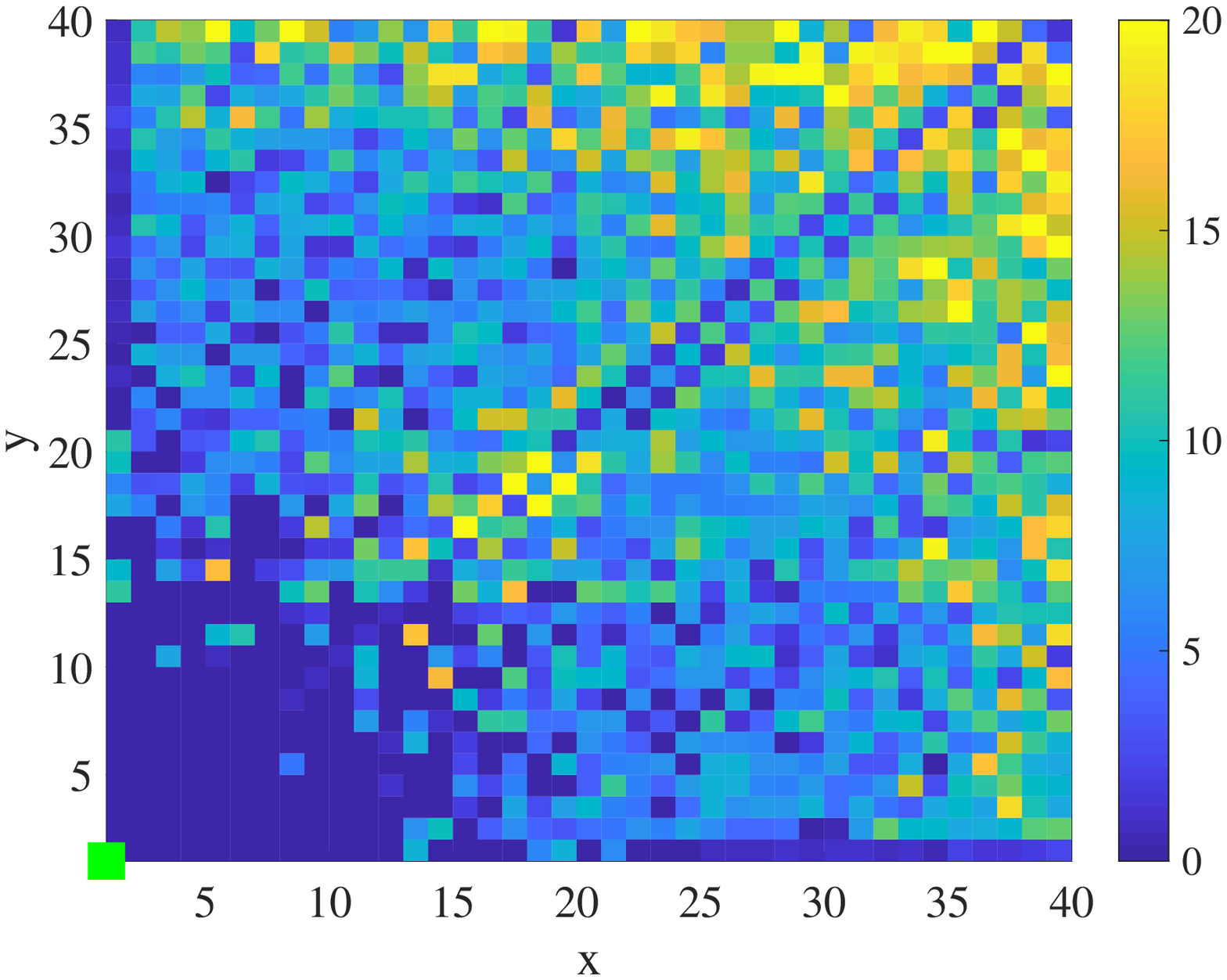}}
\centerline{
   \includegraphics[width=0.35\linewidth,draft=false]
    {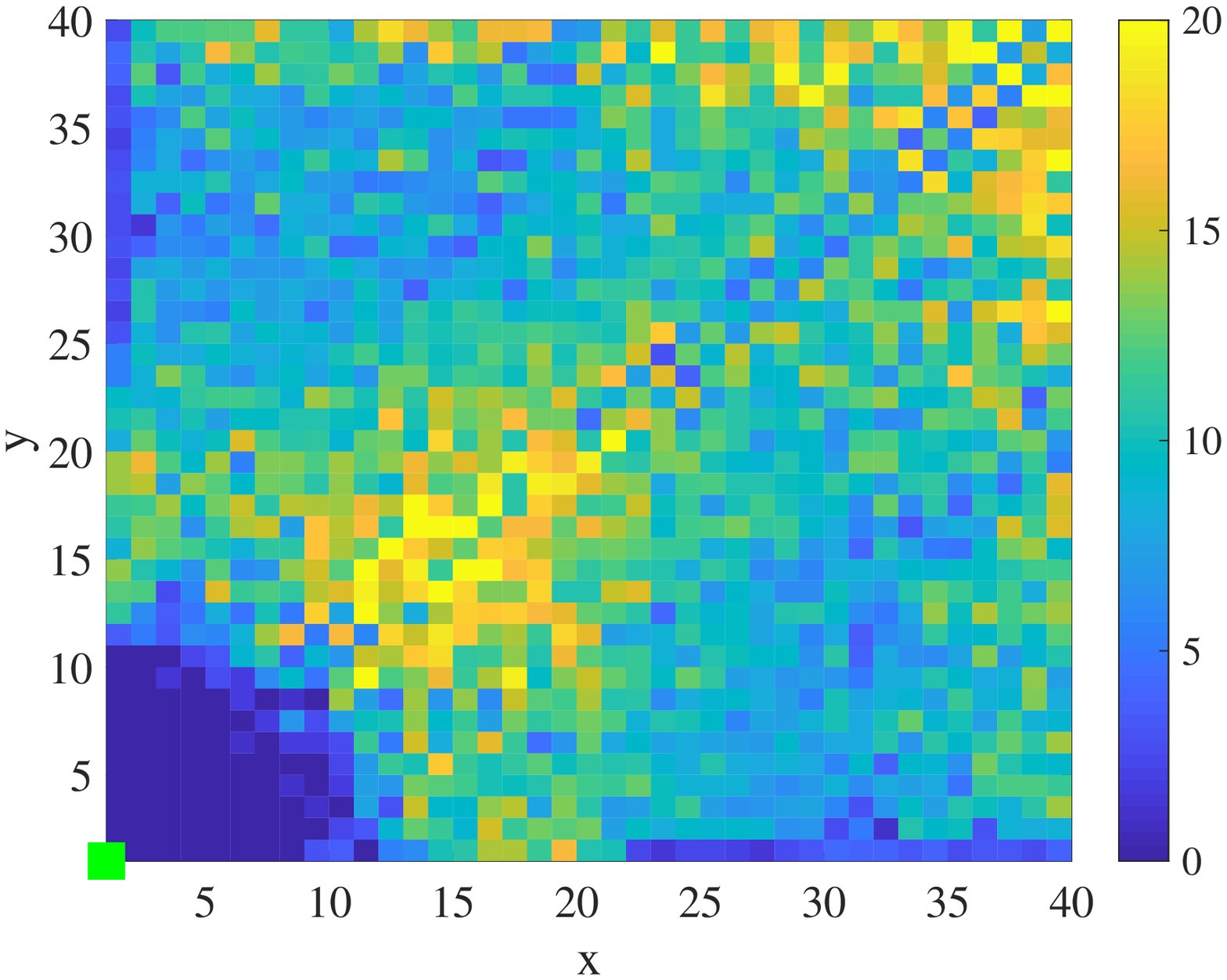}   \includegraphics[width=0.35\linewidth,draft=false]
    {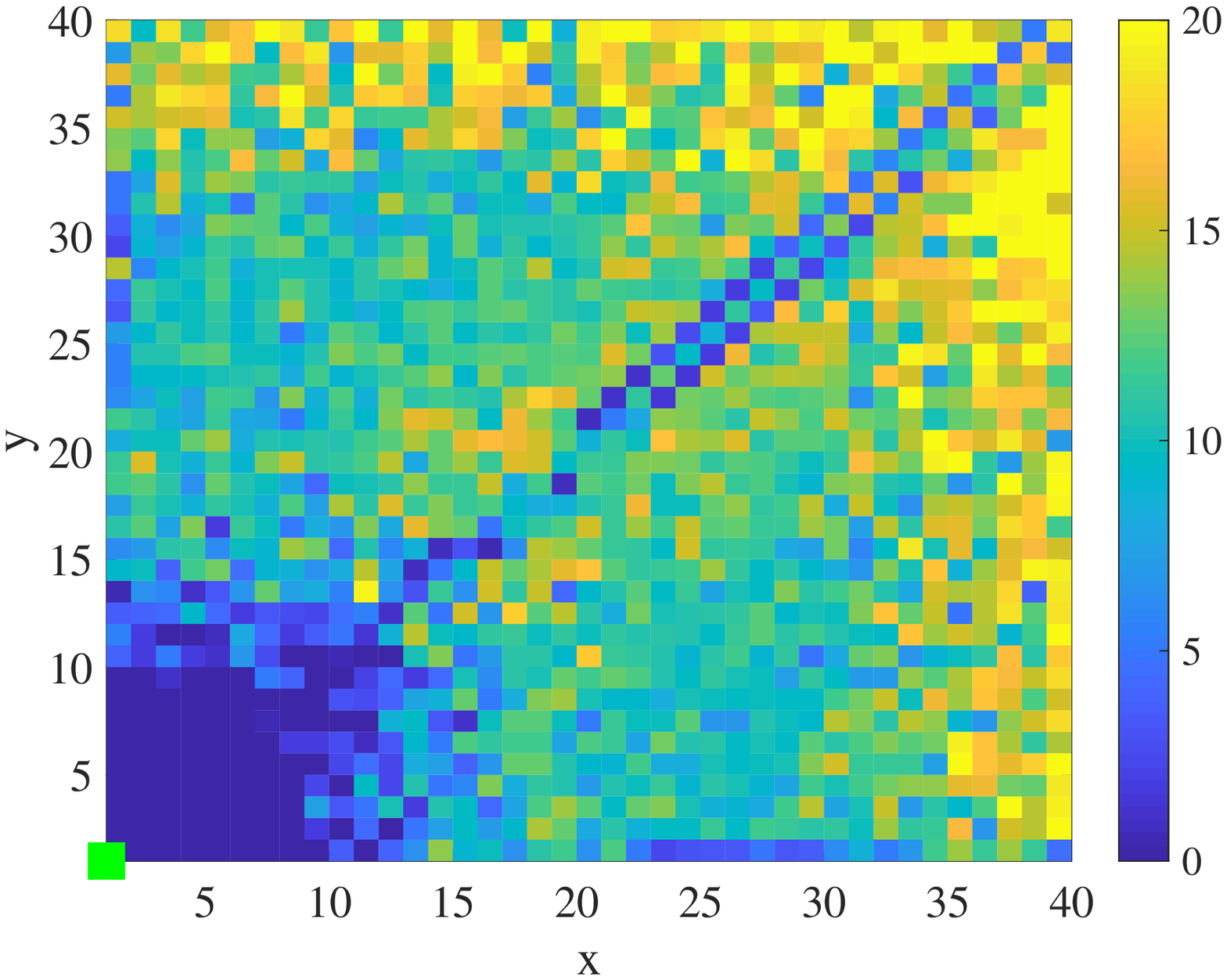}\includegraphics[width=0.35\linewidth,draft=false]
    {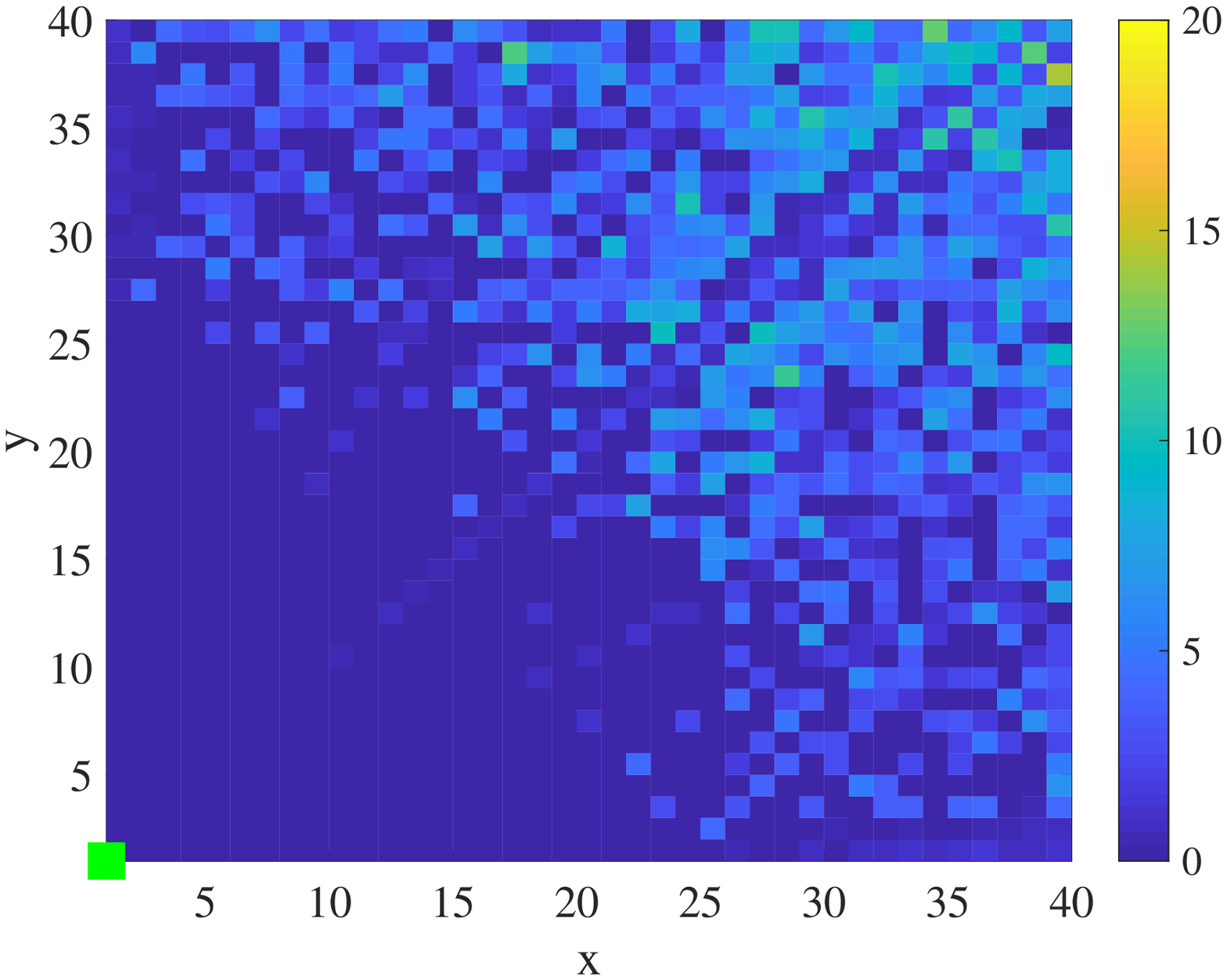}}
\centerline{
   \includegraphics[width=0.35\linewidth,draft=false]
    {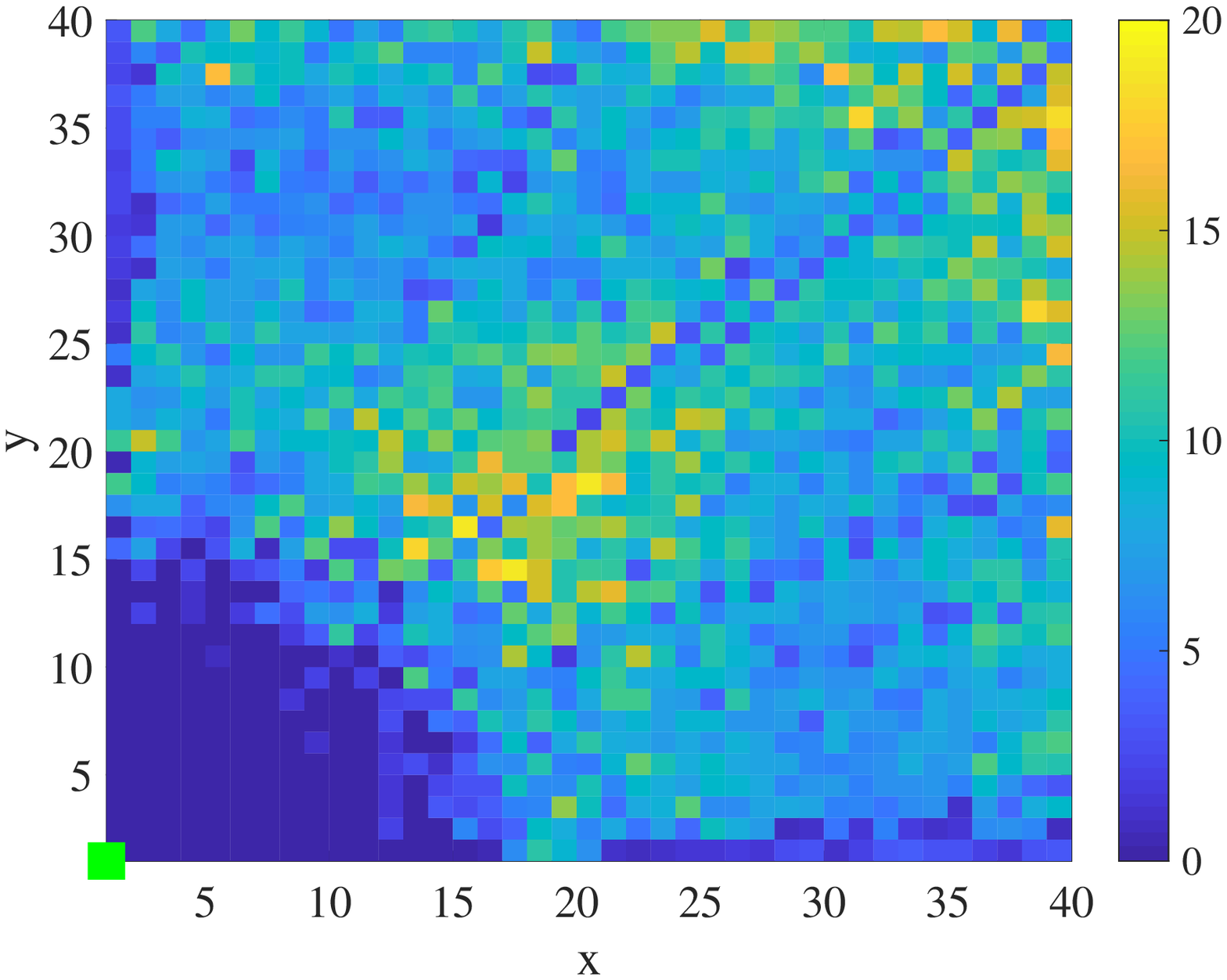}   \includegraphics[width=0.35\linewidth,draft=false]
    {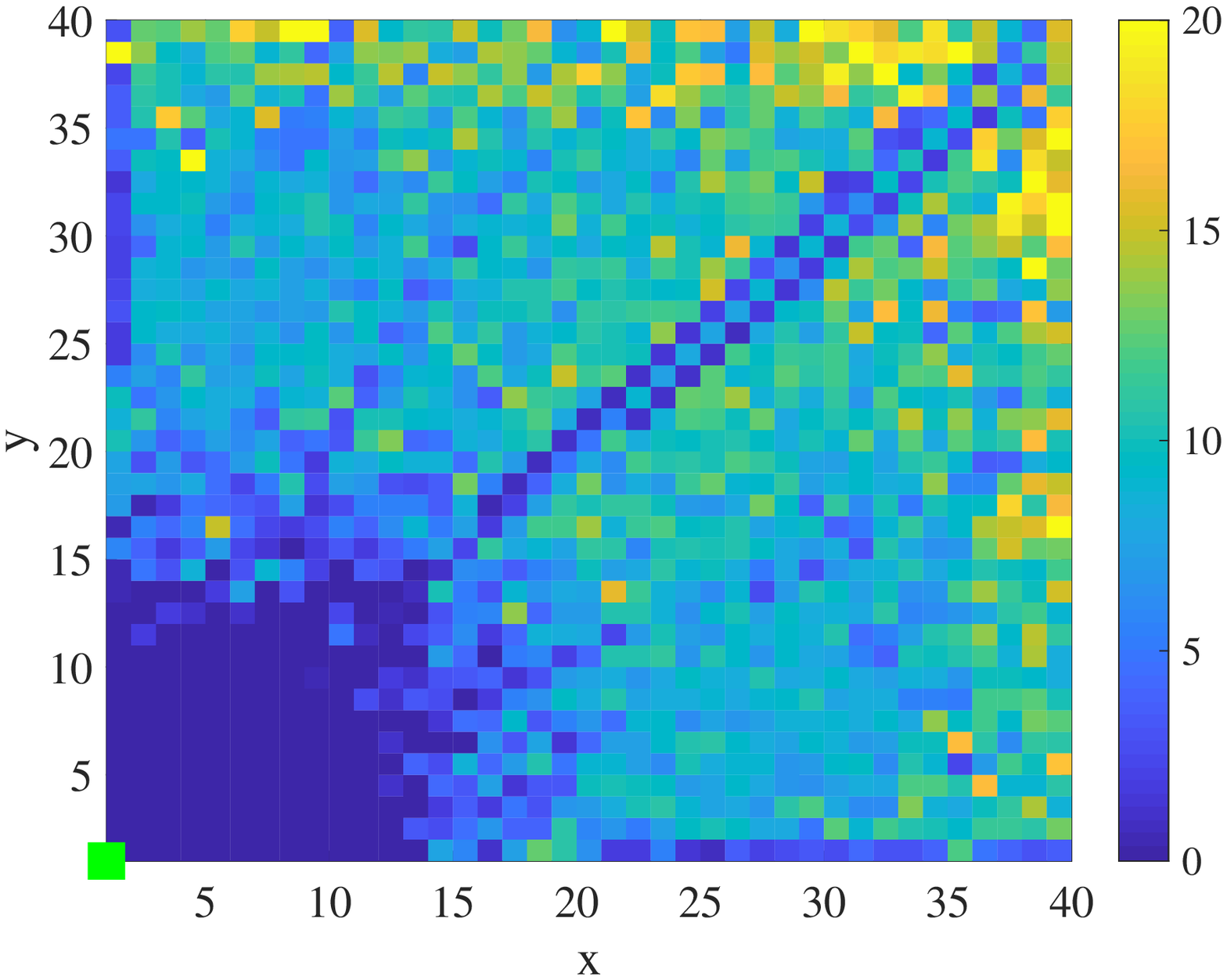}\includegraphics[width=0.35\linewidth,draft=false]
    {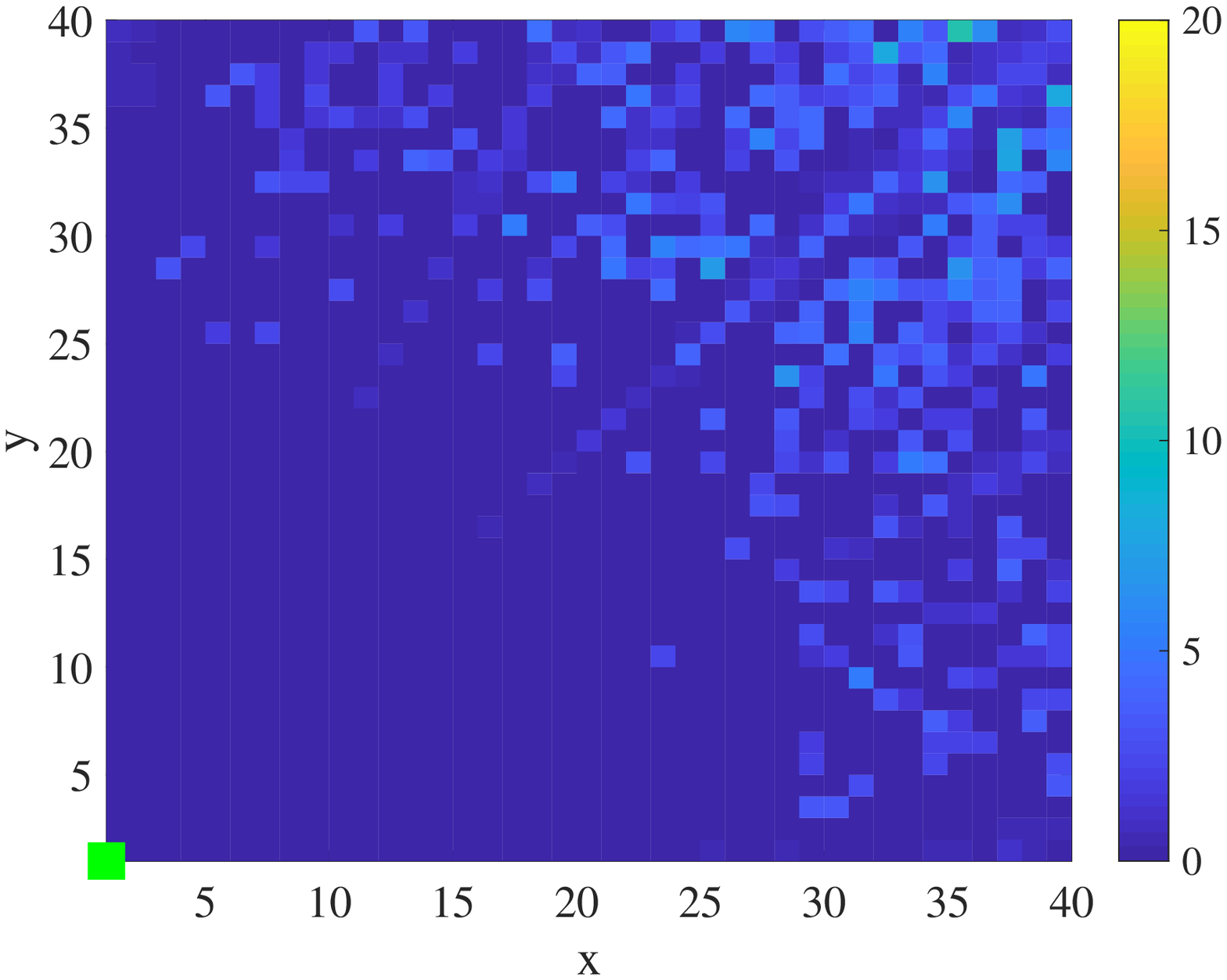}
}
\caption{RMSE maps (errors in meters) for \ac{R-lens} (left), \ac{NR-lens} (middle) and \ac{no-lens} (right), for $\Ae=100$ (top), $\Ae=150$ (middle) and $\Ae=200$ (bottom).}\label{fig:MappeScenario} 
\end{figure}
In between, the \ac{NR-lens} represents the trade-off {in terms of} RF ($\Narray$) and \ac{EM} processing (lens) complexity. This effect is more pronounced for larger distances, where the path loss increases.
In fact, as an example, for \ac{NR-lens} with $\Ae=250$ and $\Narray=81$, the positioning error is kept below $1\,$cm at $d=10\,$m, and at about $1\,$m for $d=20\,$m whereas, for the \ac{no-lens} scenario, the same error of $1\,$m is made at about $d=30\,$m.
Such a localization performance is obtained for a physical area $\Af=(10\times 2.5)\,\text{cm}^2$, $\Af=(15\times 2.5)\,\text{cm}^2$ and $\Af=(20\times 2.5)\,\text{cm}^2$, which are extremely compact and, thus, suitable for real scenarios, e.g., for an integration in future generation of access points. {Note also that the receiver
noise plays a crucial role as well. In fact, despite when the source is located at the Fraunhofer distance the wavefront can
be approximated to be planar, the receiver is still capable to infer the transmitter position if the
noise contribution is not extremely large.}

{In case of \ac{no-lens} and \ac{NR-lens}}, the performance has been obtained using the \ac{ML} estimator which, from one hand provides a useful benchmark, but from the other hand it might entail a high signal processing complexity, since it requires the search along the phase offset and the position.
In this sense, the differential approach proposed in Sec.~\ref{subsec:ortho} allows to avoid a bi-dimensional search (i.e., over $\chi$ and $\p$). 

Results are reported in Fig.~\ref{fig:RMSE-D}-right and they are encouraging, since performance is still reliable (i.e., $\mathsf{RMSE} ({\p})\approx 3\,$m at $d=20\,$m for $\Ae=200$) with the advantage that the computational complexity of the positioning algorithm can be reduced.

\paragraph{{RMSE for Different Source Angle and Distance}}
Finally, the aforementioned considerations for the different architectures are corroborated by the \ac{RMSE} maps reported in Fig.~\ref{fig:MappeScenario} for different $\Ae$. They are obtained by placing the receiver (i.e., the green square marker) in $(0,\,0)$ rotated towards the center of the area, i.e., towards the red marker, and by alternatively placing the transmitting source in all the grid points of the environment. For the considered scenario, in accordance with the results reported in Fig.~\ref{fig:RMSE-D}-left, the \ac{no-lens} provides the best performance thanks to the larger number of antennas involved in the curvature processing.

\begin{figure}[t!]
\psfrag{x}[b][c][0.7]{$x$ [m]}
\psfrag{y}[b][l][0.7]{$y$ [m]}
\psfrag{50}[b][c][0.55]{$50$}
\psfrag{40}[b][c][0.55]{$40$}
\psfrag{35}[b][c][0.55]{}
\psfrag{30}[b][c][0.55]{$30$}
\psfrag{25}[b][c][0.55]{}
\psfrag{20}[b][c][0.55]{$20$}
\psfrag{15}[b][c][0.55]{}
\psfrag{10}[b][c][0.55]{$10$}
\psfrag{5}[b][c][0.55]{}
\psfrag{1}[c][b][0.55]{$1$}
\psfrag{0}[b][c][0.55]{$0$}
\centerline{
   \includegraphics[width=0.33\linewidth,draft=false]
    {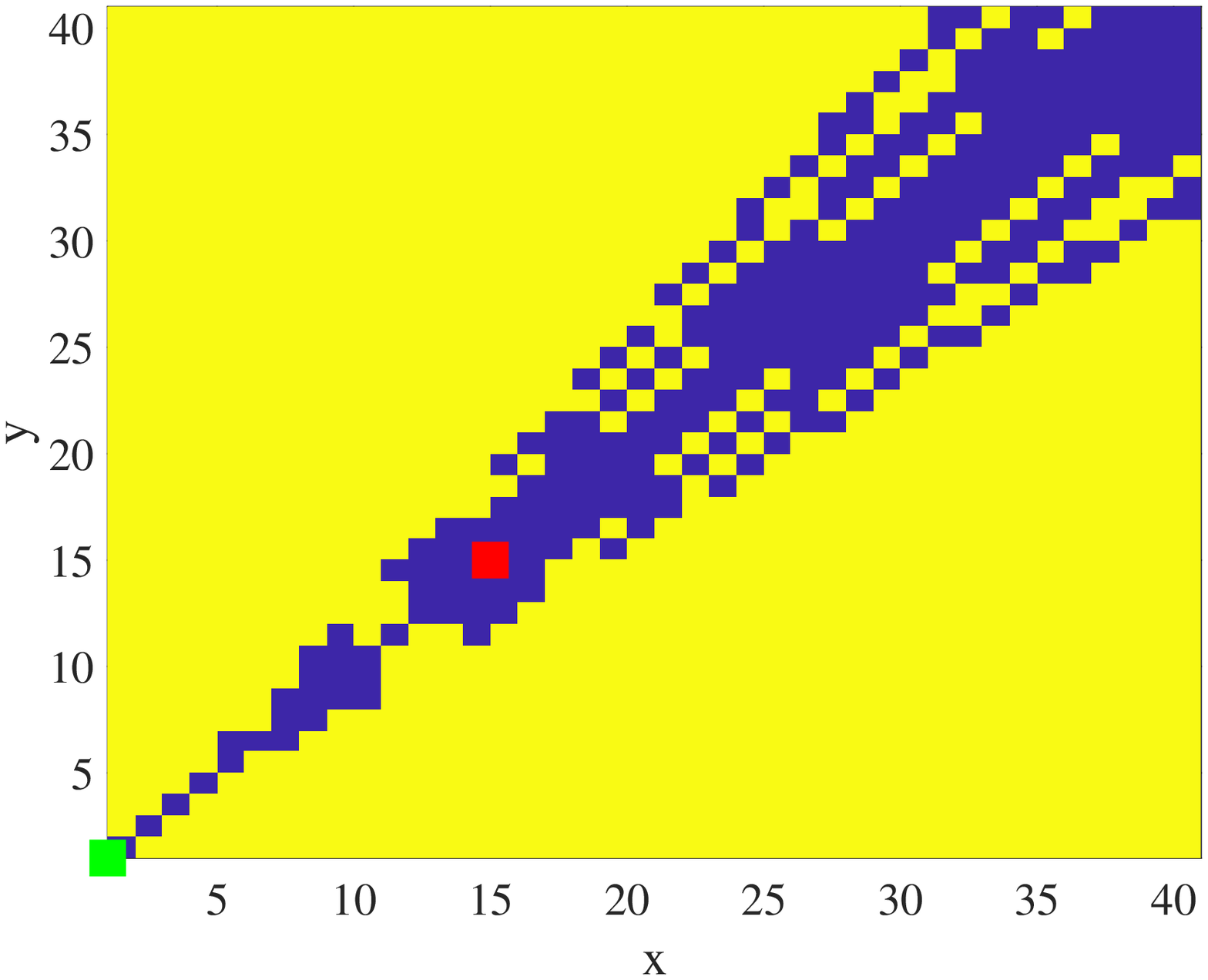}\includegraphics[width=0.33\linewidth,draft=false]
    {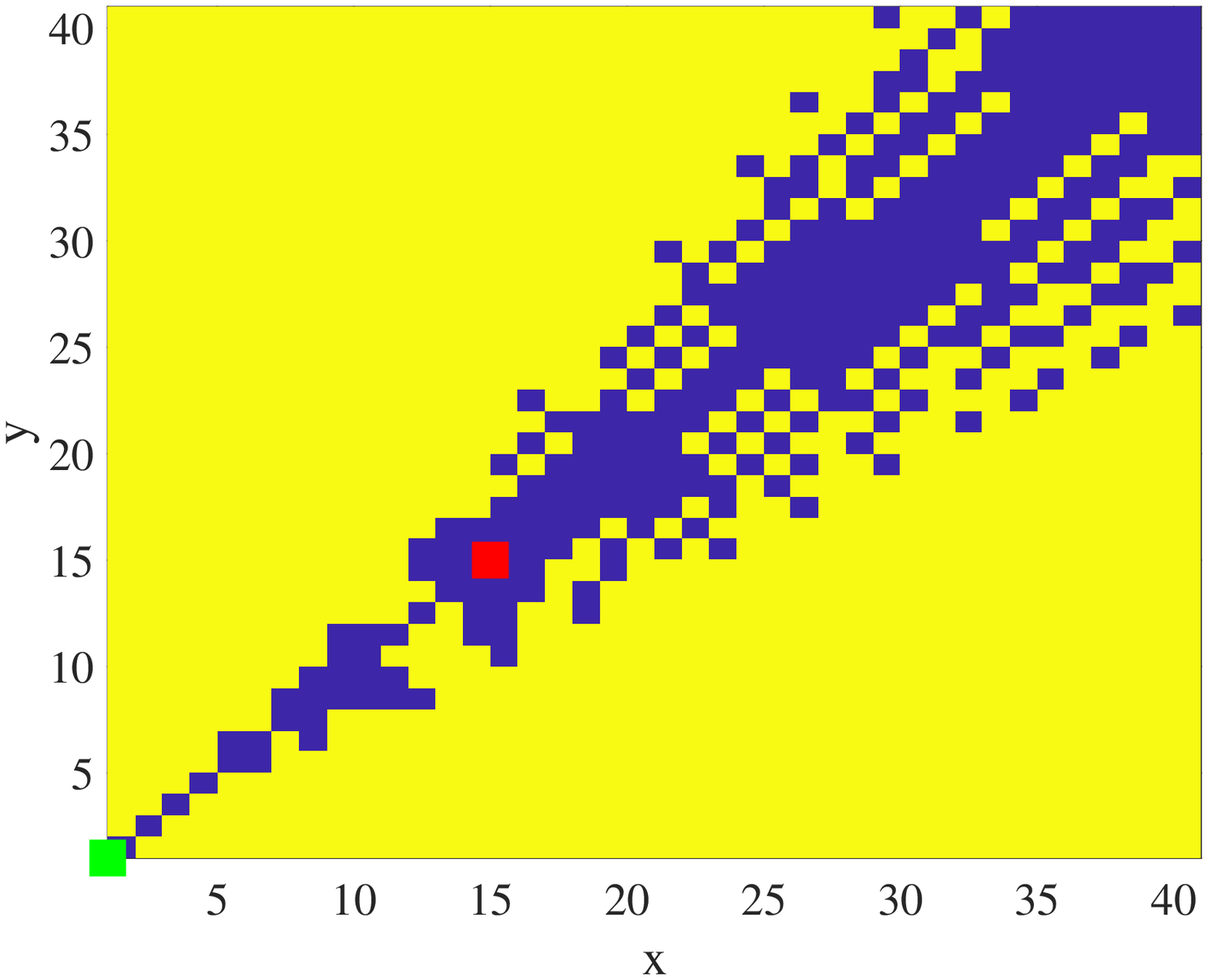}\includegraphics[width=0.33\linewidth,draft=false]
    {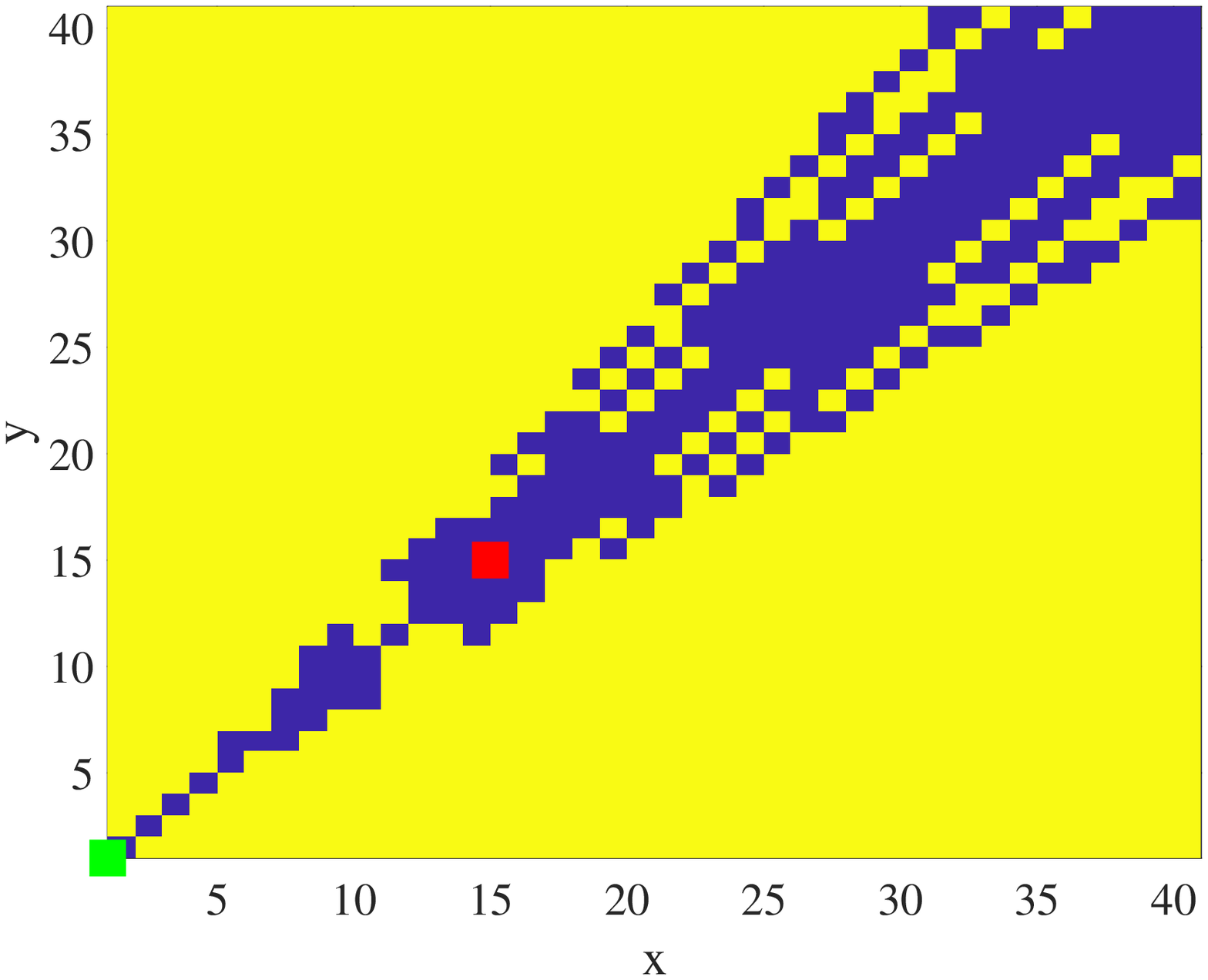}} 
\centerline{
   \includegraphics[width=0.33\linewidth,draft=false]
    {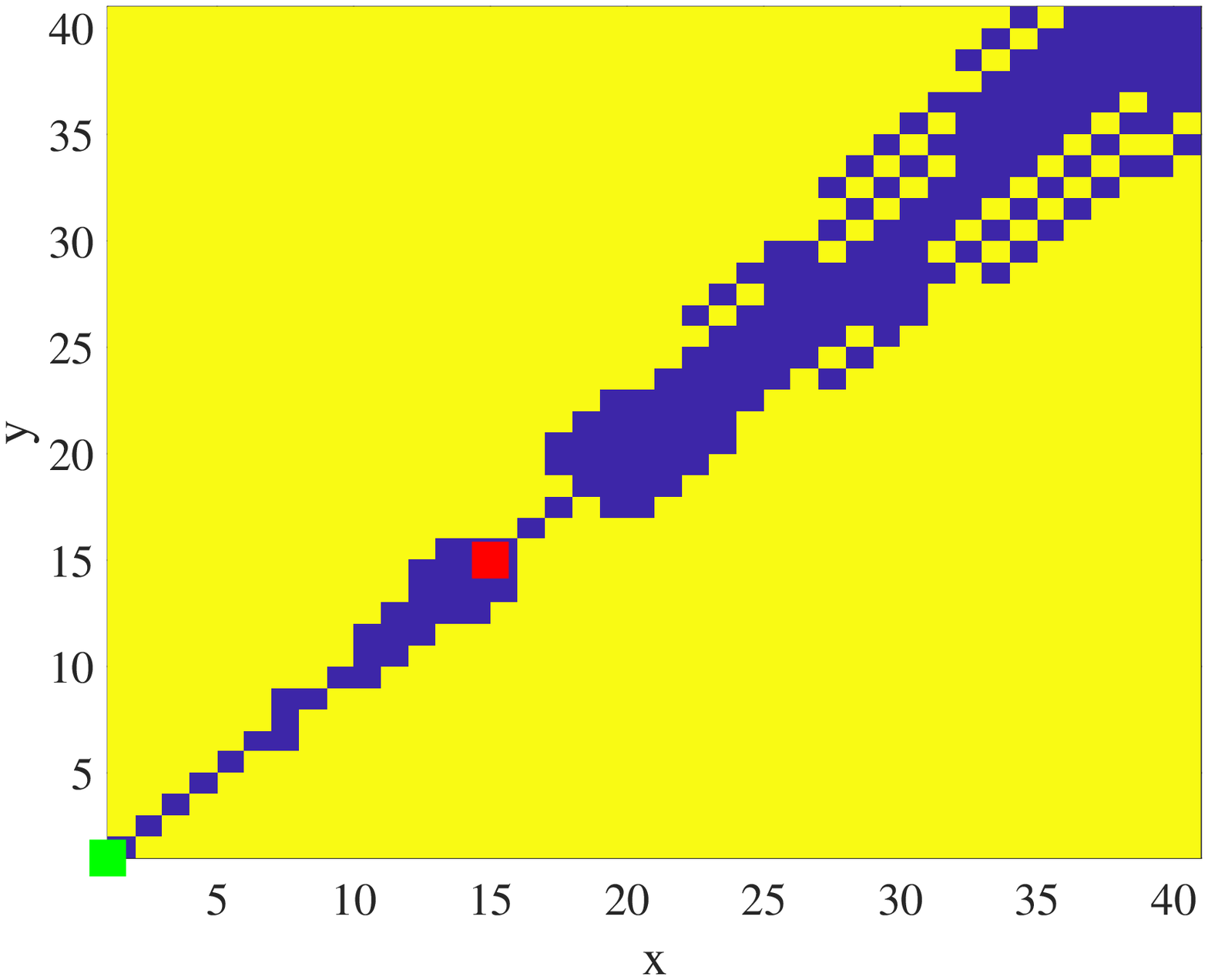}\includegraphics[width=0.33\linewidth,draft=false]
    {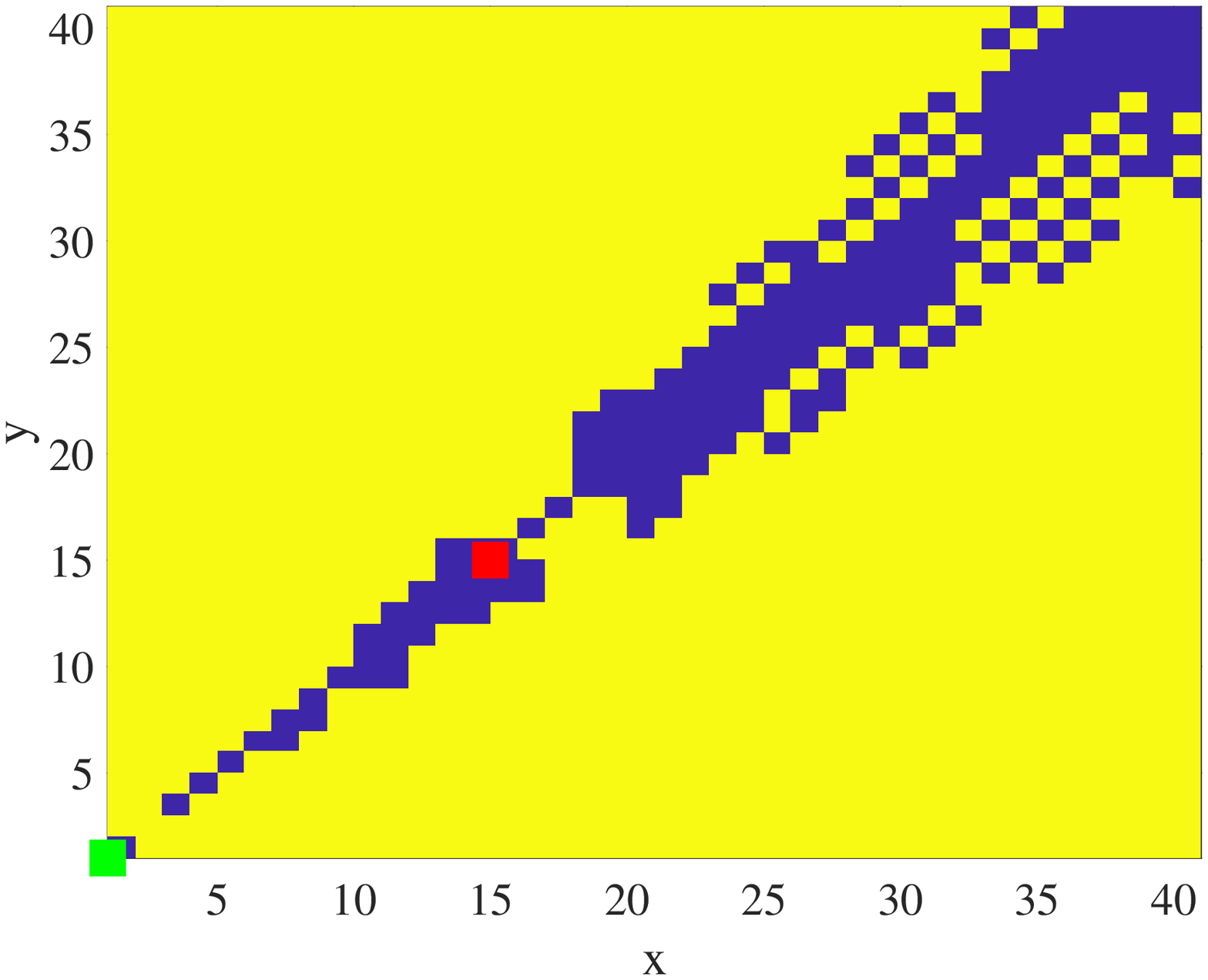}\includegraphics[width=0.33\linewidth,draft=false]
    {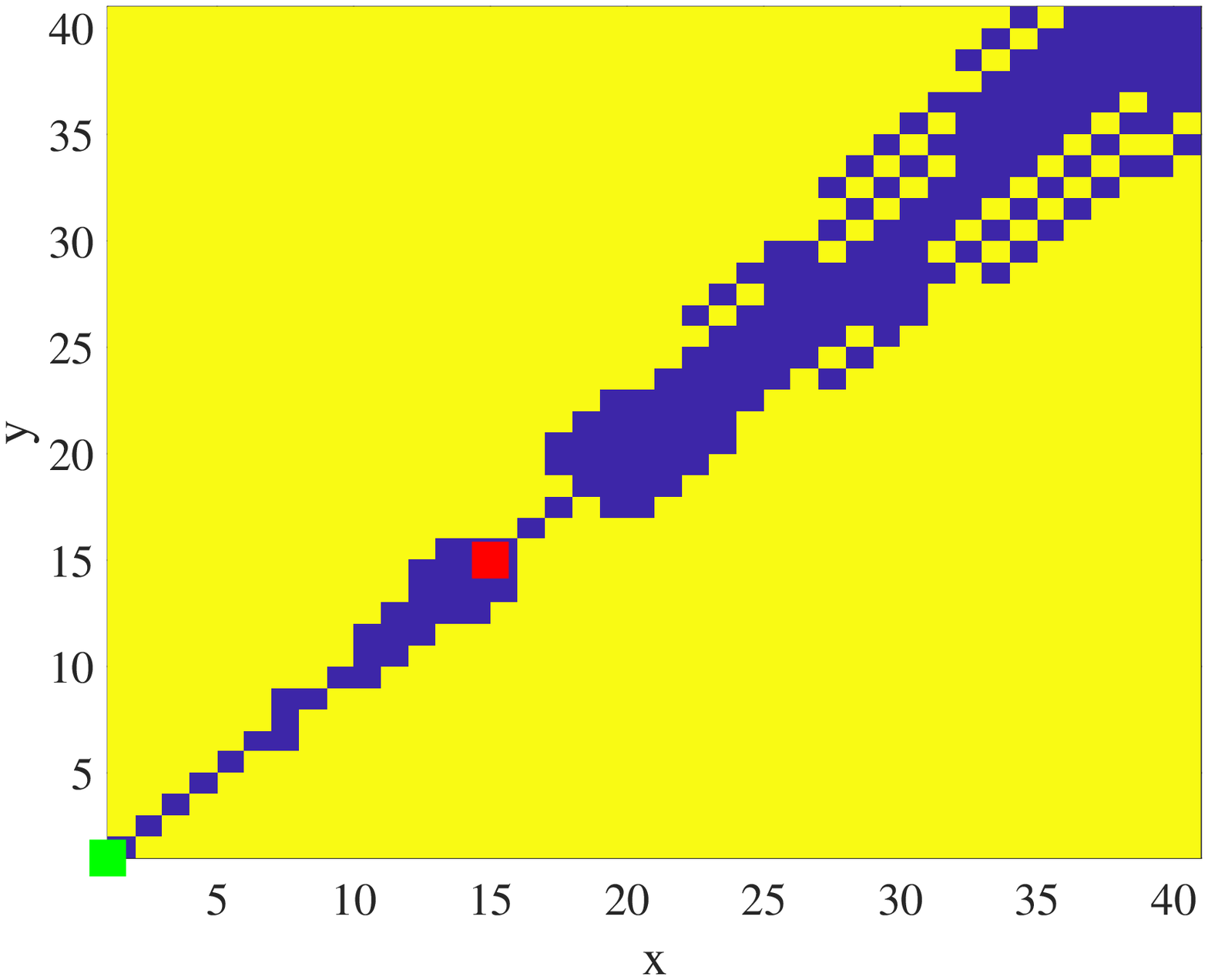}}      
\centerline{
   \includegraphics[width=0.33\linewidth,draft=false]
    {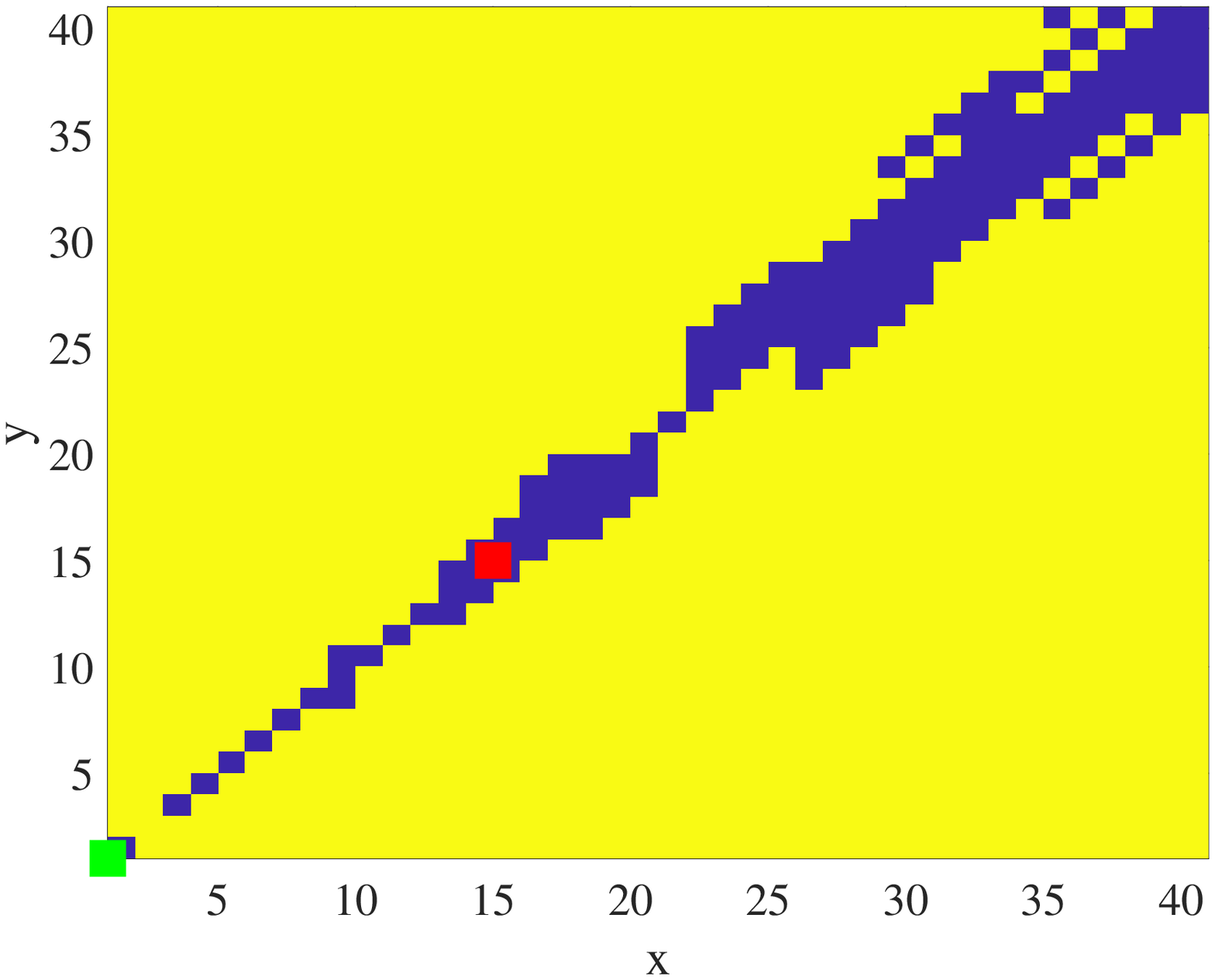}\includegraphics[width=0.33\linewidth,draft=false]
    {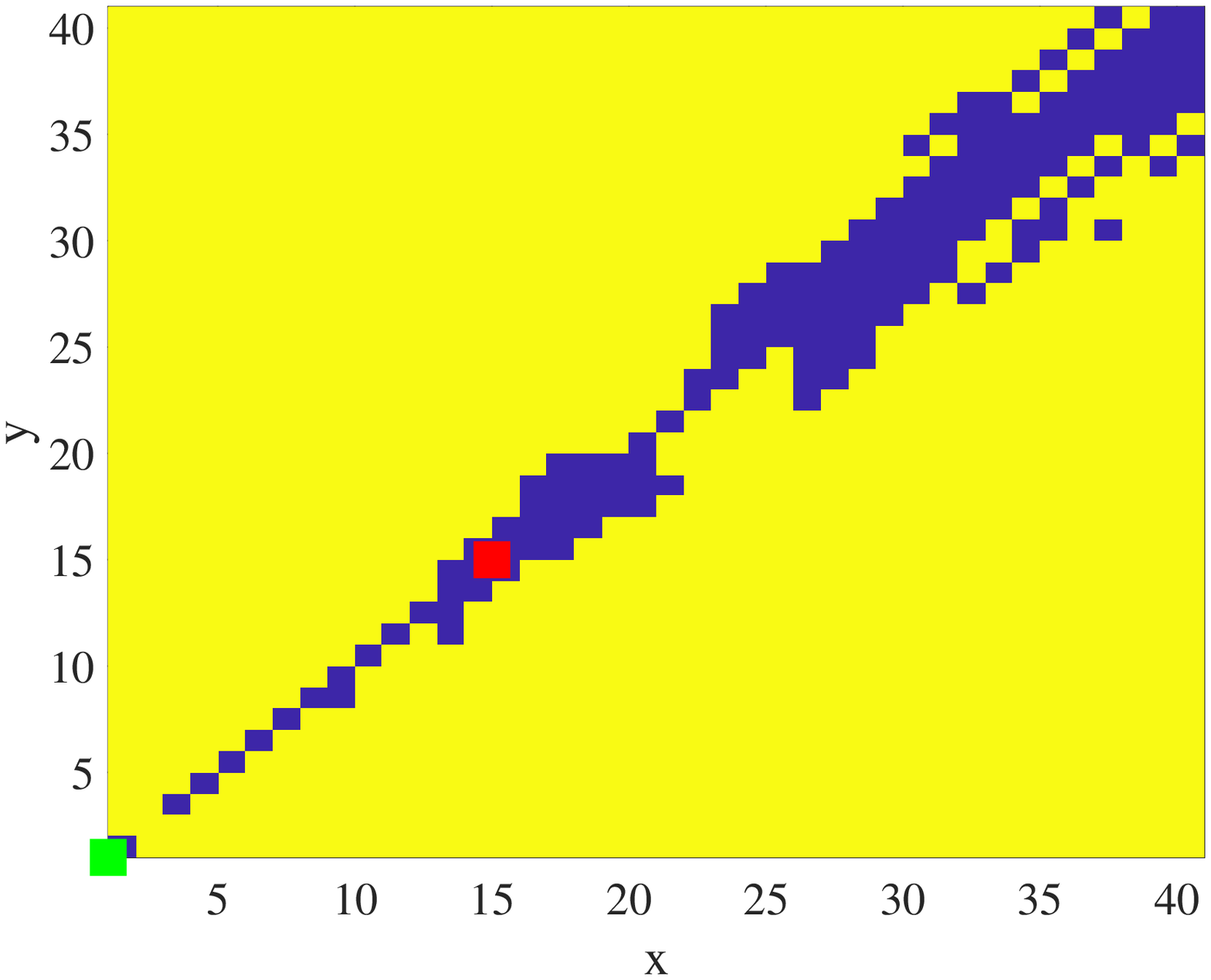}\includegraphics[width=0.33\linewidth,draft=false]
    {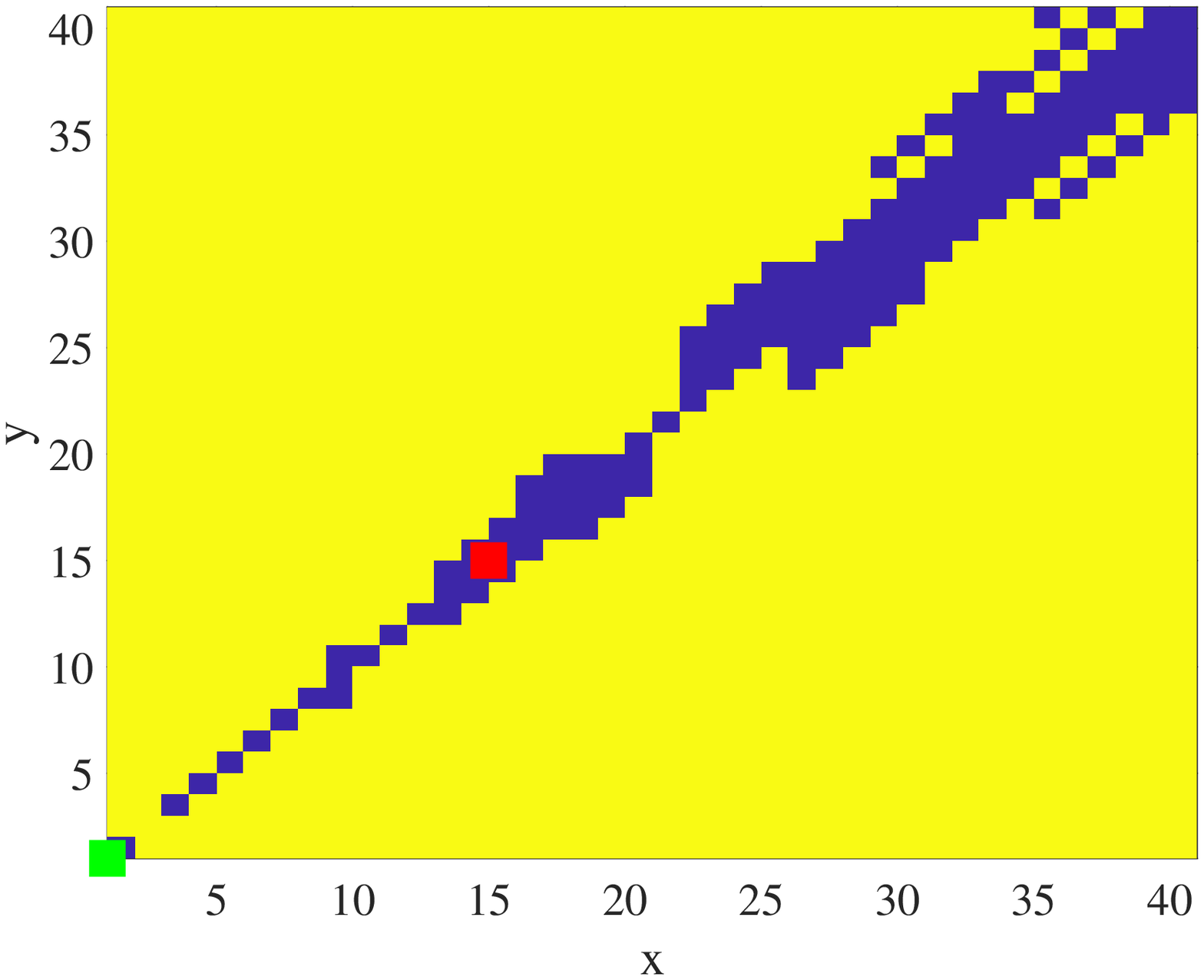}}      
\caption{Coverage for a transmitter placed at $ d\simeq 20\,$m and one interference place in different position. Blue and yellow points denote \ac{SIR} below or above $10\,$dB, respectively.  Left: \ac{no-lens}. Middle: \ac{NR-lens}. Right: \ac{R-lens}. { From top to bottom: $\Ae=100$, $\Ae=150$, $\Ae=200$.}}\label{fig:interf1}
\end{figure}

%without the need of a clock synchronization between the transmitter and the receiver. 

%\subsubsection{NR-lens - ML on the Trajectory}

%We here consider the second approach suitable for \ac{NR-lens} when the number of available processing chains is limited. 
%Again, simulations are performed accounting for the same conditions as in Sec.~\ref{sec:pos}.

%Fig.~\ref{fig:traj} shows the achieved performance, by comparing a traditional \ac{ML} position estimator to the one herein discussed. Indeed, performance is about the same, but with a dramatic reduction of the speed performance of the algorithm in case only few processing chains are available. In fact, the proposed algorithm can operate in a fast manner since only $1$ combination of the chains can be directly associated to a test position, whereas traditional \ac{ML} requires a much large variety of possible combinations to cover a single possible test position. \rednote{Non capisco: non devo di volta in volta attivare un sotto-insieme di antenne per la traiettoria testata? Quante antenne in meno sono richieste?}

%Thus, also in this case, interesting performance can be obtained with a less complex architecture, by exploiting the number of available fingers.

\subsection{{Multi-User Scenario}}
\subsubsection{Single Interference Scenario}
%\rednote{Inizierei prima con il positioning senza interferenza.Anche perche' tutti i parametri di simulazione sono introdotti dopo. }

%\rednote{Il single-user scenario non dice molto, anche per risparmiare spazio sarei per mostrare solo il multi-user.}
We first evaluate the \ac{SIR} when there is another interference source located in the environment. To that purpose, we considered a square room with size $(40\times 40)\,\text{m}^2$, represented with a grid of points with dimension $(1\times 1)\,\text{m}^2$. Then, we fixed the RX (green square marker), rotated of $45^\circ$ oriented towards the center of the room, in $(0,\,0)$ and we located the useful TX (red square marker) in $(15,\,15)$, while alternatively placing the interference in each point for computing the \ac{SIR}.
In particular, we considered a target SIR threshold $\xi^\star = 10\,$dB, which is a typical value in {multi-user} schemes, and we discriminate yellow points from the blue points if the \ac{SIR} is above the threshold (coverage). 
In this sense, results reported in Fig.~\ref{fig:interf1} show the impact of the antenna architecture, when the intended useful source is at a distance of about $20\,$m (red marker) from the receiver (green marker){, and for different $\Ae$.
Notably, the interference effect is such that it is mainly affected by the \ac{AOA} and it is not possible to discriminate the user if an interference one is on the same direction.} 
\begin{figure}[t!]
\psfrag{x}[c][c][0.7]{ $\Delta d$  [m]}
\psfrag{xx}[c][c][0.7]{$\Delta d$ [m]}
\psfrag{y}[c][c][0.7]{\ac{SIR} [dB]}
\psfrag{yy}[c][c][0.7]{\ac{SIR} [dB]}
\psfrag{data11111111111}[c][c][0.65]{\!\!\ac{R-lens}}
\psfrag{a}[c][c][0.65]{\quad \quad \quad$\Ae=200$}
\psfrag{b}[c][c][0.65]{\quad\quad \quad$\Ae=150$}
\psfrag{c}[c][c][0.65]{\quad \quad\quad$\Ae=100$}
\psfrag{data2}[c][c][0.65]{\quad\quad\quad \, \ac{NR-lens}}
\psfrag{data3}[c][c][0.65]{\quad\quad\quad \ac{no-lens}}
\psfrag{data4}[c][c][0.65]{\quad\quad\quad\quad Threshold}
\psfrag{data11111111111111111111}[c][c][0.65]{\!\!\! \!\!\!\!\! \!\!\! \ac{no-lens}, $\Ae=200$}
\psfrag{data22}[c][c][0.65]{\,\quad \quad\quad\quad\quad \quad  \ac{no-lens}, $\Ae=2500$}
\psfrag{data33}[c][c][0.65]{\!\!\quad\quad \quad \quad\quad \quad \quad\ac{no-lens}, $\Ae=10000$}
\psfrag{data44}[c][c][0.65]{ \quad \quad\quad\quad\quad\quad\,\,\ac{no-lens}, $\Ae=40000$}
\psfrag{data55}[c][c][0.65]{\quad \quad\quad\quad\quad\quad\,\,\,\,\,\ac{no-lens}, $\Ae\!=\! 160000$}
\psfrag{data66}[c][c][0.7]{\quad \quad Threshold}
\centerline{
   \includegraphics[width=0.425\linewidth,draft=false]
    {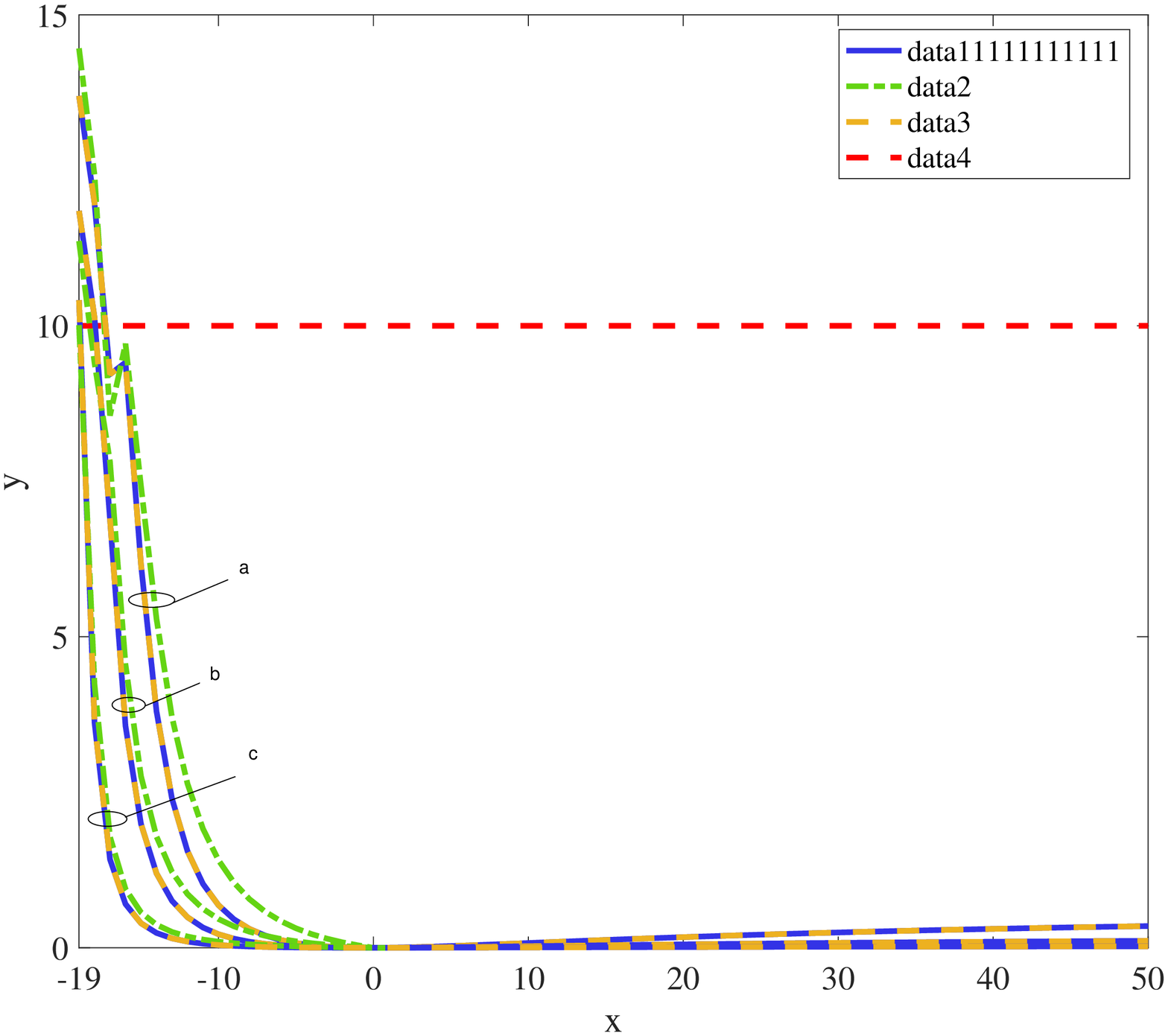}
   \includegraphics[width=0.41\linewidth,draft=false]
    {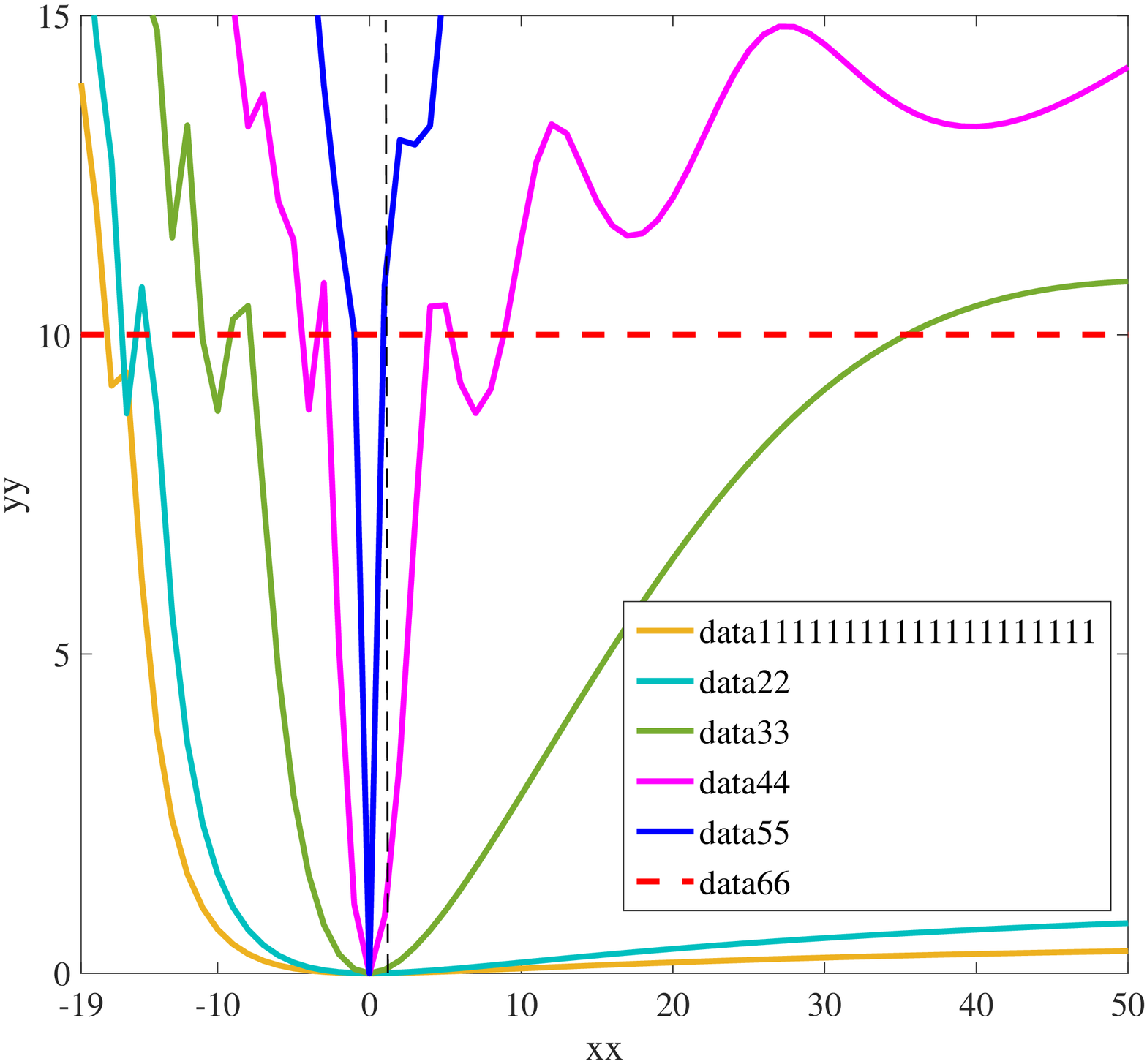}}
\caption{Interference discrimination for fixed $\theta=0^\circ$, with useful distance $d=20\,$m, interference distance $d+\Delta d$, and different $\Ae$. Left: different architectures. Right: \ac{no-lens} case.}\label{fig:int-1} 
\end{figure}

{To better investigate this detrimental effect, in Fig.~\ref{fig:int-1}-left we evaluated the \ac{SIR} by using \eqref{eq:sirgeneral-rlens}, \eqref{eq:sirnrlens_general} and \eqref{eq:sirNolens}, with the source at $d=20\,$m with $\theta=0^\circ$, while positioning the interfering user $\Delta d$ far away from the useful one. More specifically, $\Delta d=0\,$m indicates that the two users are exactly juxtaposed (i.e., $d_i=d$), whereas $\Delta d=-19\,$m means that $d_i \approx 1\,$m.
Indeed, for the three architectures the \ac{SIR} is almost always below the threshold} when the interfering user is $\Delta d$ away from the useful one. We ascribe this effect to the fact that when $d\gg D_z,\,D_y$, it is not possible to discriminate close transmitting sources. This is corroborated also by the results reported in Fig.~\ref{fig:int-1}-right obtained by using \eqref{eq:sirnemp}, where the same conditions are preserved, but $\Ae$, i.e., $\Af$, is increased such that the \ac{SIR} becomes larger than the threshold.
%while maintaining the same conditions translates into a performance improvement, as evidenced in  for the \ac{no-lens}.
%
\begin{figure}[t!]
\psfrag{x}[c][c][0.7]{$ d_i$ [m]}
\psfrag{y}[c][c][0.7]{\ac{SIR} [dB]}
\psfrag{data111111111111111111}[c][c][0.7]{\ac{no-lens}, $\Ae=200$}
\psfrag{data2}[c][c][0.7]{\quad\quad\quad \,\,\quad\quad\quad \quad \,\,\ac{NR-lens}, $\Ae=200$}
\psfrag{data3}[c][c][0.7]{\quad\quad\quad \,\,\quad\quad\quad \quad\! \ac{R-lens}, $\Ae=200$}
\psfrag{data4}[c][c][0.7]{\quad\quad\quad \quad Threshold}
\psfrag{data5}[c][c][0.7]{\quad\quad\quad \,\,\quad\quad\quad \quad \ac{no-lens}, $\Ae=150$}
\psfrag{data6}[c][c][0.7]{\quad\quad\quad \,\,\quad\quad\quad \quad \,\,\ac{NR-lens}, $\Ae=150$}
\psfrag{data7}[c][c][0.7]{\quad\quad\quad \,\,\quad\quad\quad \quad \ac{R-lens}, $\Ae=150$}
\psfrag{data8}[c][c][0.7]{\quad\quad\quad \,\,\quad\quad\quad \quad \,\,\ac{no-lens}, $\Ae=100$}
\psfrag{data9}[c][c][0.7]{\quad\quad\quad \,\,\quad\quad\quad \quad \,\,\ac{NR-lens}, $\Ae=100$}
\psfrag{data10}[c][c][0.7]{\quad\quad\quad \,\,\quad\quad\quad \quad \,\,\ac{R-lens}, $\Ae=100$}
\centerline{
   \includegraphics[width=0.41\linewidth,draft=false]
    {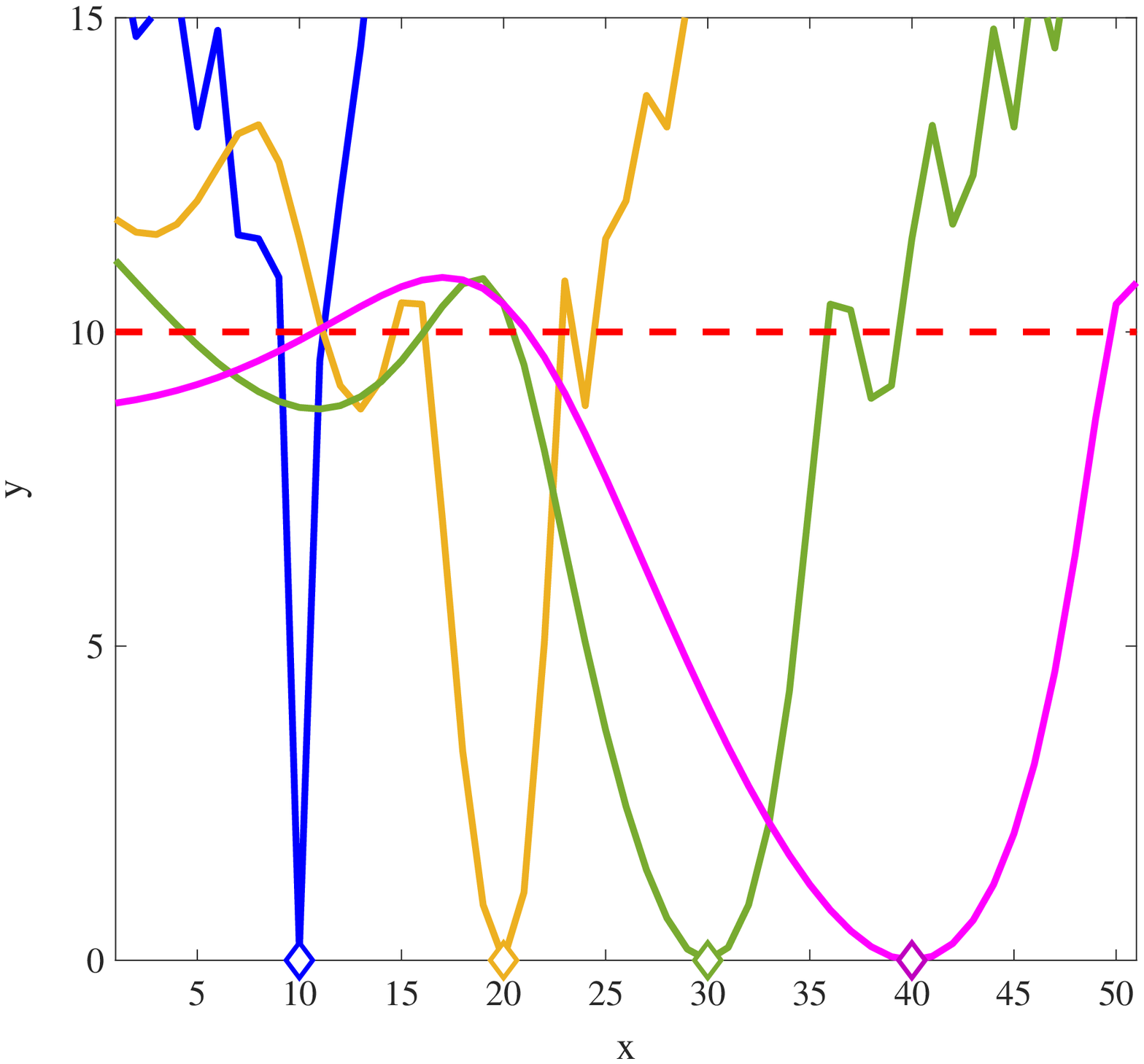} \includegraphics[width=0.41\linewidth,draft=false]
    {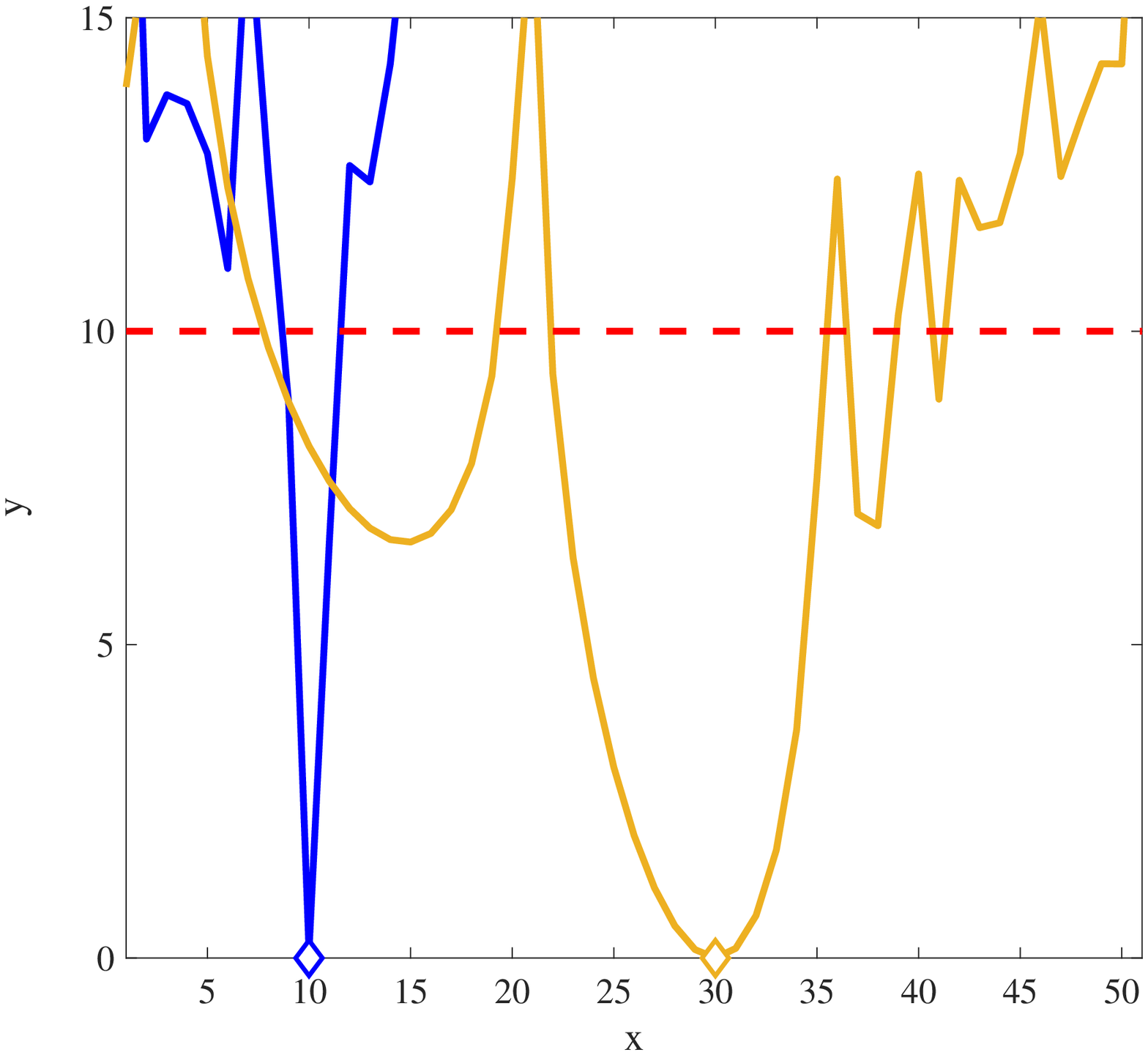}}
\caption{Interference discrimination for $\Ae=40,000$, i.e., $\Af=1\,\text{m}^2$, when the transmitter is in different positions (diamonds), for \ac{no-lens} (left) and \ac{R-lens} (right). Different colors represent the different tested distances.}\label{fig:int-2} 
\end{figure}

%While Fig.~\ref{fig:int-1} has been obtained only for $d=20\,$m in order to further investigate the coverage maps reported in Fig.~\ref{fig:interf1}, 
{To further investigate the problem, in Fig.~\ref{fig:int-2}-left and Fig.~\ref{fig:int-2}-right we considered the approximated expressions in \eqref{eq:sirnemp} (\ac{no-lens}) and in \eqref{eq:vulnRLENS} (\ac{R-lens}), respectively, for $\Af=1\,\text{m}^2$ ($\Ae=40,000$). In particular, we accounted for different values of the useful source distances (diamond markers), while the distance of one interfering source is varied.
As an example, the point of the green curve in $d_i=20\,$m represents the \ac{SIR} value when the useful source is at $d=30\,$m, and $\Delta d=10\,$m.}   
%{Then, we considered the \ac{no-lens} (left) and the \ac{R-lens} (right), with a large area, i.e., $\Af=1\,\text{m}^2$ ($\Ae=40,000$).}
According to Fig.~\ref{fig:int-2}-left, if $d_i=10\,$m and $d +\Delta d>11\,$m or $d- \Delta d<9\,$m, for the \ac{no-lens} the receiver is capable to discriminate the two users only through the position with high precision, since a \ac{SIR} larger than $10\,$dB is guaranteed.  The same considerations are valid for the \ac{NR-lens} in Fig.~\ref{fig:int-2}-right. 
When instead $d=30\,$m, in both figures a larger $\Delta d$ does not automatically imply robustness against the interference. In fact, for $d+\Delta d=15\,$m, it holds  $\mathrm{SIR}<10\,$dB. 

%When $\Narray = 201$ (three maps in the bottom), the performance for the \ac{NR-lens} (bottom-middle) is outperformed by the other two schemes.  

\subsubsection{Multi-Interference Scenario}
Now we consider a scenario with multiple interferers, and the intended useful source alternatively placed in each grid point of the environment. Notably, for each TX position and for each Monte Carlo iteration, interference is generated and distributed in the considered environment according to a \ac{PPP} with intensity $\lambda_p=5$.

\begin{figure}[t!]
\psfrag{x}[b][c][0.7]{$x$ [m]}
\psfrag{y}[b][l][0.7]{$y$ [m]}
\psfrag{50}[b][c][0.55]{$50$}
\psfrag{40}[b][c][0.55]{$40$}
\psfrag{35}[b][c][0.55]{}
\psfrag{30}[b][c][0.55]{$30$}
\psfrag{25}[b][c][0.55]{}
\psfrag{20}[b][c][0.55]{$20$}
\psfrag{15}[b][c][0.55]{}
\psfrag{10}[b][c][0.55]{$10$}
\psfrag{5}[b][c][0.55]{}
\psfrag{1}[c][b][0.55]{$1$}
\psfrag{0.8}[b][c][0.55]{\,\,$0.8$}
\psfrag{0.6}[b][c][0.55]{\,\,$0.6$}
\psfrag{0.4}[b][c][0.55]{\,\,$0.4$}
\psfrag{0.2}[b][c][0.55]{\,\,$0.2$}
\psfrag{0}[b][c][0.55]{$0$}
%\centerline{
%   \includegraphics[width=0.33\linewidth,draft=false]
%    {figures/New/Interference/451_nEMP_Poisson.eps}\includegraphics[width=0.33\linewidth,draft=false]
%    {figures/New/Interference/451_NRL_Poisson.eps}\includegraphics[width=0.33\linewidth,draft=false]
%    {figures/New/Interference/451_RL_Poisson.eps}}  
%\centerline{
%   \includegraphics[width=0.33\linewidth,draft=false]
%    {figures/New/Interference/671_nEMP_Poisson.eps}\includegraphics[width=0.33\linewidth,draft=false]
%    {figures/New/Interference/671_NRL_Poisson.eps}\includegraphics[width=0.33\linewidth,draft=false]
%    {figures/New/Interference/671_RL_Poisson.eps}}  
\centerline{
   \includegraphics[width=0.33\linewidth,draft=false]
    {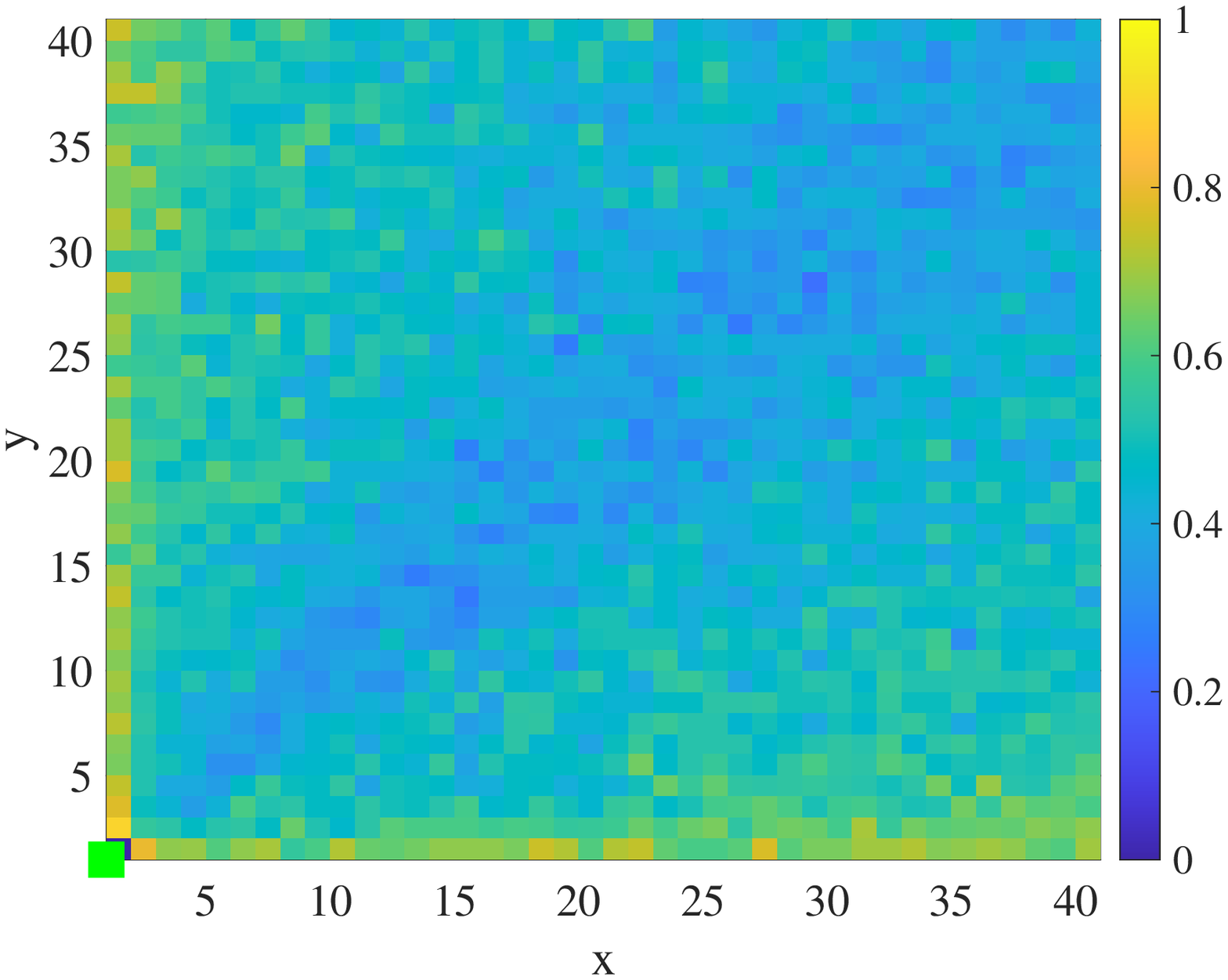}\includegraphics[width=0.33\linewidth,draft=false]
    {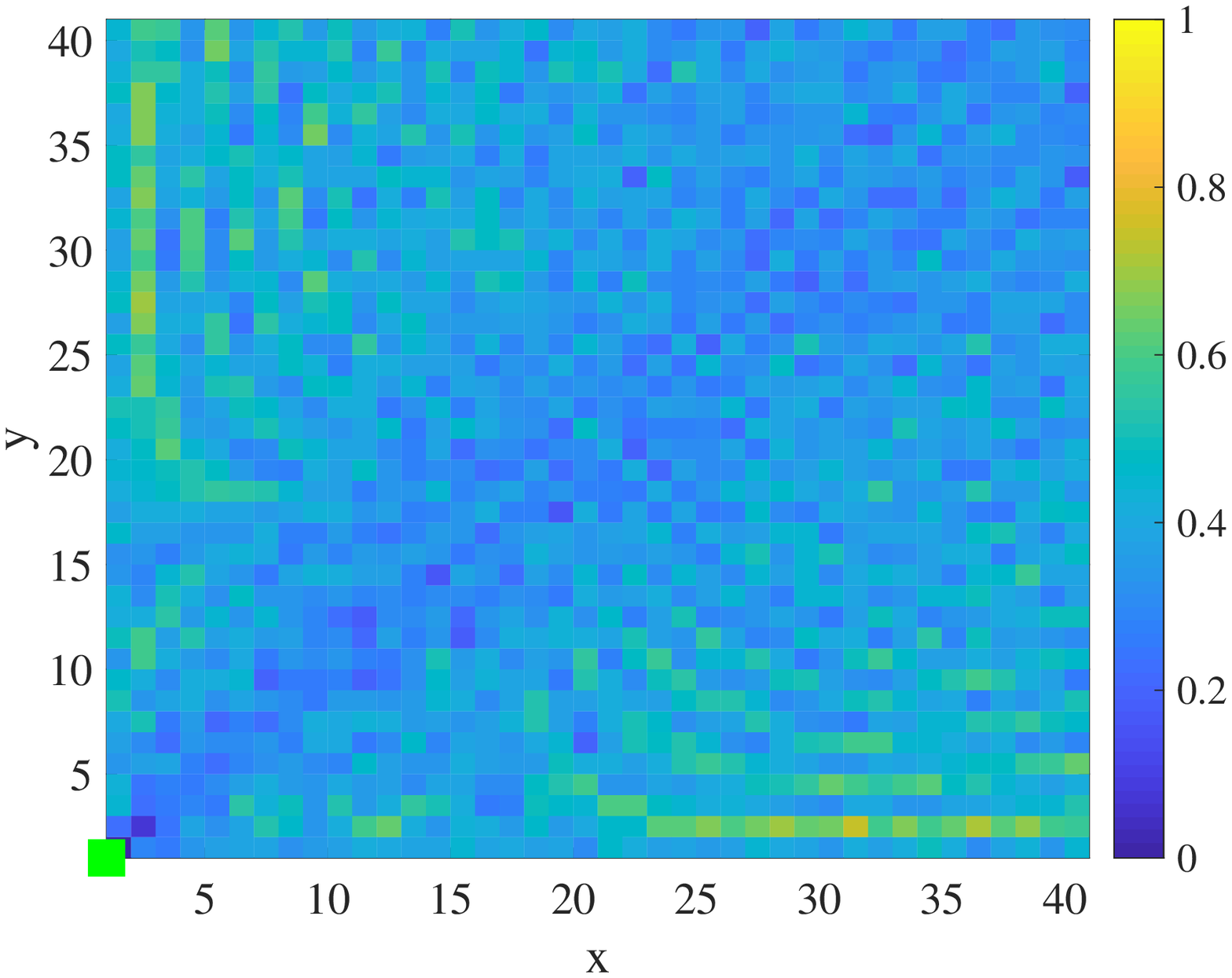}\includegraphics[width=0.33\linewidth,draft=false]
    {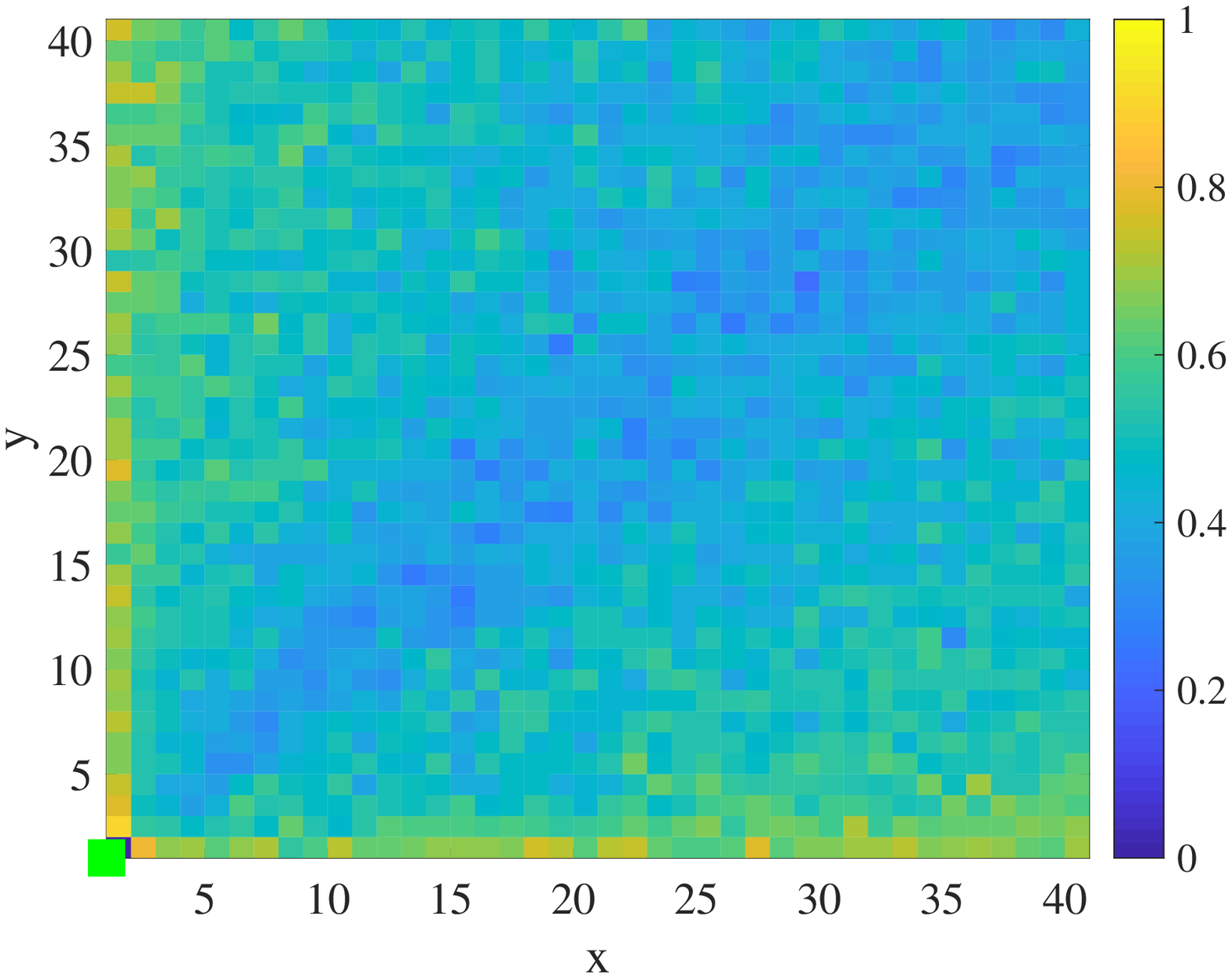}}
    \centerline{
   \includegraphics[width=0.33\linewidth,draft=false]
    {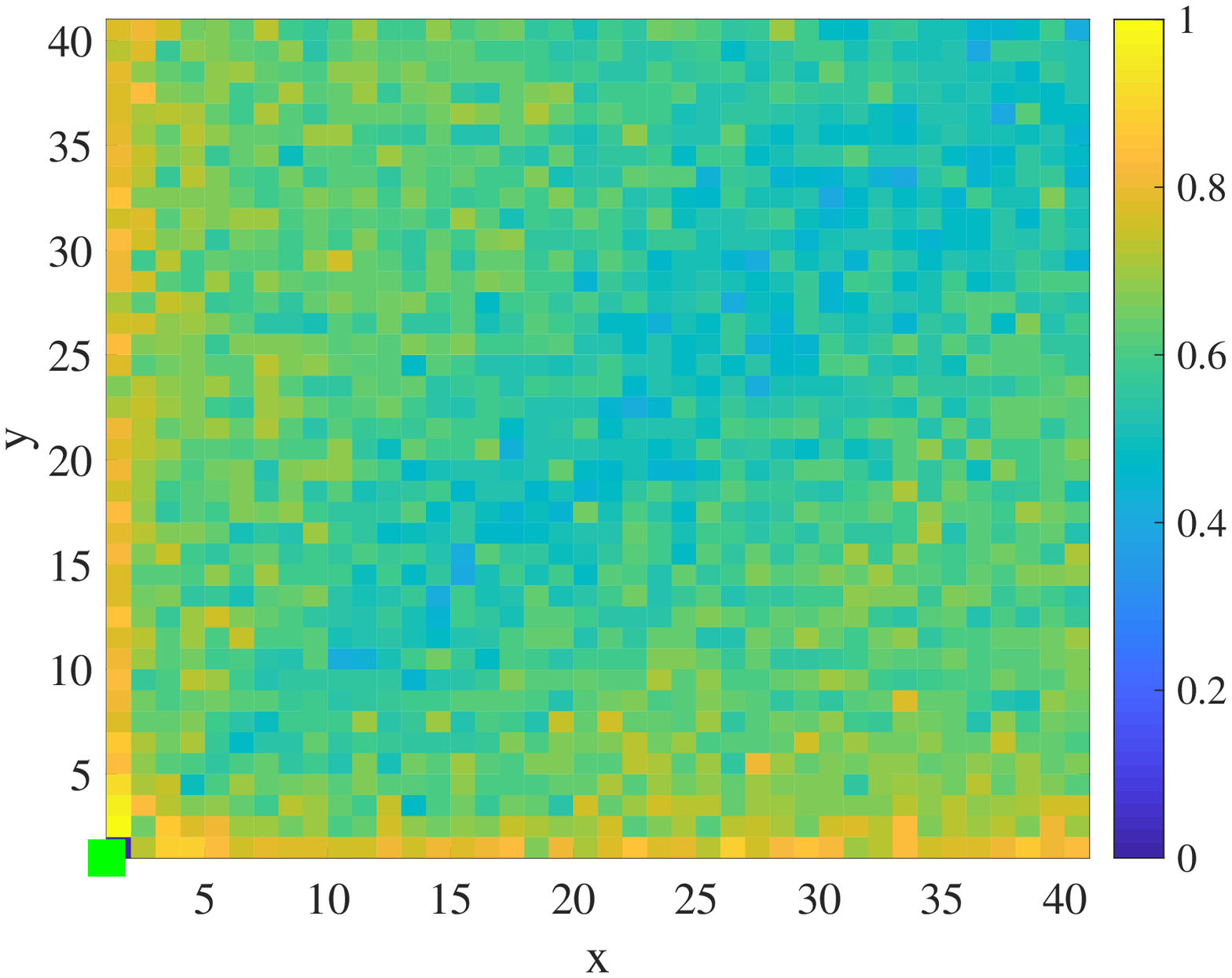}\includegraphics[width=0.33\linewidth,draft=false]
    {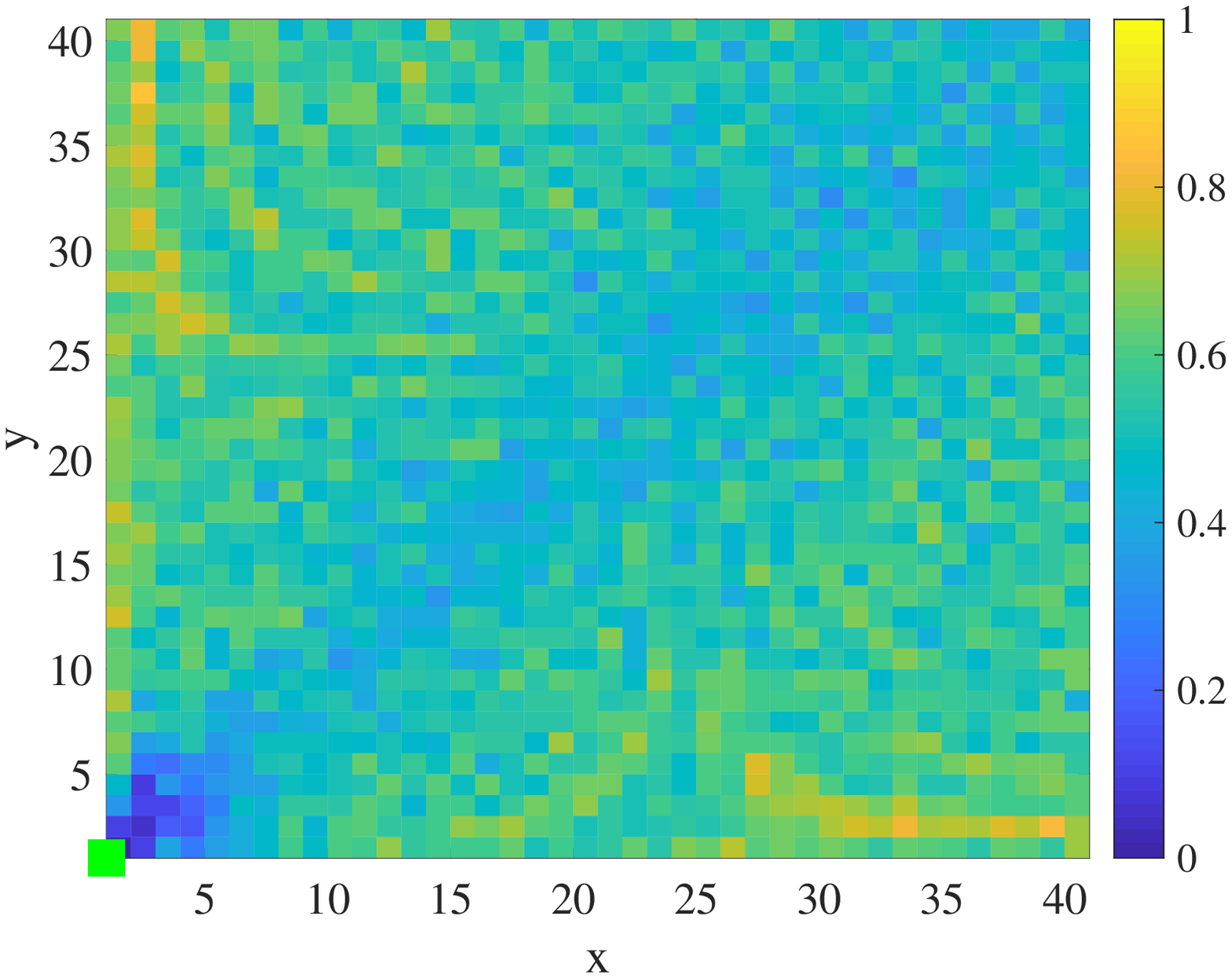}\includegraphics[width=0.33\linewidth,draft=false]
    {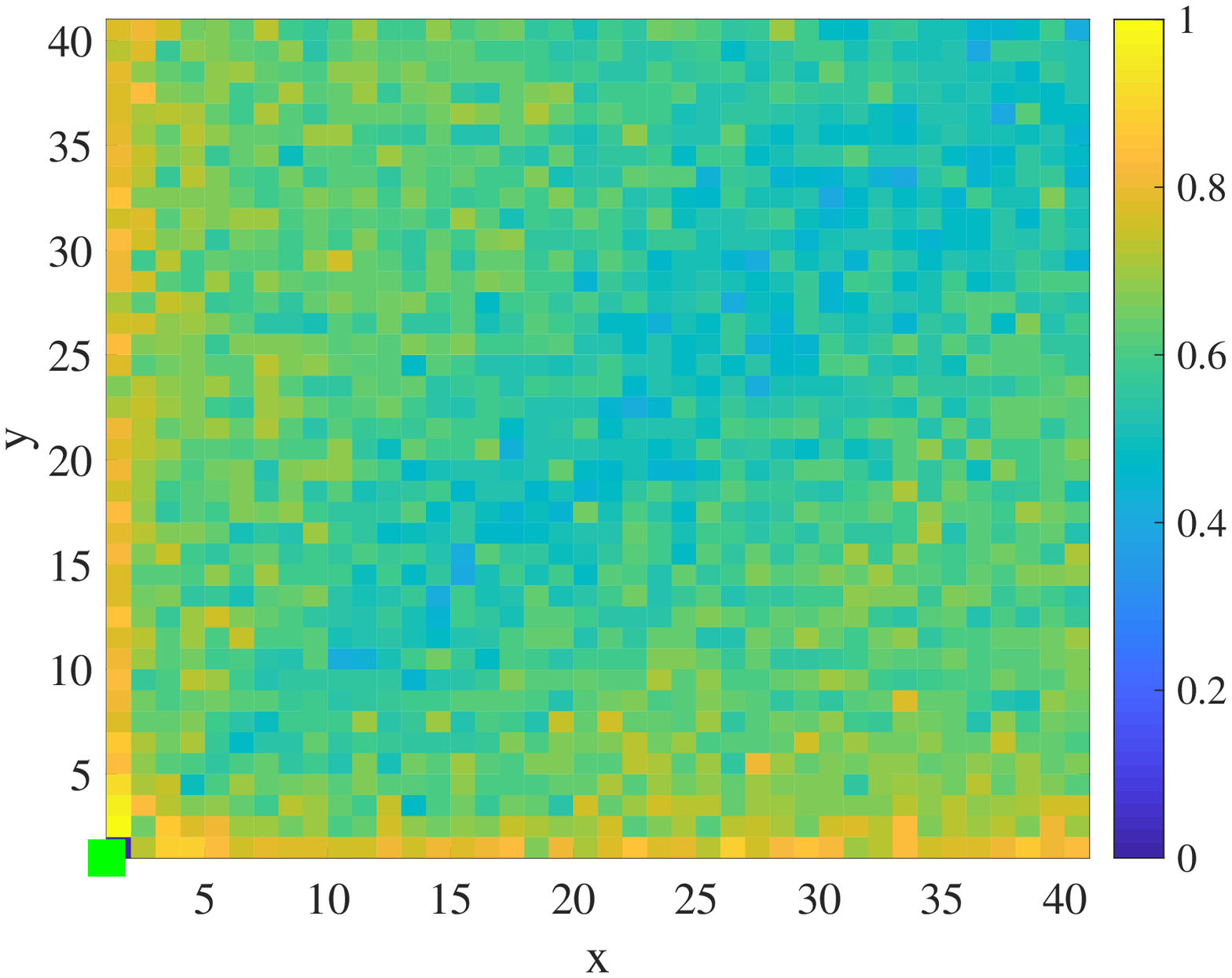}}
     \centerline{
   \includegraphics[width=0.33\linewidth,draft=false]
    {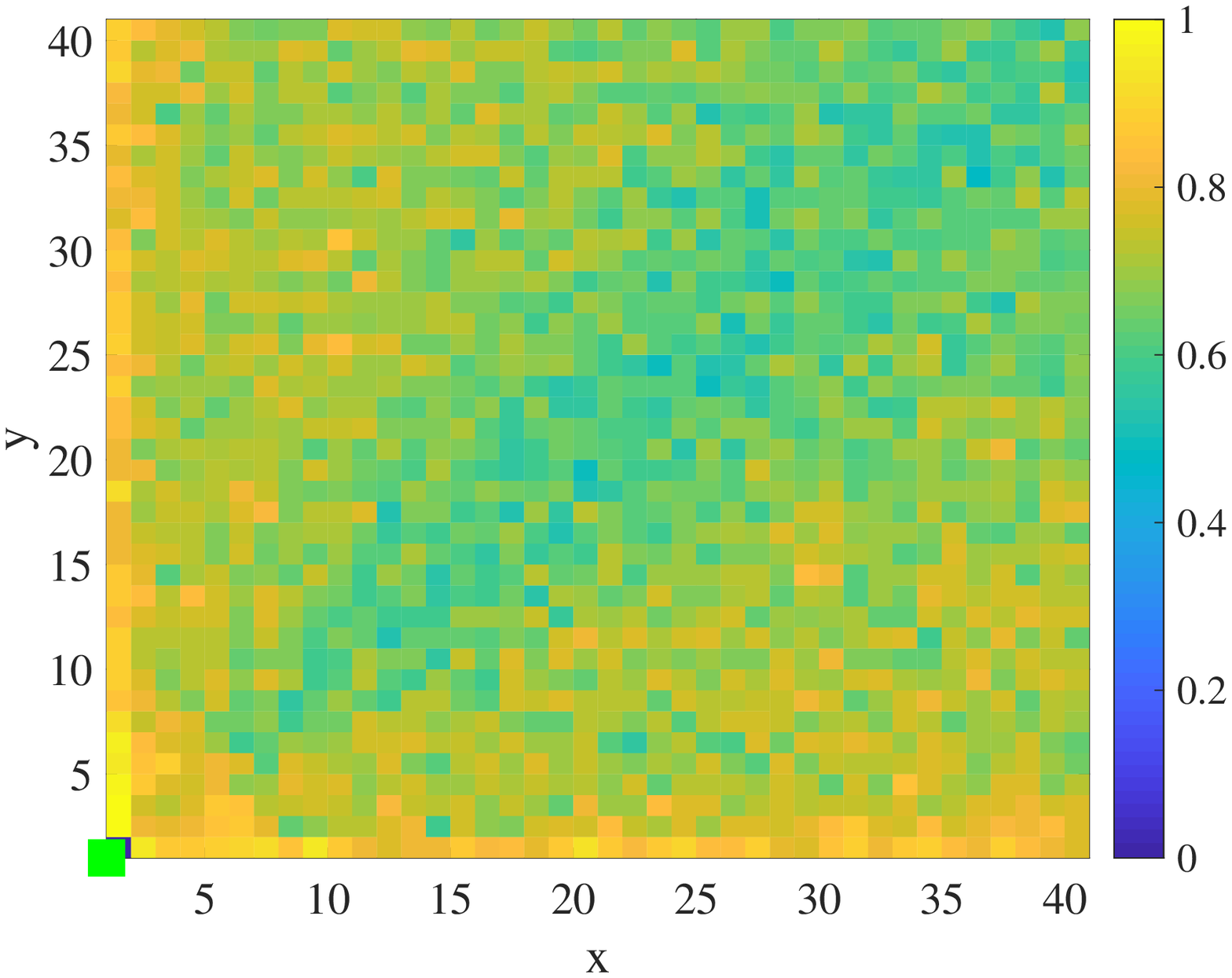}\includegraphics[width=0.33\linewidth,draft=false]
    {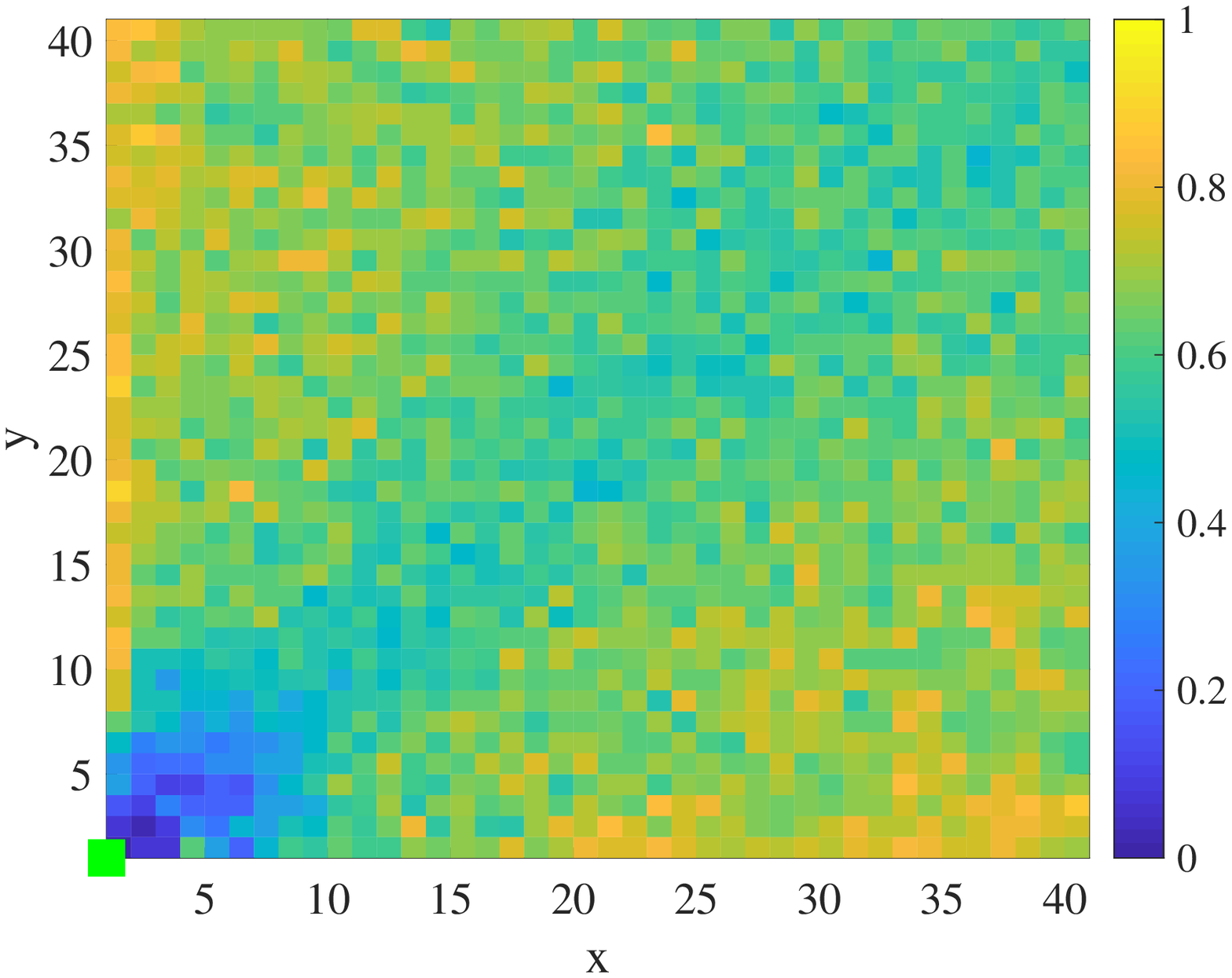}\includegraphics[width=0.33\linewidth,draft=false]
    {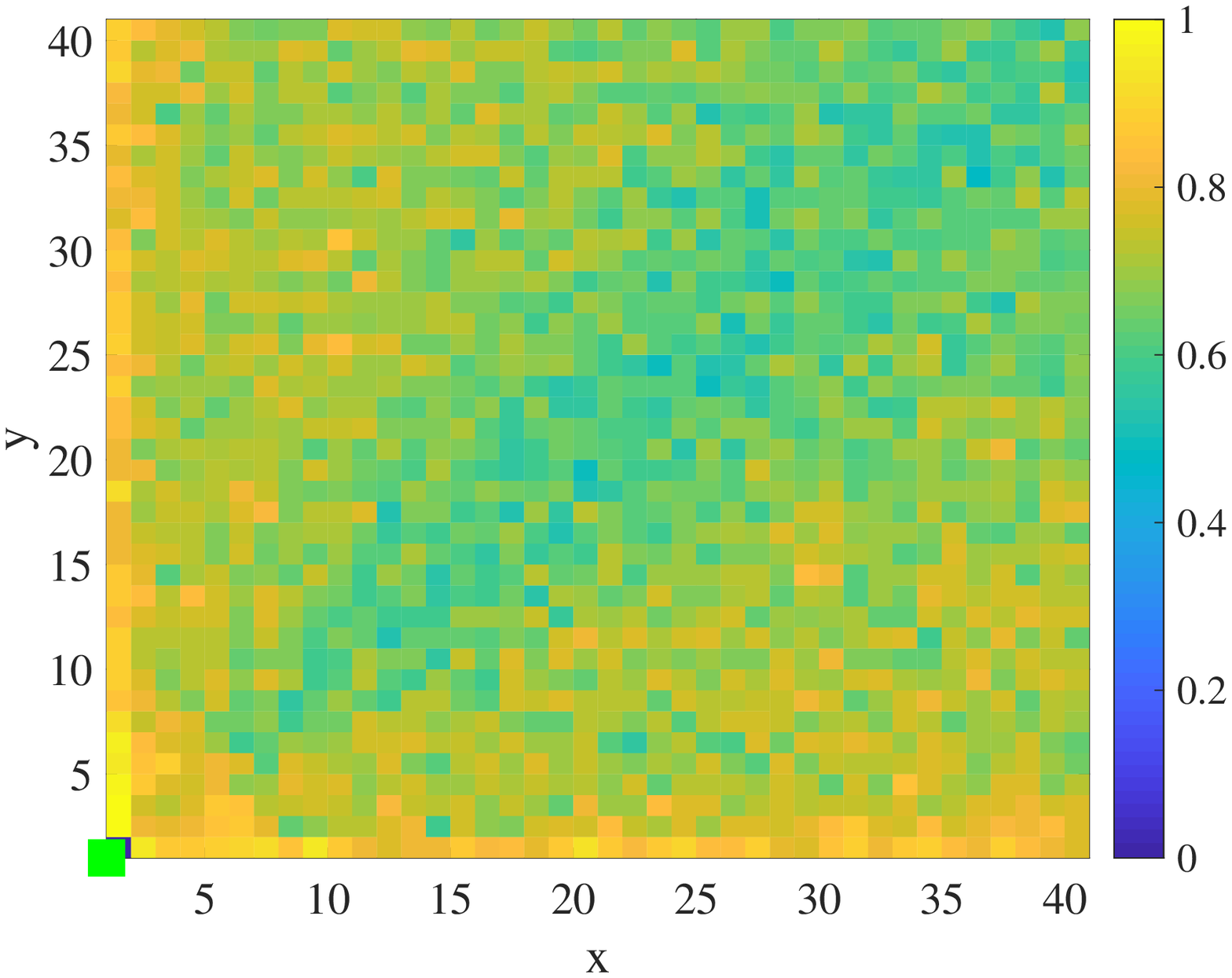}}  
     \centerline{
   \includegraphics[width=0.33\linewidth,draft=false]
    {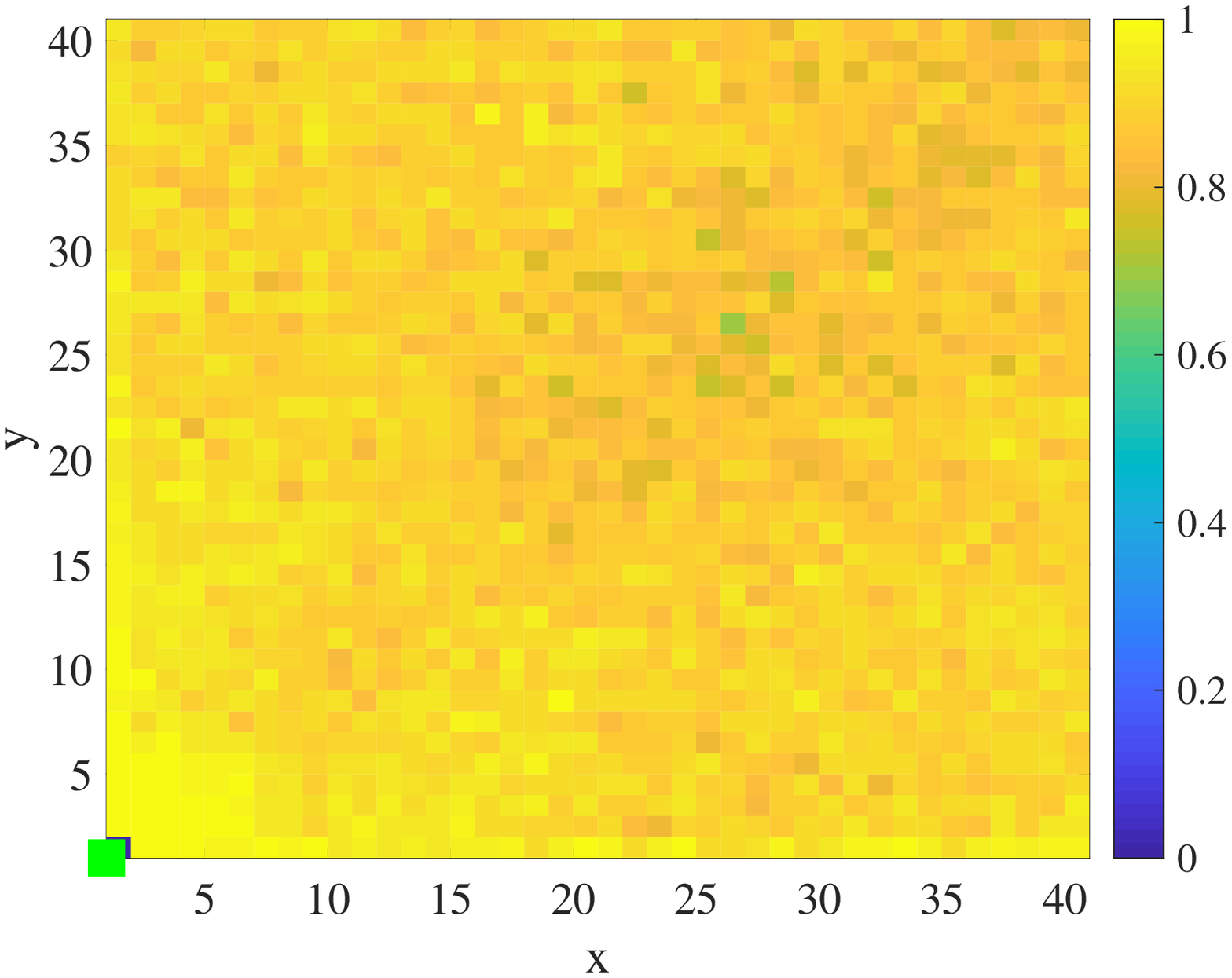}\includegraphics[width=0.33\linewidth,draft=false]
    {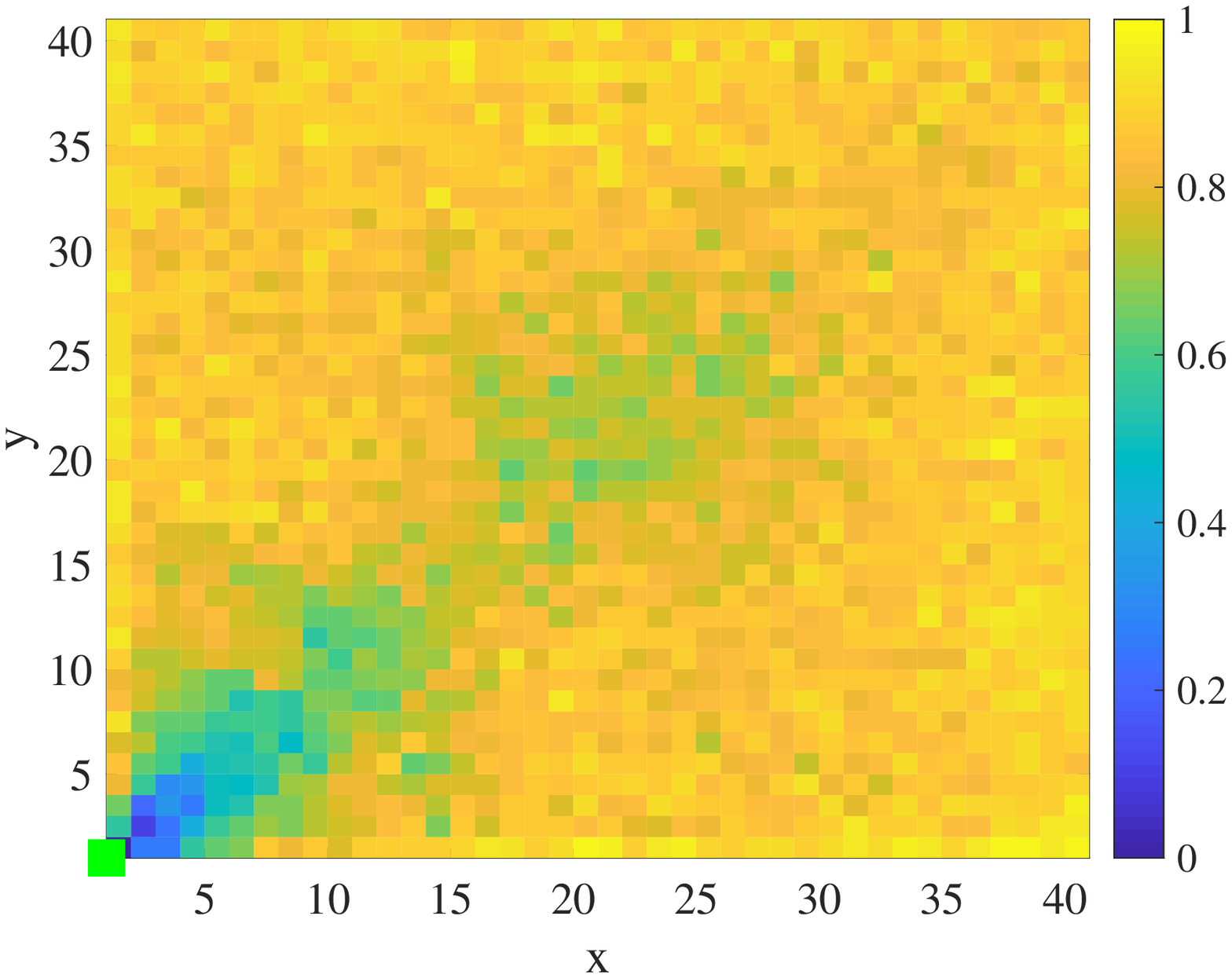}\includegraphics[width=0.33\linewidth,draft=false]
    {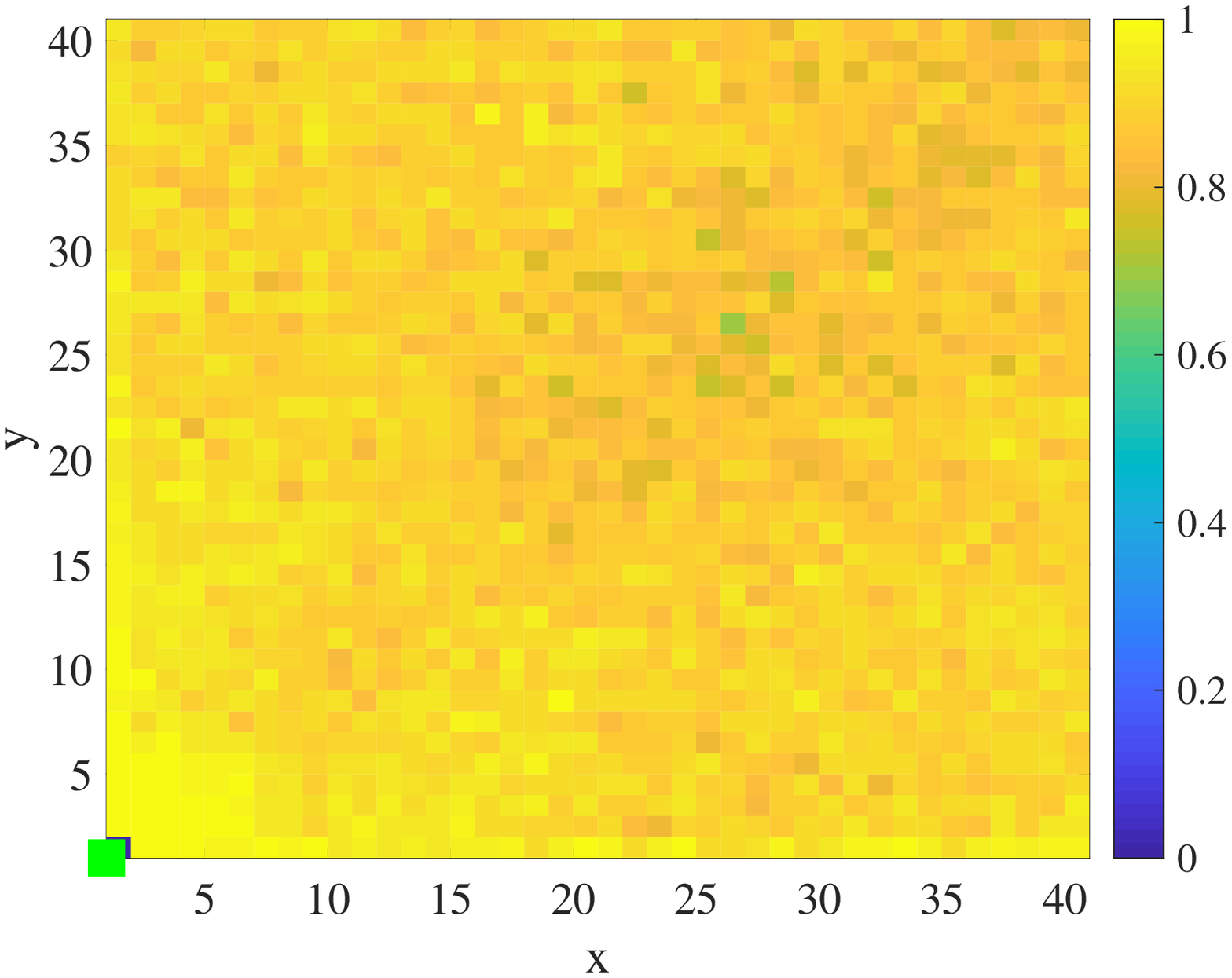}}      
\caption{Coverage rate for interferers distributed as a \ac{PPP}. Left: \ac{no-lens}. Middle: \ac{NR-lens}. Right: \ac{R-lens}. From top to bottom: $\Ae=100$, $\Ae=150$, $\Ae=200$ and $\Ae=10,000$. }\label{fig:intppp}
 \end{figure}

To this purpose, Fig.~\ref{fig:intppp} shows maps for the three architectures (from the left to the right) and for different $\Ae$ (from top to bottom). The value of the points of the map refers to the coverage rate at which $\mathrm{SIR}>\xi^\star=10\,$dB. 
Indeed, now the impact of $\Ae$ is much more evident as before, and there is a certain improvement when solutions based on \ac{R-lens} or \ac{no-lens} are employed, with respect to the \ac{NR-lens}.
The solution with $\Ae=10,000$, i.e., $\Af=1\,\text{m}^2$ denotes a robust interference rejection, even for distances up to $50\,$m.
{These results are very promising towards the possibility of realizing multi-user schemes by taking advantage only on the incident spherical wavefront.}

\section{Conclusion}\label{sec:con}
%\textcolor{red}{DA FARE ALLA FINE}

 In this paper, the possibility to infer the transmitter position from the impinging spherical wavefront curvature at mm-wave has been investigated. This is of crucial importance whenever the TX and the RX are not in far-field, with the advantage that they do not need any ad-hoc synchronization procedure and only a single anchor node is sufficient to estimate the position. To this purpose, we proposed a general model that accounts for the \ac{EM} processing of the wavefront curvature through a lens (if any), followed by an antenna array scheme.

Then, we analyzed the possibility to employ an \ac{EM} lens that can be either reconfigurable or fixed, showing how the performance varies when complexity is put in the lens or in the number of employed antennas. 
In addition, we evaluated also the impact of the multiple-source interference, showing what happens both in a single interference scenario, and in a multiple-interference scenario generated through a \ac{PPP}.
Results show that the considered solutions allow robust source localization using a single receiver, while guaranteeing robustness against multiple-source interference.
Furthermore, our framework allows to determine the trade-off between processing the signal at EM or signal level as a function of the target performance and complexity. 

{Future works will consider also scenarios entailing the presence of multipath or NLOS conditions, where it might become critical to retrieve the position from the spherical wavefront.
}


%\bibliographystyle{IEEEtran}
%\bibliography{Biblio}

\end{document}